\documentclass[
superscriptaddress,
 amsmath,amssymb,
 aps,
floatfix, showkeys
]{revtex4-2}
\usepackage{lineno}
\usepackage{todonotes}
\usepackage{graphicx}
\usepackage{dcolumn}
\usepackage{bm}
\usepackage{subcaption}
\usepackage{ragged2e}
\usepackage{amsthm}
\usepackage{natbib}
\usepackage{booktabs} 
\usepackage{siunitx}
\usepackage{float}

\newtheorem{theorem}{Theorem}
\newtheorem{corollary}[theorem]{Corollary}
\newtheorem{lemma}[theorem]{Lemma}

\AtEndEnvironment{theorem}{\hfill{\Large $\diamond$}}
\AtEndEnvironment{corollary}{\hfill{\Large $\diamond$}}
\AtEndEnvironment{lemma}{\hfill{\Large $\diamond$} }

\newcommand{\rev}[1]{{\color{black}#1}} 

\usepackage[
 colorlinks=true,        
    linkcolor=blue,        
    citecolor=blue,         
    filecolor=black,      
    urlcolor=black
]{hyperref}

\begin{document}
\preprint{APS/123-QED}

\title{Nonlinear input-output analysis of transitional shear flows \\ using small-signal finite-gain $\mathcal{L}_p$ stability}

\author{Zhengyang Wei}
\author{Chang Liu}%
 \email{Corresponding Email: chang$\_$liu@uconn.edu}
\affiliation{%
 School of Mechanical, Aerospace, and Manufacturing Engineering\\
 University of Connecticut, Storrs, 06269, Connecticut
}%

\date{\today}

\begin{abstract} 
Input-output analysis has been widely used to predict the transition to turbulence in wall-bounded shear flows, but it typically does not capture the full nonlinear effects involved. This work employs the Small-Signal Finite-Gain (SSFG) $\mathcal{L}_p$ stability theorem to assess nonlinear input-output stability. This SSFG $\mathcal{L}_p$ stability theorem can predict permissible forcing amplitudes below which a finite nonlinear input-output gain can be maintained. Our analysis employs Linear Matrix Inequalities (LMI) and Sum-of-Squares (SOS) as the primary tools to search for a quadratic Lyapunov function of an unforced nonlinear system. The resulting Lyapunov function can certify the SSFG $\mathcal{L}_p$ stability of a nonlinear input-output system. We demonstrate the applicability of the SSFG $\mathcal{L}_p$ stability theorem using a nine-mode shear flow model with a random body force. The predicted nonlinear input-output $\mathcal{L}_p$ gain (amplification) is consistent with numerical simulations; the $\mathcal{L}_p$ norm of the output from numerical simulations remains bounded by the theoretical prediction from the SSFG $\mathcal{L}_p$ stability theorem, with the gap between simulated and theoretical bounds narrowing as $p \rightarrow \infty$. The input-output gain obtained from the nonlinear SSFG $\mathcal{L}_p$ stability theorem is higher than the linear $\mathcal{L}_p$ gain, which suggests that nonlinearity can significantly amplify small disturbances. Both nonlinear $\mathcal{L}_p$ gain and linear $\mathcal{L}_p$ gain are valid for each $p\in [1,\infty]$, and such generalizability leads to much higher upper bounds on input-output gain than those predicted by linear $\mathcal{L}_2$ gain ($\mathcal{H}_\infty$ norm of transfer function). The SSFG $\mathcal{L}_p$ stability theorem requires the input forcing to be smaller than a permissible forcing amplitude to maintain finite input-output gain, which is an inherently nonlinear behavior that cannot be predicted by linear input-output analysis. We also identify such permissible forcing amplitude using numerical simulations and bisection search, where below such forcing amplitude, the output norm at any time will be lower than a given threshold value. The permissible forcing amplitude identified from the SSFG $\mathcal{L}_p$ stability theorem is conservative but also consistent with that obtained by numerical simulations and bisection search. This study demonstrates that the nonlinear SSFG $\mathcal{L}_p$ stability theorem is a powerful tool for analyzing nonlinear input-output systems such as transitional shear flows with external disturbance. 
\end{abstract}

\maketitle

\section{\label{sec:level1}Introduction}

Understanding and predicting the transition to turbulence of wall-bounded shear flows is essential for a wide range of applications, from improving aerodynamic efficiency to controlling mixing processes in industrial flows \citep{nienow1997mixing, ghoreyshi2014reduced}. One major challenge is characterizing the conditions under which shear flows remain in the laminar state or transition to turbulence. Linear stability analysis, while effective for identifying stability to infinitesimal disturbances, falls short in capturing the transition behavior that arises from finite-amplitude disturbances—a phenomenon often observed in experiments and simulations of wall-bounded shear flows \citep{schmid2012stability, grossmann2000onset, waleffe1995transition,reynolds1883xxix}. This limitation is because linear stability analysis only considers infinitesimal disturbances, neglecting other finite-amplitude disturbances that can trigger transition to turbulence. Finite-amplitude disturbances can be induced by transient growth due to the non-normality of the linearized Navier-Stokes operator, the potential disturbance growth driven by nonlinear dynamics, and external disturbances such as background noise, free-stream disturbances, and experimental uncertainties. Transition to turbulence is strongly sensitive to external disturbances; e.g., the onset of transition to turbulence can be shifted from Reynolds number $\text{Re}=2000$ to $\text{Re}=13,000$ by reducing the inlet disturbance in pipe flow experiments \citep{reynolds1883xxix}.

Input-output analysis \citep{farrell1993stochastic,bamieh2001energy,jovanovic2005componentwise,illingworth2018estimating,vadarevu2019coherent,madhusudanan2019coherent,symon2021energy,liu2020input_convective,liu2019convective} captures the system's transient energy growth mechanisms \citep{schmid2012stability,Reddy_Henningson_1993} and sensitivity to external disturbances \citep{farrell1993stochastic,bamieh2001energy,jovanovic2005componentwise,jovanovic2004modeling} when the base flow is linearly stable. For example, input-output analysis based on singular value decomposition of the resolvent operator or frequency response operators has been used to identify dominant transition-inducing modes \citep{trefethen2020spectra,farrell1993stochastic,bamieh2001energy,jovanovic2005componentwise} and to predict coherent structures in turbulent regimes \citep{mckeon2010critical,mckeon2013experimental,mckeon2017engine}.  Traditional linear input-output analysis is usually computed based on $\mathcal{H}_2$ or $\mathcal{H}_\infty$ norm of spatio-temporal frequency response operator \citep{jovanovic2005componentwise,jovanovic2021bypass}, \rev{which assumes} idealized forcing spectra; e.g., spatio-temporal white noise or harmonic forcing. \rev{In the turbulent regime, the nonlinear effect of small-scale turbulence on large-scale diffusion can be modeled as eddy viscosity within the input-output analysis framework, which allows us to accurately predict the spanwise spacing of near-wall streaks \citep{ALAMO2006, pujals2009note, hwang2010linear, symon2023use}, and estimate large-scale structures in the outer layer at high Reynolds numbers \citep{morra2019relevance,madhusudanan2019coherenta}. However, such eddy viscosity enhancement is not applicable for the transitional regime, and thus, it remains critical to include nonlinearity properly to analyze finite-amplitude disturbances in the transitional regime \citep{Liu2020PRE,kalur2021nonlinear,liu2021structured}. }

Prior finite-amplitude stability analyses of transitional shear flows, such as those based on linear matrix inequalities (LMI), have provided foundational insights into determining regions of attraction  \citep{kalur2021nonlinear,Liu2020PRE,kalur2020stability,kalur2022estimating,kafshgarkolaei2025local,liao2022quadratic}. These approaches constrain nonlinear interactions via conservative quadratic constraints by utilizing componentwise sector bounds on nonlinear terms and the energy conserving property of nonlinearity, allowing us to characterize the region of attraction of the laminar state in shear flows. However, these approaches analyze nonlinear initial value problems \citep{Liu2020PRE,kalur2021nonlinear} without analyzing the sensitivity to external disturbances originating from sustained environmental noise or active control inputs. Structured input-output analysis \citep{liu2021structured,liu2022structured,liu2023structured,shuai2023structured,rath2024structured,bhattacharjee2024structured,mushtaq2023structured,mushtaq2024exact,mushtaq2024structured,frank2025input} provides a tractable approach to model the nonlinearity and identifies transition-inducing disturbances consistent with experiments, direct numerical simulations, and nonlinear optimal perturbations. However, this approach uses structured feedback forcing to model nonlinearity and does not consider the input-output amplification due to external forcing. Recent advances in nonlinear input-output analysis \citep{rigas2021nonlinear} address this gap by solving harmonic-balanced Navier-Stokes equations to identify the nonlinear minimal forcing seed (optimal transition-inducing forcing) in the frequency domain, while the nonlinear input-output gain and input-output stability have not been fully analyzed. Dissipation inequalities and closely related LMI have been used to analyze input-output amplification of nonlinear wall-bounded shear flows \citep{ahmadi2019framework}, while the permissible forcing amplitude was not analyzed in this work.

To bridge this gap, we will employ the Small-Signal Finite-Gain (SSFG) $\mathcal{L}_p$ stability theorem \citep[Theorem 5.1]{Khalil2002} to assess the nonlinear input-output amplification of input disturbances in the presence of the system's nonlinearity. The nonlinear SSFG $\mathcal{L}_p$ stability theorem provides a framework for quantifying a finite input-output gain (amplification) when input forcing amplitude is smaller than a permissible forcing amplitude (i.e., small-signal). Such a finite permissible forcing amplitude is an inherently nonlinear phenomenon, which cannot be predicted by linear input-output analysis due to linear superposition. The SSFG $\mathcal{L}_p$ stability theorem allows for the assessment of system behavior under finite initial disturbances and finite forcing amplitude, making it a valuable tool for analyzing nonlinear input-output systems such as transitional shear flows that are sensitive to initial disturbance and external forcing. This paper aims to demonstrate the applicability of the SSFG $\mathcal{L}_p$ stability theorem in transitional shear flows to identify a permissible forcing amplitude below which a finite input-output gain is guaranteed. 

We will analyze a nine-mode shear flow model \citep{moehlis2004low,Moehlis2005} describing flow between infinite parallel free-slip walls driven by a sinusoidal body force. This nine-mode shear flow model offers tractable alternatives to model shear flows, enabling accurate reproduction of subcritical transitions, intermittent turbulence, and exponential lifetime distributions observed experimentally \citep{moehlis2004low}. This nine-mode shear flow model is instrumental in analyzing subcritical transition to turbulence, which has been used to quantify the permissible perturbation amplitude to sustain the laminar state \citep{Joglekar2015,Liu2020PRE,kalur2022estimating} and study the geometric structure of chaotic phase-space boundaries in shear flows \citep{Joglekar2015}. The bifurcation structure of this shear flow model \citep{Moehlis2005} further highlights the emergence of unstable periodic orbits and connections between transient chaos and turbulent lifetimes. We will add a random body force in this nine-mode shear flow model to represent external disturbance for nonlinear input-output stability analysis.

We will use linear matrix inequalities (LMI) and sum-of-squares (SOS) to search for Lyapunov functions of unforced systems to certify nonlinear SSFG $\mathcal{L}_p$ stability. LMI has been widely used in nonlinear stability analysis of shear flows \citep{kalur2021nonlinear,Liu2020PRE,kalur2020stability,kalur2022estimating,ahmadi2019framework}. We will use quadratic constraints to describe the local upper bounds of nonlinearity as Ref. \citep{Liu2020PRE}, which can then be directly included in LMI to certify the nonlinear input-output $\mathcal{L}_p$ stability against a finite forcing amplitude. The Lyapunov function of the unforced system can also be found by SOS optimization, which explicitly accounts for the quadratic nonlinearity of fluid systems. The SOS method has been previously used for rigorous verification of global stability beyond conservative energy-based stability analysis \cite{chernyshenko2014polynomial,GOULART2012692,huang2015sum,fuentes2022global}. Here, we will search the Lyapunov function of unforced systems by LMI and SOS to certify the nonlinear SSFG $\mathcal{L}_p$ stability, identifying the permissible forcing amplitude and finite input-output $\mathcal{L}_p$ gain.

The paper is organized as follows. Sec. \ref{sec:Problem setup} introduces the nine-mode shear flow model and the nonlinear SSFG $\mathcal{L}_p$ stability theorem to determine permissible forcing amplitudes below which a finite input-output $\mathcal{L}_p$ gain can be maintained. It also presents the methods used to search for a Lyapunov function of the unforced nonlinear system to certify SSFG $\mathcal{L}_p$ stability, including LMI and SOS approaches. In Sec. \ref{sec:nonlinear}, we validate our SSFG $\mathcal{L}_p$ stability prediction with numerical simulations and compare input-output gain over the Reynolds number based on SSFG $\mathcal{L}_p$ stability with that predicted by various linear input-output analyses (linear finite-gain $\mathcal{L}_p$ stability and linear $\mathcal{L}_2$ stability). We also compare the permissible forcing amplitude from the SSFG $\mathcal{L}_p$ stability theorem with that found through extensive numerical simulations and bisection search. Finally, the conclusions are presented in Sec. \ref{sec:conclusion}.

\section{\label{sec:Problem setup}Problem setup}

\begin{figure}[!htbp]
\centering
\includegraphics[width=0.8\linewidth]{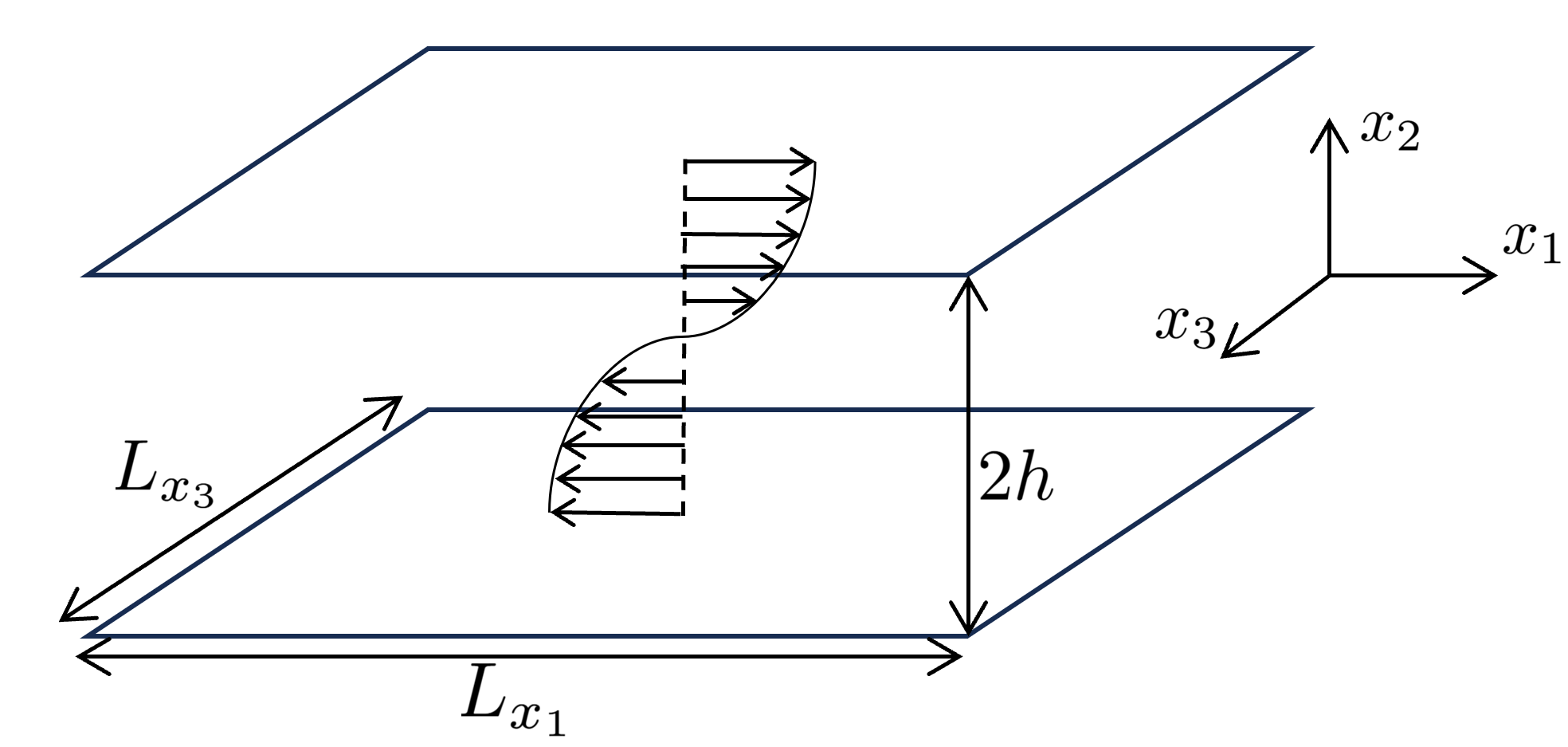} 
\caption{\label{fig:shear_flow} %
Illustration of sinusoidal shear flow following \citet{moehlis2004low,Moehlis2005}.}
\end{figure}

The geometry for sinusoidal shear flow is shown in Figure \ref{fig:shear_flow} \citep{waleffe1997self, moehlis2004low, Moehlis2005}. We take coordinates with $x_1$ in the streamwise direction, $x_2$ in the wall-normal direction, and $x_3$ in the spanwise direction \citep{Moehlis2005}. We use the same model coefficients as in Ref. \citep{moehlis2004low} associated with the dimensionless domain size of $L_{x_{1}}/h$ = 1.75$\pi$ and $L_{x_{3}}/h$ = 1.2$\pi$ normalized by the channel half height $h$. This nine-mode shear flow model \citep{moehlis2004low,Moehlis2005} extends the eight-mode shear flow model presented in Waleffe \citep{waleffe1997self} by incorporating a ninth mode, which accounts for changes to the mean velocity profile caused by turbulent effects. After projecting the Navier-Stokes equations onto the basis defined by \citet{moehlis2004low,Moehlis2005} and adding a forcing term $\boldsymbol{f}$ modeling external disturbance, we can express the governing equations of the nonlinear input-output systems as:
\begin{align}
    \begin{split}
\dot{\boldsymbol{a}}=& -\frac{\mathbf{\Xi}}{\mathrm{Re}} \boldsymbol{a} + \boldsymbol{J}(\bar{\boldsymbol{a}})\boldsymbol{a} + \boldsymbol{J}(\boldsymbol{a}) \bar{\boldsymbol{a}} + \boldsymbol{J}(\boldsymbol{a}) \boldsymbol{a} + \boldsymbol{f}  \\
\mathrel{\mathop:} =& \boldsymbol{\mathcal{N}}(t, \boldsymbol{a}, \boldsymbol{f}). 
    \end{split}
    \label{eq:dynamical_eq_2}
\end{align}
In equation \eqref{eq:dynamical_eq_2}, $\boldsymbol{a} \in \mathbb{R}^n$ represents the state variable, $\dot{\boldsymbol{a}}$ means the time derivative of $\boldsymbol{a}$, $\bar{\boldsymbol{a}}$ denotes the laminar base flow solution, $-\frac{\mathbf{\Xi}}{\mathrm{Re}} \boldsymbol{a}$ is the viscous term, $\boldsymbol{J}(\bar{\boldsymbol{a}})\boldsymbol{a}$ and $\boldsymbol{J}(\boldsymbol{a})\bar{\boldsymbol{a}}$ respectively represent the advection by the laminar mean flow and the mean shear term, and $\boldsymbol{J}(\boldsymbol{a})\boldsymbol{a}$ is the nonlinear term \citep{Liu2020PRE}. The details of these terms are provided in Appendix \ref{appendix:a}, which also includes the derivation of equation \eqref{eq:dynamical_eq_2}. The body force term $\boldsymbol{f} \in \mathbb{R}^{n}$ is our system input, which can be used to model experimental uncertainty, background noise, or free-stream disturbance. Combining equation \eqref{eq:dynamical_eq_2} with an output variable $\boldsymbol{y}(t)\in \mathbb{R}^n$, we can write the nonlinear input-output systems as:
\begin{subequations}
\label{eq:simp_sys}
\begin{align}
\dot{\boldsymbol{a}} =& \boldsymbol{\mathcal{N}}(t, \boldsymbol{a}, \boldsymbol{f}),
\label{eq:simp_sys_1} \\
\boldsymbol{y} =& \boldsymbol{h}(t, \boldsymbol{a}, \boldsymbol{f}) := \boldsymbol{a},
\label{eq:simp_sys_2}
\end{align}
\end{subequations}
where $\boldsymbol{y} \in \mathbb{R}^n$ is the output variable, which is chosen to be the same as the state variable $\boldsymbol{a}$ as shown in equation \eqref{eq:simp_sys_2}. The initial condition is denoted as $\boldsymbol{a}(t=0)=\boldsymbol{a}_0$. We also have $\boldsymbol{a} = \boldsymbol{0}$ representing the laminar state \citep{moehlis2004low,Moehlis2005} that is an equilibrium point of the unforced system:
\begin{align}
\dot{\boldsymbol{a}} = \boldsymbol{\mathcal{N}}(t, \boldsymbol{a}, \boldsymbol{0}). 
\label{eq:initial}
\end{align}

We can group the linear term in \eqref{eq:dynamical_eq_2} together to have a linear operator  $\boldsymbol{L} \in \mathbb{R}^{n\times n}$ such that $\boldsymbol{L}\boldsymbol{a}= -\frac{\mathbf{\Xi}}{\mathrm{Re}} \boldsymbol{a} + \boldsymbol{J}(\bar{\boldsymbol{a}})\boldsymbol{a} + \boldsymbol{J}(\boldsymbol{a}) \bar{\boldsymbol{a}} $, which allows us to rewrite \eqref{eq:simp_sys} as 
\begin{subequations}
\label{eq:simp_eq}    
\begin{align}
   \dot{\boldsymbol{a}}=& \boldsymbol{L}\boldsymbol{a}+\boldsymbol{\Upsilon}(\boldsymbol{a}) + \boldsymbol{f},\label{eq:simp_eq_1} \\
   \boldsymbol{y}=&\boldsymbol{a},
    \label{eq:simp_eq_2}
\end{align}
\end{subequations}
where $\boldsymbol{\Upsilon}(\boldsymbol{a}) \in \mathbb{R}^{n} $ is the nonlinear term such that $\boldsymbol{\Upsilon}(\boldsymbol{a}):= \boldsymbol{J}(\boldsymbol{a})\boldsymbol{a}$. Here, $\boldsymbol{L}$ is assumed to be Hurwitz; i.e., all eigenvalues have negative real parts, as commonly observed in subcritical regimes of transitional shear flows.

To introduce the small-signal finite-gain (SSFG) $\mathcal{L}_p$ stability theorem, we first need to define the following notations. The truncated input variable $\boldsymbol{f}_\tau$ is a truncation of the input variable $\boldsymbol{f}$ defined by
\begin{align}
\boldsymbol{f}_\tau(t)=\left\{\begin{array}{cc}
\boldsymbol{f}(t), & 0 \leq t \leq \tau \\
\boldsymbol{0}, & t>\tau.
\end{array}\right.
\label{eq:trucated}
\end{align}
Throughout this paper, we select $\tau$ = $7\times 10^4$ that corresponds to the maximal simulation time considered here, which is large enough to reach statistically steady states as shown later in \S \ref{sec:results}. We then define the vector norm $\| \boldsymbol{f}_{\tau}(t)\|$ as
\begin{align}
    \|\bm{f}_{\tau}(t)\|:= \sqrt{\sum_{i=1}^{n} |f_{\tau,i}(t)|^2},
    \label{eq:vector_norm}
\end{align}
where $f_{\tau,i}(t)$ is the i$^\text{th}$ element of $\bm{f}_{\tau}(t)$. The $\mathcal{L}_p$ norm ($1 \leq p\leq \infty$) of the truncated input $\boldsymbol{f}_\tau$ is defined as 
\begin{align}
\|\boldsymbol{f}_{\tau}\|_{\mathcal{L}_p} := \left( \int_{0}^{\infty} \|\boldsymbol{f}_{\tau}(t)\|^p \, dt \right)^{1/p}, \;\; p \in [1,\infty),
\label{eq:Lpnorm}
\end{align}
and the $\mathcal{L}_\infty$ norm of the $\boldsymbol{f}_{\tau}$ is defined as:
\begin{align}
\|\boldsymbol{f}_{\tau}\|_{\mathcal{L}_{\infty}} :=\sup _{t \geq 0}\|\boldsymbol{f}_{\tau}(t)\|.
\label{eq:L_infty}
\end{align}
\rev{If a signal has a bounded $\mathcal{L}_p$ norm ($\|\boldsymbol{f}_\tau\|_{\mathcal{L}_p}<\infty$), it belongs to the corresponding $\mathcal{L}_p$ space, meaning that it satisfies specific integrability or boundedness conditions.} For example, if $\|\boldsymbol{f}_\tau\|_{\mathcal{L}_\infty}<\infty$, then we have $\boldsymbol{f}_\tau$ is in the space of piecewise continuous, bounded functions ($\mathcal{L}_\infty$ space). Moreover, $\|\boldsymbol{f}_\tau\|_{\mathcal{L}_2}<\infty$ will define the space of piecewise continuous, square-integrable functions ($\mathcal{L}_2$ space). Throughout this paper, we explicitly add the subscript $\mathcal{L}_p$ to denote the $\mathcal{L}_p$ norm as defined in \eqref{eq:Lpnorm}-\eqref{eq:L_infty}, while $\|\cdot\|$ without the subscript will refer to the vector norm in \eqref{eq:vector_norm}. 

We define the $\mathcal{L}_p$ space as the set of all piecewise continuous functions $\boldsymbol{f}_\tau(t)$: $[0,\infty)\rightarrow \mathbb{R}^n$ with a finite $\mathcal{L}_p$ norm: 
\begin{align}
    \mathcal{L}_p:=\{\boldsymbol{f}_\tau \; | \; \|\boldsymbol{f}_\tau\|_{\mathcal{L}_p}<\infty \}. 
\end{align}
We can also define the extended space $\mathcal{L}_{pe}$ as
\begin{align}
\mathcal{L}_{pe}=\left\{\bm{f} \mid \bm{f}_\tau \in \mathcal{L}_p, \forall \tau \in[0, \infty)\right\},
\end{align}
where the extended space $\mathcal{L}_{pe}$ is a linear space that contains the unextended space $\mathcal{L}_p$ as a subset. The extended space $\mathcal{L}_{pe}$ allows us to deal with unbounded ``ever-growing" signals in the SSFG $\mathcal{L}_p$ theorem \cite{Khalil2002}. For example, $f(t)=t$ does not belong to the space $\mathcal{L}_\infty$, but the truncated signal $f_\tau(t)$ belongs to $\mathcal{L}_\infty$ for every finite $\tau$. Thus, we have $f(t)=t$ belongs to the extended space $\mathcal{L}_{\infty e}$. To define the truncated output variable $\boldsymbol{y}_\tau$, the vector norm of output $\|\boldsymbol{y}_\tau\|$, and the $\mathcal{L}_p$ norm of the output $\|\boldsymbol{y}_{\tau}\|_{\mathcal{L}_p}$ ($p\in[0,\infty]$), we can repeat the above procedure as equations \eqref{eq:trucated}-\eqref{eq:L_infty} by replacing $\boldsymbol{f}$ with $\boldsymbol{y}$.

\subsection{\label{sec:NonlinearSSFG} Nonlinear SSFG $\mathcal{L}_p$ stability theorem}

This subsection will introduce the nonlinear SSFG $\mathcal{L}_p$ stability theorem \cite{Khalil2002}. We will then use linear matrix inequalities \cite{Liu2020PRE} to search for a Lyapunov function to certify nonlinear SSFG $\mathcal{L}_p$ stability \cite{Khalil2002}. This nonlinear SSFG $\mathcal{L}_p$ stability theorem will provide a permissible forcing amplitude below which a finite input-output amplification is maintained, as presented in  Theorem~\ref{thm:small_signal_stability} below:
\begin{theorem} \cite[Theorem 5.1]{Khalil2002}
\label{thm:small_signal_stability}
Consider the nonlinear input-output system in \eqref{eq:simp_sys_1}-\eqref{eq:simp_sys_2} and take $\delta>0$ and $r_u>0$ such that $\{\|\boldsymbol{a}\| \leq \delta\} \subset D$ and $\left\{\|\boldsymbol{f}\| \leq r_u\right\} \subset D_u$. Suppose that 
\begin{itemize}
    \item $\boldsymbol{a}=\boldsymbol{0}$ is an exponentially stable equilibrium point of unforced system in \eqref{eq:initial}, and there is a Lyapunov function $V(t, \boldsymbol{a})$ that satisfies
\begin{subequations}
\label{eq:theorem1_Lyapunov}
\begin{align}
c_1 \|\boldsymbol{a}\|^2 \leq V(t, \boldsymbol{a}) \leq& c_2 \|\boldsymbol{a}\|^2,\label{eq:theorem1_Lyapunov_a}
\\
\frac{\partial V}{\partial t} + \frac{\partial V}{\partial \boldsymbol{a}} \boldsymbol{\mathcal{N}}(t, \boldsymbol{a}, \boldsymbol{0}) \leq& - c_3 \|\boldsymbol{a}\|^2, \label{eq:theorem1_Lyapunov_b}
\\
\left\| \frac{\partial V}{\partial \boldsymbol{a}} \right\| \leq& c_4 \|\boldsymbol{a}\|,\label{eq:theorem1_Lyapunov_c}
\end{align}
\end{subequations}
for all $(t, \boldsymbol{a}) \in[0, \infty) \times D$ for some positive constants $c_1$, $c_2$, $c_3$, and $c_4$.
\item $\boldsymbol{\mathcal{N}}$ and $\boldsymbol{h}$ in equations \eqref{eq:simp_sys_1}-\eqref{eq:simp_sys_2} satisfy the inequalities
\begin{subequations}
\begin{align}
\|\boldsymbol{\mathcal{N}}(t, \boldsymbol{a}, \boldsymbol{f})-\boldsymbol{\mathcal{N}}(t, \boldsymbol{a}, \boldsymbol{0})\| \leq& K\|\boldsymbol{f}\|,\label{eq:theorem1_input}
\\
\|\boldsymbol{h}(t, \boldsymbol{a}, \boldsymbol{f})\| \leq& \eta_1\|\boldsymbol{a}\|+\eta_2\|\boldsymbol{f}\|,\label{eq:theorem1_output}
\end{align} 
\end{subequations}
for all $(t, \boldsymbol{a}, \boldsymbol{f}) \in[0, \infty) \times D \times D_u$ for some nonnegative constants $K$, $\eta_1$, and $\eta_2$. 
\end{itemize}
Then, for each $\boldsymbol{a}_0$ with $\|\boldsymbol{a}_0\| \leq \delta \sqrt{c_1 / c_2}$, the system \eqref{eq:simp_sys_1}-\eqref{eq:simp_sys_2} is small-signal finite-gain $\mathcal{L}_p$ stable for each $p \in[1, \infty]$. In particular, for each $\bm{f} \in \mathcal{L}_{p e}$ with 
\begin{align}
 \sup _{0 \leq t \leq \tau}\|\boldsymbol{f}(t)\| \leq \min \left\{r_u, c_1 c_3 \delta /\left(c_2 c_4 K\right)\right\},
\end{align}
the state variable $\boldsymbol{a}(t)$ satisfies 
\begin{align}
    \|\boldsymbol{a}(t)\|\leq \delta,\;\;\forall t\in [0,\tau]
    \label{eq:upper_bound_state_variable}
\end{align}
and the output $\boldsymbol{y}(t)$ satisfies
\begin{equation}
\left\|\boldsymbol{y}_\tau\right\|_{\mathcal{L}_p} \leq \gamma\left\|\boldsymbol{f}_\tau\right\|_{\mathcal{L}_p}+\beta,
\label{eq:ineq_output}
\end{equation}
for all $\tau \in[0, \infty)$, with
\begin{equation}
\begin{split}
\gamma=\eta_2+\frac{\eta_1 c_2 c_4 K}{c_1 c_3},\;\; \beta=\eta_1 \|\boldsymbol{a}_0\| \sqrt{\frac{c_2}{c_1}} \rho, \;\;
\text {where }\;\; \rho= \begin{cases}1, & \text {if } p=\infty \\ \left(\frac{2 c_2}{c_3 p}\right)^{1 / p}, & \text {if } p \in[1, \infty).\end{cases}
\label{eq:gamma_beta}
\end{split}
\end{equation}
\end{theorem}   

In Theorem \ref{thm:small_signal_stability}, the conclusion of the state variable in \eqref{eq:upper_bound_state_variable} is not explicitly included in the statement of \cite[Theorem 5.1]{Khalil2002}, but is included as an intermediate result in the proof of \cite[Theorem 5.1]{Khalil2002}. From \eqref{eq:upper_bound_state_variable}, we can see that $\delta$ is a preset upper bound for state variable amplitude $\|\boldsymbol{a}(t)\|$. The main challenge of applying Theorem \ref{thm:small_signal_stability} is searching for the Lyapunov function $V(t,\boldsymbol{a})$ of the unforced system satisfying \eqref{eq:theorem1_Lyapunov}. Here, we will focus on a quadratic Lyapunov function candidate in the form of ${V}=\boldsymbol{a}^\text{T}\boldsymbol{P}\boldsymbol{a}$ and use linear matrix inequalities to certify whether they satisfy the requirements in the nonlinear SSFG $\mathcal{L}_p$ stability theorem in \eqref{eq:theorem1_Lyapunov}.

To formulate linear matrix inequalities (LMI) to search for a Lyapunov function candidate satisfying \eqref{eq:theorem1_Lyapunov}, we will use the same procedure as Ref. \citep{Liu2020PRE} to construct quadratic constraints to describe the nonlinear term $\boldsymbol{\Upsilon}(\boldsymbol{a})$ in a local region $\|\boldsymbol{a}\|\leq \delta$. As we focus on shear flow models, we consider quadratic nonlinearity such that each component of $\boldsymbol{\Upsilon}$ can be written into a quadratic form; i.e.,  $\Upsilon_m:=\boldsymbol{e}_m^\text{T}\boldsymbol{\Upsilon}$ can be written as $\Upsilon_m=\boldsymbol{a}^\text{T}\boldsymbol{R}_m\boldsymbol{a}$. Here, $\boldsymbol{e}_m$ is a column unit vector with the m$^\text{th}$ element equal to one, and all other elements equal to zero. Then in a local region $\|\boldsymbol{a}\|\leq \delta$, we have the upper bound of $\Upsilon_m$ as
\begin{align}
    \Upsilon_m^2\leq \delta^2 \boldsymbol{a}^\text{T}\boldsymbol{R}_m\boldsymbol{R}_m\boldsymbol{a}. 
    \label{eq:local_upper_bound_upsilon_m}
\end{align}
The proof of equation \eqref{eq:local_upper_bound_upsilon_m} is shown in Appendix \ref{appendix:b}. For the nonlinear system in \eqref{eq:simp_eq} obtained from Galerkin projection of the Navier-Stokes equations, the nonlinear term is energy-conserving, or satisfies the lossless condition:  
\begin{align}
\boldsymbol{a}^\text{T} \boldsymbol{\Upsilon}= 0. 
\label{eq:energy_conserving_constraint}
\end{align} 
If we define $\boldsymbol{M}_0:=\boldsymbol{I}$ as the identity matrix, we can also write \eqref{eq:energy_conserving_constraint} as $\boldsymbol{a}^\text{T}\boldsymbol{M}_0\boldsymbol{\Upsilon}=0$. We will also explore the orthogonal complement $\boldsymbol{n}$ satisfying $\boldsymbol{n}^\text{T}\boldsymbol{\Upsilon}=0$, which directly allows us to construct the quadratic constraints of the nonlinear term as
\begin{subequations}
    \label{eq:orthogonal_complement}
\begin{align}
\boldsymbol{a}^\text{T}\boldsymbol{M}_i\boldsymbol{f}=&0,\;\;i=1,2,...,n,\\
\boldsymbol{\Upsilon}^\text{T}\boldsymbol{T}_j\boldsymbol{\Upsilon}=&0,\;\;j=1,2,...,n, 
\end{align}
\end{subequations}
where 
\begin{subequations}
    \begin{align}
    \boldsymbol{M}_i := &\boldsymbol{e}_i \boldsymbol{n}^\text{T},\\
    \boldsymbol{T}_j :=& \boldsymbol{e}_j \boldsymbol{n}^\text{T} + \boldsymbol{n} \boldsymbol{e}_j^\text{T},
\end{align}
\end{subequations}

We then apply Theorem \ref{thm:small_signal_stability} to our nonlinear input-output system in \eqref{eq:simp_eq_1}-\eqref{eq:simp_eq_2} utilizing the quadratic constraints of nonlinearity $\boldsymbol{\Upsilon}$ as in equations \eqref{eq:local_upper_bound_upsilon_m}-\eqref{eq:orthogonal_complement}. 
\begin{theorem}
\label{thm:Liu2020PRE}
Consider the nonlinear input-output system in \eqref{eq:simp_eq_1}-\eqref{eq:simp_eq_2} and take $\delta>0$ such that $\{\|\boldsymbol{a}\| \leq \delta\} \subset D$. If there exists a symmetric matrix $\boldsymbol{P} \in \mathbb{R}^{n \times n}$ satisfying
\begin{equation} 
\begin{split}
\boldsymbol{P} - \varepsilon \boldsymbol{I} \succeq 0,\;\;\varepsilon > 0, \;\;\boldsymbol{G} \preceq 0, \\
\;\;s_m \geq 0, \quad m=1, \ldots, n,
\label{eq:sdp_P}
\end{split}
\end{equation}
where $(\cdot) \succeq 0$ and $(\cdot) \preceq 0$, respectively, represent positive and negative semi-definiteness of the associated matrix, and $\boldsymbol{G}$ is defined as:
 \begin{align}
 \boldsymbol{G}:=\left[\begin{array}{cc}
 \boldsymbol{L}^\text{T} \boldsymbol{P}+\boldsymbol{P} \boldsymbol{L}+\varepsilon \boldsymbol{I}+\sum\limits_{m=1}^n s_m \delta^2 \boldsymbol{R}_m \boldsymbol{R}_m & \boldsymbol{P}+\sum\limits_{i=0}^n \mu_i \boldsymbol{M}_i \\
\boldsymbol{P}+\sum\limits_{i=0}^n \mu_i \boldsymbol{M}_i^\text{T} & -\sum\limits_{m=1}^n s_m \boldsymbol{e}_m \boldsymbol{e}_m^\text{T}+\sum\limits_{j=1}^n \kappa_j \boldsymbol{T}_j
\end{array}\right],
\label{eq:LMI}
\end{align}
for all $(t,\boldsymbol{a})\in [0,\infty)\times D$. Then, for each $\boldsymbol{a}_0$ with 
\begin{align}
 \|\boldsymbol{a}_0\|\leq \delta\sqrt{\frac{\lambda_{\text{min}}(\boldsymbol{P})}{\lambda_{\text{max}}(\boldsymbol{P})}}=:\delta_f,
 \label{eq:delta_f_LMI}
\end{align}
the system \eqref{eq:simp_eq_1}-\eqref{eq:simp_eq_2} is small-signal finite-gain $\mathcal{L}_p$ stable for each $p \in[1, \infty]$, where $\lambda_{\text{min}}(\boldsymbol{P})$ and $\lambda_{\text{max}}(\boldsymbol{P})$ are the minimal and maximal eigenvalues of the matrix $\boldsymbol{P}$, respectively. In particular, for each $\bm{f} \in \mathcal{L}_{p e}$ with 
\begin{align}
 \sup _{0 \leq t \leq \tau}\|\boldsymbol{f}(t)\| \leq \frac{\lambda_{\text{min}}(\boldsymbol{P}) \varepsilon \delta }{\lambda_{\text{max}}(\boldsymbol{P}) \|2\boldsymbol{P}\| }=:f_{\text{LMI}},
 \label{eq:forcing_upper_bound}
\end{align}
the state variable $\boldsymbol{a}(t)$ satisfies 
\begin{align}
    \|\boldsymbol{a}(t)\|\leq \delta,\;\;\forall t\in [0,\tau]
    \label{eq:upper_bound_state_variable_LMI}
\end{align}
and the output $\boldsymbol{y}(t)$ satisfies
\begin{equation}
\left\|\boldsymbol{y}_\tau\right\|_{\mathcal{L}_p} \leq \gamma\left\|\boldsymbol{f}_\tau\right\|_{\mathcal{L}_p}+\beta,
\label{eq:gamma_upper_bound}
\end{equation}
for all $\tau \in[0, \infty)$, with
\begin{subequations}
\label{eq:gamma_beta}
\begin{align}
\gamma=&\frac{\lambda_{\text{max}}(\boldsymbol{P}) \|2\boldsymbol{P}\|}{\lambda_{\text{min}}(\boldsymbol{P}) \varepsilon},\;\;\label{eq:gamma_value} \\
\beta= &\|\boldsymbol{a}_0\| \sqrt{\frac{\lambda_{\text{max}}(\boldsymbol{P})}{\lambda_{\text{min}}(\boldsymbol{P})}} \rho, \;\;
\text {where }\;\; \rho= \begin{cases}1, & \text {if } p=\infty \\ \left(\frac{2 \lambda_{\text{max}}(\boldsymbol{P})}{\varepsilon\; p}\right)^{1 / p}, & \text {if } p \in[1, \infty).\end{cases}
\label{eq:beta_value}
\end{align}
\end{subequations}
\end{theorem}

\begin{proof} Because $\boldsymbol{L}$ in \eqref{eq:simp_eq} is assumed as Hurwitz, we have $\boldsymbol{a}=\boldsymbol{0}$ is an exponentially stable equilibrium of the unforced system in \eqref{eq:initial}. When inequalities in equation \eqref{eq:sdp_P} are feasible, $\boldsymbol{P}$ can be used to define a positive definite function $V:=\boldsymbol{a}^\text{T} \boldsymbol{P a} \geqslant \varepsilon \boldsymbol{a}^\text{T} \boldsymbol{a}>0, \quad \forall \boldsymbol{a} \neq \boldsymbol{0}$. Thus, $V$ satisfies \eqref{eq:theorem1_Lyapunov_a} with $c_1=\lambda_{\text{min}}(\boldsymbol{P})$ and $c_2=\lambda_{\text{max}}(\boldsymbol{P})$ based on the Rayleigh quotient theorem (see, e.g., Theorem 4.2.2 in Ref. \citep{horn2012matrix}). We now demonstrate that $V$ satisfies \eqref{eq:theorem1_Lyapunov_b} for the unforced system in Eq.~\eqref{eq:initial} in the region $\|\boldsymbol{a}\| \leqslant \delta$. According to equation \eqref{eq:local_upper_bound_upsilon_m} and Appendix \ref{appendix:b}, we have $\delta^2 \boldsymbol{a}^\text{T} \boldsymbol{R}_m \boldsymbol{R}_m \boldsymbol{a}-\Upsilon_m^2 \geqslant 0, m=1,2, \ldots, n$ in the region $\|\boldsymbol{a}\| \leqslant \delta$, and therefore, we have:
\begin{subequations}
\label{eq:dVdt}
\begin{align}
\frac{\partial V}{\partial t} + \frac{\partial V}{\partial \boldsymbol{a}} \boldsymbol{\mathcal{N}}(t, \boldsymbol{a}, \boldsymbol{0}) \leqslant & \frac{\partial V}{\partial t} + \frac{\partial V}{\partial \boldsymbol{a}} \boldsymbol{\mathcal{N}}(t, \boldsymbol{a}, \boldsymbol{0}) +\sum_{m=1}^n s_m\left(\delta^2 \boldsymbol{a}^\text{T} \boldsymbol{R}_m \boldsymbol{R}_m \boldsymbol{a}-\Upsilon_m^2\right) \label{eq:dVdt_a} \\
= & {\left[\begin{array}{l}
\boldsymbol{a} \\
\boldsymbol{\Upsilon}
\end{array}\right]^\text{T} \boldsymbol{G}\left[\begin{array}{l}
\boldsymbol{a} \\
\boldsymbol{\Upsilon}
\end{array}\right]-\varepsilon \boldsymbol{a}^\text{T} \boldsymbol{a} }  \label{eq:dVdt_b}\\
\leqslant & -\varepsilon \boldsymbol{a}^\text{T} \boldsymbol{a}<0, \label{eq:dVdt_c}
\end{align}
\end{subequations}
where we also utilize \eqref{eq:energy_conserving_constraint} and \eqref{eq:orthogonal_complement} when computing $\left[\begin{array}{l}
\boldsymbol{a} \\
\boldsymbol{\Upsilon}
\end{array}\right]^\text{T} \boldsymbol{G}\left[\begin{array}{l}
\boldsymbol{a} \\
\boldsymbol{\Upsilon}
\end{array}\right]$ in \eqref{eq:dVdt_b}. Thus, we have $V$ as a Lyapunov function of the unforced system satisfying \eqref{eq:theorem1_Lyapunov_b} with $c_3=\varepsilon$. We can also compute $\frac{\partial V}{\partial \boldsymbol{a}}=(\boldsymbol{P}+\boldsymbol{P}^\text{T})\boldsymbol{a}$ and $\left\|\frac{\partial V}{\partial \boldsymbol{a}}\right\|=\|2\boldsymbol{P}\boldsymbol{a}\|\leq \|2\boldsymbol{P}\|\,\|\boldsymbol{a}\|$ (by the sub-multiplicative property of matrix norms) to show that $V$ satisfies \eqref{eq:theorem1_Lyapunov_c} with $c_4=\|2\boldsymbol{P}\|=2\lambda_{\text{max}}(\boldsymbol{P})$. 

Based on the form of our nonlinear dynamical system in \eqref{eq:simp_eq}, we have $\mathcal{N}(t,\boldsymbol{a},\boldsymbol{f})-\mathcal{N}(t,\boldsymbol{a},\boldsymbol{0})=\boldsymbol{f}$. Thus, we have $\mathcal{N}$ satisfies \eqref{eq:theorem1_input} with $K=1$ for any $\boldsymbol{f}\in \mathbb{R}^n$ leading to $r_u=\infty$ in Theorem \ref{thm:small_signal_stability}. Based on the form of output $\boldsymbol{y}=\boldsymbol{h}(t,\boldsymbol{a},\boldsymbol{f})=\boldsymbol{a}$ as we defined in \eqref{eq:simp_sys_2}, we have $\boldsymbol{h}$ satisfies \eqref{eq:theorem1_output} with $\eta_1=1$ and $\eta_2=0$. 

Thus, all requirements of Theorem \ref{thm:small_signal_stability} are satisfied, and we can substitute values of $c_1$, $c_2$, $c_3$, $c_4$, $K$, $\eta_1$, $\eta_2$, and $r_u$ into the conclusion of Theorem \ref{thm:small_signal_stability} to obtain the conclusion of Theorem \ref{thm:Liu2020PRE}. 
\end{proof}

\subsection{\label{sec:SOSopt}Sum-of-squares (SOS) optimization}

We can also use the sum-of-squares optimization to certify the Lyapunov function of the unforced nonlinear system. SOS-based programs directly characterize the polynomial nonlinearity instead of describing nonlinearity using local upper bounds in the form of quadratic constraints. \rev{While SOS allows for a wider range of higher-order polynomial Lyapunov function candidates that can reduce the conservatism of the results, this greater flexibility comes with a significantly higher computational cost; see e.g. \citep{GOULART2012692,chernyshenko2014polynomial,huang2015sum,fuentes2022global}. For this study, we focus on quadratic Lyapunov functions to be consistent with the LMI method and to balance the computational efficiency and accuracy. The SOS framework provides a clear pathway for future work to explore higher-order Lyapunov function candidates and potentially obtain tighter bounds.} Here, we modify \eqref{eq:sdp_P} in Theorem \ref{thm:Liu2020PRE} as shown below:
\begin{theorem}
\label{thm:SOS}
Consider the nonlinear input-output system in \eqref{eq:simp_eq_1}-\eqref{eq:simp_eq_2} and take $\delta>0$ such that $\{\|\boldsymbol{a}\| \leq \delta\} \subset D$. If there exists a $V_{\text{SOS}}:=\boldsymbol{a}^\text{T}\boldsymbol{P}_{\text{SOS}}\boldsymbol{a}$ with symmetric matrix $\boldsymbol{P}_{\text{SOS}} \in \mathbb{R}^{n \times n}$ satisfying
\begin{subequations} \label{eq:SOS_constraints}
\begin{align}
V_{\text{SOS}}-\varepsilon \boldsymbol{a}^\text{T} \boldsymbol{a}&\geq 0,\;\;
\varepsilon > 0,\label{eq:SOS_constraints_a}
\\
-\left(\frac{\partial V_{\text{SOS}}}{\partial t} + \frac{\partial V_{\text{SOS}}}{\partial \boldsymbol{a}} \boldsymbol{\mathcal{N}}(t, \boldsymbol{a}, \boldsymbol{0})+\left(\delta^2-\boldsymbol{a}^\text{T} \boldsymbol{a}\right) \boldsymbol{a}^\text{T} \boldsymbol{Q} \boldsymbol{a}+\varepsilon \boldsymbol{a}^\text{T} \boldsymbol{a}\right) &\geq 0,\label{eq:SOS_constraints_b}
\\
\boldsymbol{Q} &\succeq 0,\label{eq:SOS_constraints_c}
\end{align}
\end{subequations}
for all $(t,\boldsymbol{a})\in [0,\infty)\times D$. Then, for each $\boldsymbol{a}_0$ with 
\begin{align}
 \|\boldsymbol{a}_0\|\leq \delta\sqrt{\frac{\lambda_{\text{min}}(\boldsymbol{P}_{\text{SOS}})}{\lambda_{\text{max}}(\boldsymbol{P}_{\text{SOS}})}}=:\delta_{f,\text{SOS}},   
 \label{eq:delta_f_SOS}
\end{align}
the system \eqref{eq:simp_eq_1}-\eqref{eq:simp_eq_2} is small-signal finite-gain $\mathcal{L}_p$ stable for each $p \in[1, \infty]$. In particular, for each $\bm{f} \in \mathcal{L}_{p e}$ with 
\begin{align}
 \sup _{0 \leq t \leq \tau}\|\boldsymbol{f}(t)\| \leq \frac{\lambda_{\text{min}}(\boldsymbol{P}_{\text{SOS}}) \varepsilon \delta}{\lambda_{\text{max}}(\boldsymbol{P}_{\text{SOS}}) \|2\boldsymbol{P}_{\text{SOS}}\| }=:f_{\text{SOS}},
 \label{eq:forcing_upper_bound_SOS}
\end{align}
the state variable $\boldsymbol{a}(t)$ satisfies 
\begin{align}
    \|\boldsymbol{a}(t)\|\leq \delta,\;\;\forall t\in [0,\tau]
\label{eq:upper_bound_state_variable_SOS}
\end{align}
and the output $\boldsymbol{y}(t)$ satisfies
\begin{equation}
\left\|\boldsymbol{y}_\tau\right\|_{\mathcal{L}_p} \leq \gamma_{\text{SOS}}\left\|\boldsymbol{f}_\tau\right\|_{\mathcal{L}_p}+\beta_{\text{SOS}},
\label{eq:gamma_upper_bound_SOS}
\end{equation}
for all $\tau \in[0, \infty)$, with
\begin{subequations}
\label{eq:gamma_beta_SOS}
\begin{align}
\gamma_{\text{SOS}}=&\frac{\lambda_{\text{max}}(\boldsymbol{P}_{\text{SOS}}) \|2\boldsymbol{P}_{\text{SOS}}\|}{\lambda_{\text{min}}(\boldsymbol{P}_{\text{SOS}}) \varepsilon},\;\;\label{eq:gamma_value_SOS} \\
\beta_{\text{SOS}}= &\|\boldsymbol{a}_0\| \sqrt{\frac{\lambda_{\text{max}}(\boldsymbol{P}_{\text{SOS}})}{\lambda_{\text{min}}(\boldsymbol{P}_{\text{SOS}})}} \rho, \;\;
\text {where }\;\; \rho= \begin{cases}1, & \text {if } p=\infty \\ \left(\frac{2 \lambda_{\text{max}}(\boldsymbol{P}_{\text{SOS}})}{\varepsilon\; p}\right)^{1 / p}, & \text {if } p \in[1, \infty).\end{cases}
\label{eq:beta_value_SOS}
\end{align}
\end{subequations}
\end{theorem}

\begin{proof}
    Based on \eqref{eq:SOS_constraints_c}, we have $\left(\delta^2-\boldsymbol{a}^\text{T} \boldsymbol{a}\right) \boldsymbol{a}^\text{T} \boldsymbol{Q} \boldsymbol{a}\geq 0$ when $\|\boldsymbol{a}\|^2\leq \delta^2$, and thus according to \eqref{eq:SOS_constraints_b}, we have $\frac{\partial V_{\text{SOS}}}{\partial t} + \frac{\partial V_{\text{SOS}}}{\partial \boldsymbol{a}} \boldsymbol{\mathcal{N}}(t, \boldsymbol{a}, \boldsymbol{0}) \leq -\varepsilon \boldsymbol{a}^\text{T} \boldsymbol{a}$ satisfies \eqref{eq:theorem1_Lyapunov_b} with $c_3=\varepsilon$ in the region $\|\boldsymbol{a}\|\leq \delta$. Then, following the same proof of Theorem \ref{thm:Liu2020PRE} with the $\boldsymbol{P}_{\text{SOS}}$ matrix obtained from SOS, we obtain the conclusion of Theorem \ref{thm:SOS}. 
\end{proof}

Here, $\frac{\partial V}{\partial t}$ in equation \eqref{eq:dVdt_a} and $\frac{\partial V_{\text{SOS}}}{\partial t}$ in equation \eqref{eq:SOS_constraints_b} are both zero because we assume a time-independent $\boldsymbol{P}$ or $\boldsymbol{P}_{\text{SOS}}$ matrix in the Lyapunov function candidate. However, this term may be important for determining the Lyapunov function of time-varying (non-autonomous) systems \citep{Khalil2002}, where we can introduce explicit time dependence of Lyapunov function candidates $V$ and $V_{\text{SOS}}$; see e.g. Ref. \citep{wei2025upper,kochnev2025stability}.

\section{\label{sec:nonlinear} Nonlinear input-output stability analysis}

\label{sec:results}
We implement Linear Matrix Inequalities (LMI) stated in \eqref{eq:sdp_P} from Theorem~\ref{thm:Liu2020PRE} using YALMIP version R20230622 (A Toolbox for Modeling and Optimization in MATLAB) \citep{Lofberg2004} within MATLAB R2022a and the semidefinite programming (SDP) solver Mosek 10.2 \citep{Mosek2024}. For SOS in Theorem \ref{thm:SOS}, we employ SOSTOOLS version 3.03 \citep{papachristodoulou2021sostoolsversion400sum} within MATLAB R2018a to test whether the left-hand side of inequality \eqref{eq:SOS_constraints} is sum-of-squares, which will imply positive semi-definiteness. We have to use SeDuMi \citep{sturm1999using} v1.3 as the SDP solver for SOS, as SOSTOOLS in this version does not support Mosek. Throughout this paper, the parameter $\varepsilon$ in inequality \eqref{eq:sdp_P} is chosen as $10^{-4}$ for LMI and $\varepsilon$ in inequality \eqref{eq:SOS_constraints} is chosen as $10^{-2}$ for SOS. When changing $\varepsilon$, the matrix $\boldsymbol{P}$ can be scaled by a positive constant, and such a scaling constant will cancel out in the conclusion of Theorems \ref{thm:Liu2020PRE} and \ref{thm:SOS}. However, we find that a larger $\varepsilon$ (e.g., $\varepsilon=10^{-2}$ for LMI and $\varepsilon=10^{-1}$ for SOS) will lead to numerical issues in high Reynolds number regimes leading to more infeasible solutions, and a smaller $\varepsilon$ (e.g., $\varepsilon=10^{-9}$ for LMI and $\varepsilon=10^{-5}$ for SOS) may not effectively ensure that $\boldsymbol{P}$ matrix is positive definite leading to unreasonable results. For feasible solutions, we validate that $\varepsilon\in [10^{-8}, 10^{-2}]$ does not significantly change our results obtained from LMI, as shown in Appendix \ref{appendix:c}, and also validate that $\varepsilon\in [10^{-4}, 10^{-1}]$ does not significantly change our SOS results. We also validated our LMI and SOS implementation by reproducing the permissible perturbation amplitudes for the same nine-mode shear flow model in Figure 5 of Ref. \cite{Liu2020PRE}.

We test the feasibility of inequalities in \eqref{eq:sdp_P} and \eqref{eq:SOS_constraints} across 16 logarithmically spaced Reynolds numbers, Re $ \in [200, 2000]$, where the Reynolds number influences the matrix $\boldsymbol{L}$ associated with the linearized system. At each Reynolds number, we test the feasibility of LMI in \eqref{eq:sdp_P} of Theorem \ref{thm:Liu2020PRE} and the feasibility of SOS in \eqref{eq:SOS_constraints} of Theorem \ref{thm:SOS} over 400 logarithmically spaced $\delta$ within $ \delta\in [10^{-6}, 1]$. Thus, results from Theorems \ref{thm:Liu2020PRE} and \ref{thm:SOS} are functions over both $\text{Re}$ and $\delta$, which include the upper bound of input-output amplification $\gamma(\text{Re},\delta)$ and $\gamma_{\text{SOS}}(\text{Re},\delta)$ as well as the permissible forcing amplitude $f_{\text{LMI}}(\text{Re}, \delta)$ and $f_{\text{SOS}}(\text{Re}, \delta)$ below which a finite input-output amplification ($\gamma$ and $\gamma_{\text{SOS}}$) can be sustained.

\begin{figure}[!htbp]
\centering
    
    \begin{subfigure}{0.485\textwidth}
        \centering
        \includegraphics[width=\textwidth]{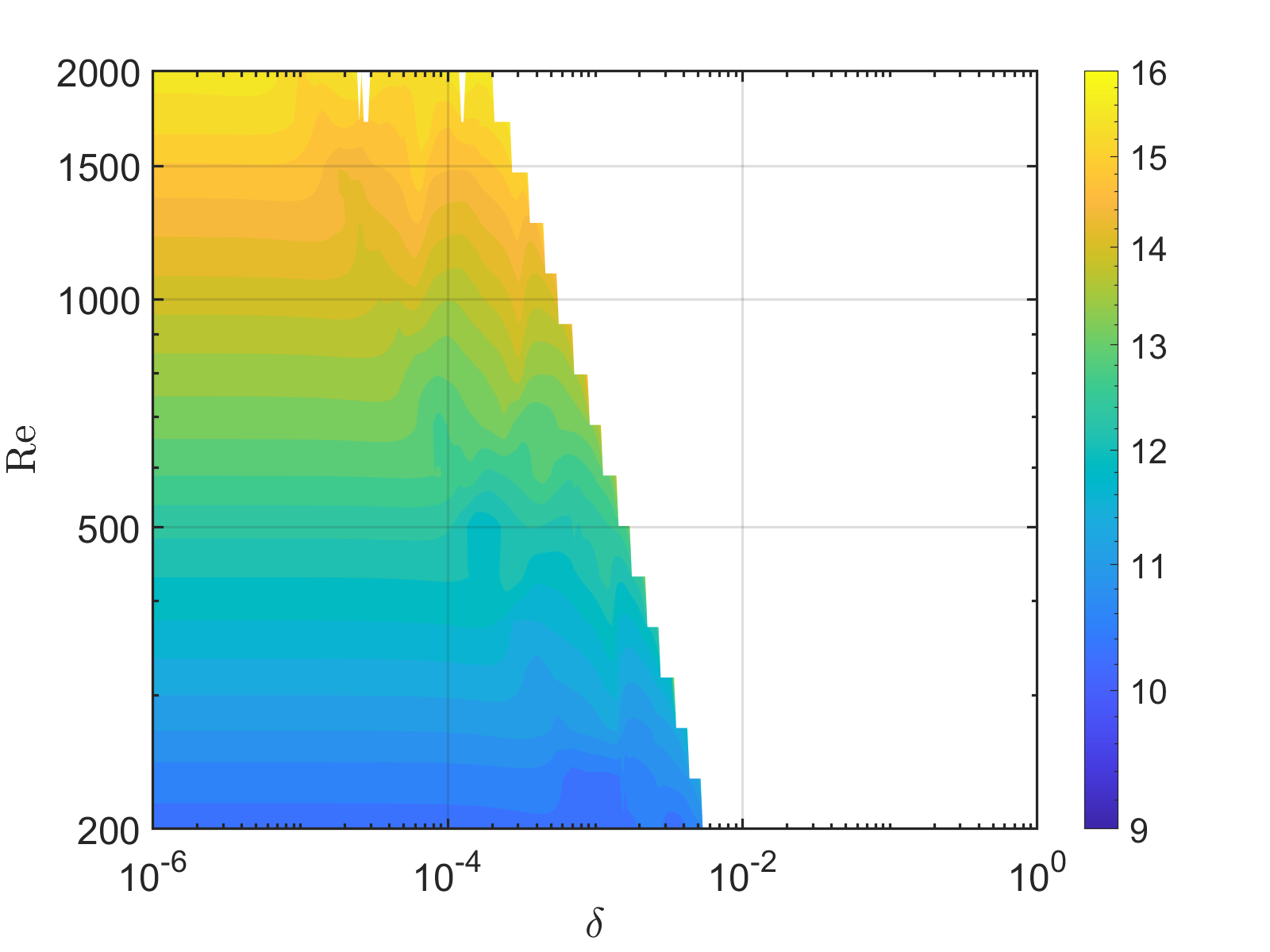}
        \caption{$\text{log}_{10}[\gamma(\text{Re},\delta)]$ based on LMI method in Theorem \ref{thm:Liu2020PRE}. }
        \label{fig:contour_gamma}
    \end{subfigure}
    \begin{subfigure}{0.485\textwidth}
        \centering
        \includegraphics[width=\textwidth]{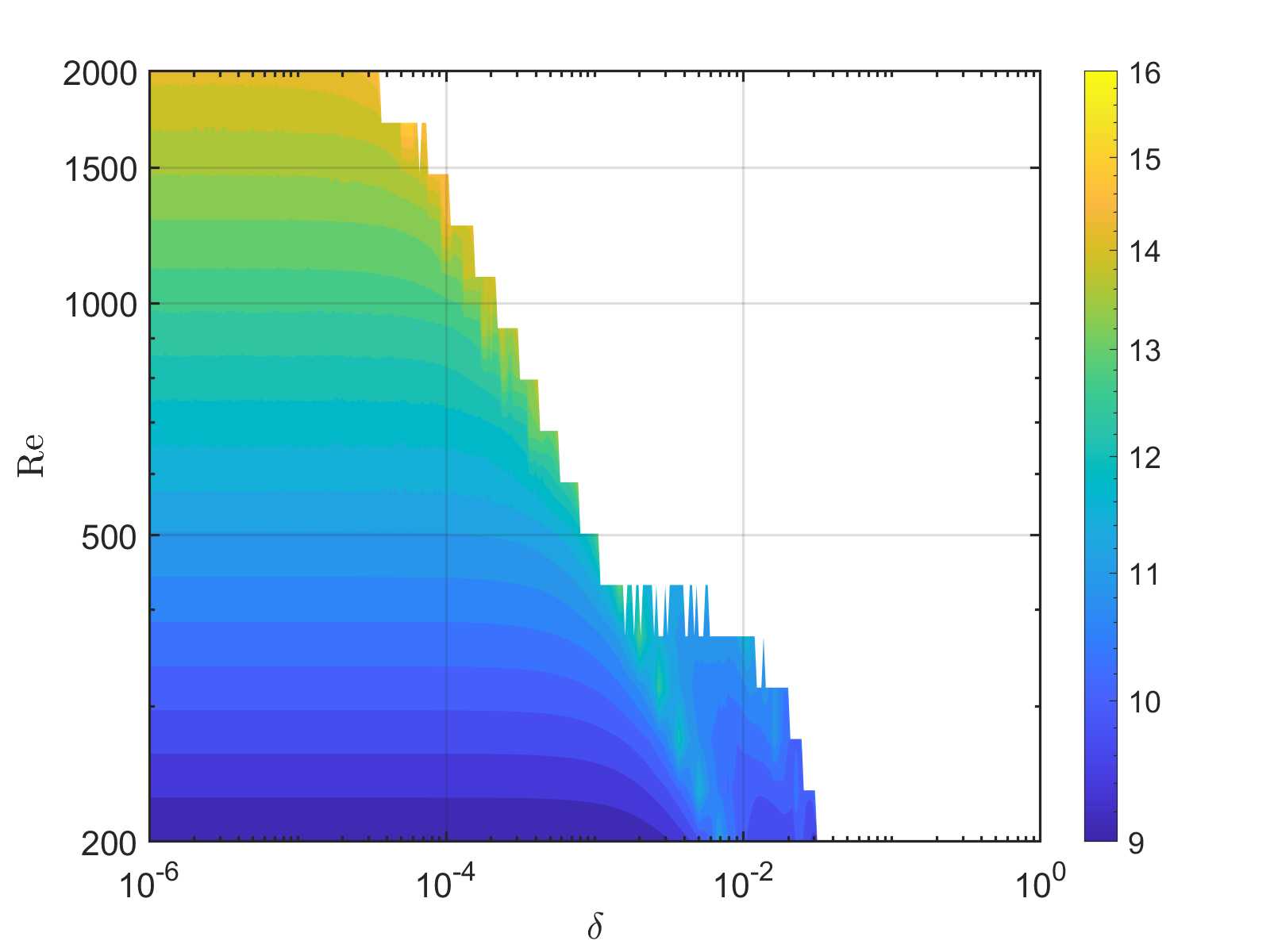}
        \caption{$\text{log}_{10}[\gamma_{\text{SOS}}(\text{Re},\delta)]$ based on SOS method in Theorem \ref{thm:SOS}. }
        \label{fig:contour_gamma_SOS}
    \end{subfigure}

    \begin{subfigure}[t]{0.48\textwidth}
    \centering
    \includegraphics[width=\linewidth]{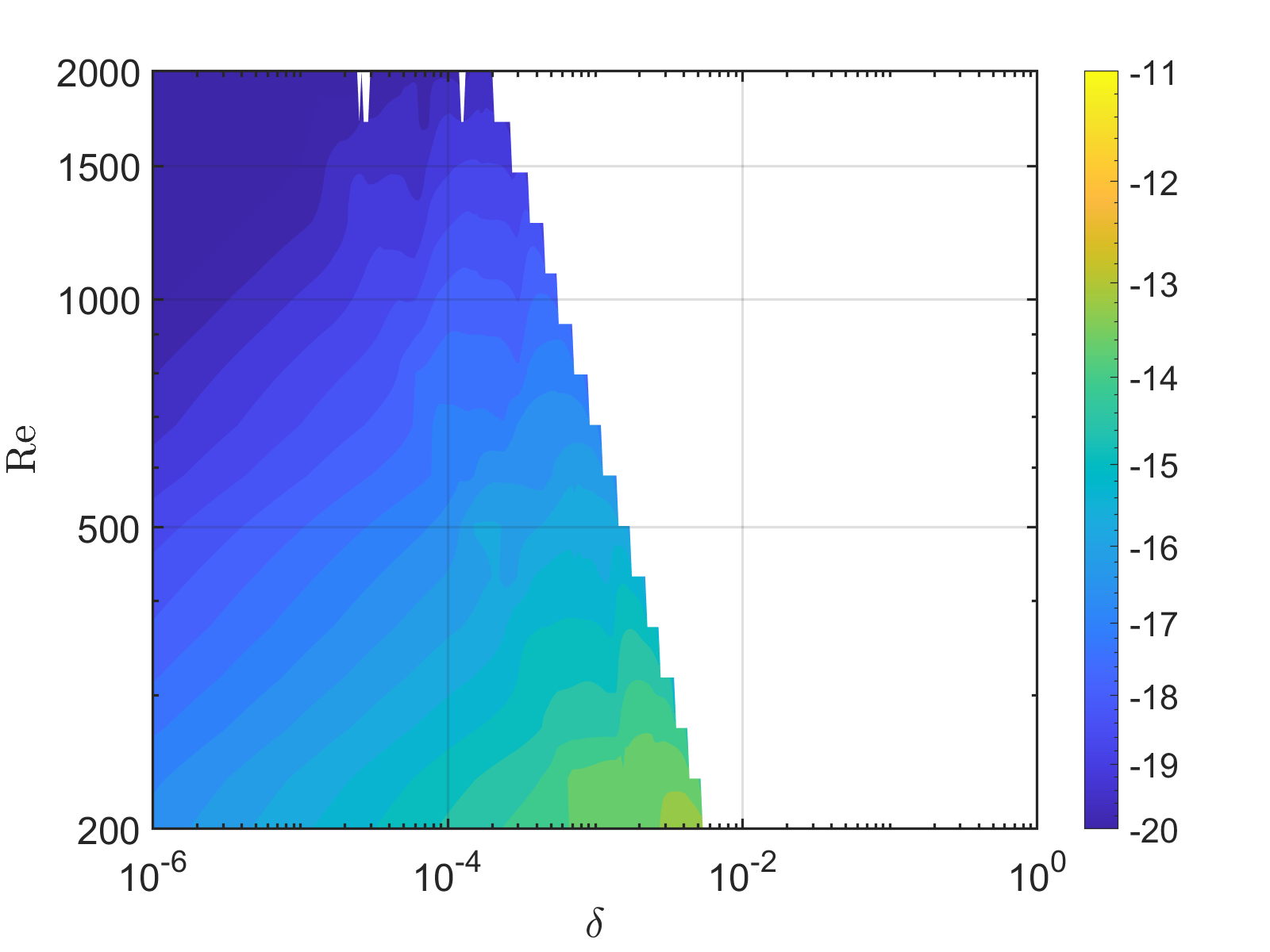}
    \caption{$\text{log}_{10}[f_{\text{LMI}}(\text{Re},\delta)]$ based on LMI method in Theorem \ref{thm:Liu2020PRE}.}
    \label{fig:contour_lmi}  
    \end{subfigure}
    \hfill
    \begin{subfigure}[t]{0.48\textwidth}
    \centering
    \includegraphics[width=\linewidth]{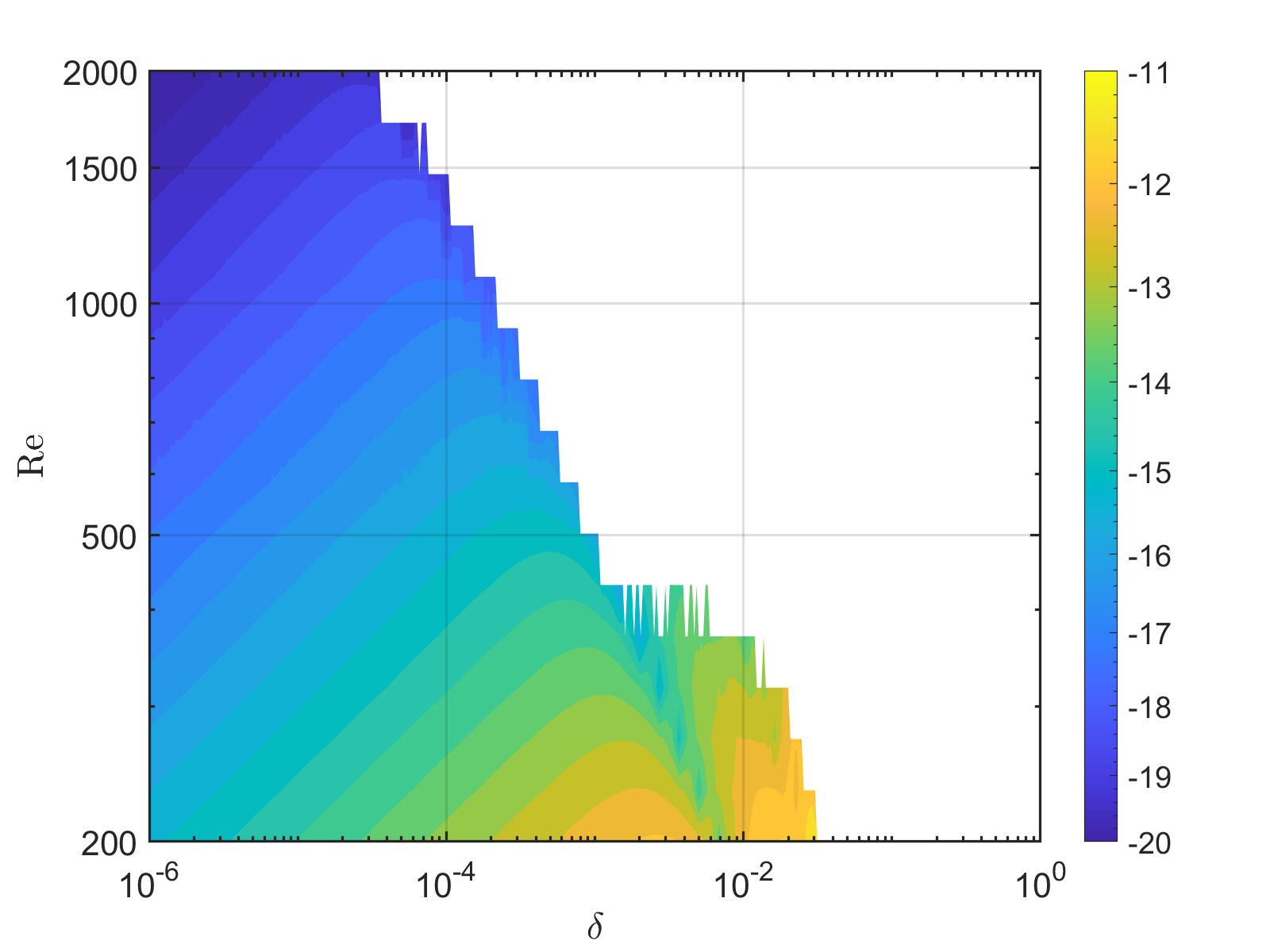}
    \caption{$\text{log}_{10}[f_{\text{SOS}}(\text{Re},\delta)]$ based on SOS method in Theorem \ref{thm:SOS}.}
    \label{fig:contour_SOS} 
    \end{subfigure}
\caption{\justifying  (a) $\text{log}_{10}[\gamma(\text{Re},\delta)]$, (b) $\text{log}_{10}[\gamma_{\text{SOS}}(\text{Re},\delta)]$, (c) $\text{log}_{10}[f_{\text{LMI}}(\text{Re},\delta)]$, and (d) $\text{log}_{10}[f_{\text{SOS}}(\text{Re},\delta)]$. The white region indicates parameter regimes where the LMI or SOS is infeasible.}
\label{fig:gamma_forcing_upper_bound}
\end{figure}

Figure \ref{fig:gamma_forcing_upper_bound} shows $\gamma(\text{Re},\delta)$,  $\gamma_{\text{SOS}}(\text{Re},\delta)$,  $f_{\text{LMI}}(\text{Re}, \delta)$, and $f_{\text{SOS}}(\text{Re}, \delta)$ as a function over $\text{Re}$ and $\delta$. The white regions in Figure \ref{fig:gamma_forcing_upper_bound} are associated with the infeasible parameter regimes, where we cannot find a feasible matrix $\boldsymbol{P}$ for the LMI problem in \eqref{eq:sdp_P} of Theorem \ref{thm:Liu2020PRE} or a feasible $\boldsymbol{P}_{\text{SOS}}$ for the SOS problem in \eqref{eq:SOS_constraints} of Theorem \ref{thm:SOS}. Note that Theorem \ref{thm:Liu2020PRE} and Theorem \ref{thm:SOS} are sufficient but not necessary conditions for SSFG $\mathcal{L}_p$ stability. Thus, these white regimes with infeasible LMI or SOS cannot show instability, but only mean that these methods fail to find a feasible Lyapunov function to certify SSFG $\mathcal{L}_p$ stability for these parameter regimes. Here, the nonlinear $\mathcal{L}_p$ input-output gain $\gamma$ or $\gamma_{\text{SOS}}$ both increase as we increase the Reynolds number $\text{Re}$, and it is nearly independent of $\delta$ at low $\delta$. Closer to the boundary of the infeasible parameter regimes, $\gamma$ and $\gamma_{\text{SOS}}$ show stronger dependence on $\delta$. For the permissible forcing amplitude, both $f_{\text{LMI}}$ and $f_{\text{SOS}}$ will decrease as we increase the $\text{Re}$, indicating that we require a smaller forcing amplitude to maintain a finite input-output $\mathcal{L}_p$ gain. Note that such permissible forcing amplitude is an inherently nonlinear property, while for a linear system, the input-output gain is independent of the forcing amplitude. As a larger $\delta$ allows a larger amplitude of state variables (e.g., $\|\boldsymbol{a}(t)\|\leq \delta$ as equation  \eqref{eq:upper_bound_state_variable}), we have an increase in the permissible forcing amplitude $f_{\text{LMI}}$ or $f_{\text{SOS}}$ as shown in Figures \ref{fig:contour_lmi}-\ref{fig:contour_SOS}. Figure \ref{fig:gamma_forcing_upper_bound} shows that the input-output gain obtained by SOS $\gamma_{\text{SOS}}(Re,\delta)$ is smaller than that obtained by LMI $\gamma(Re,\delta)$ in most parameter regimes, and SOS provides a larger permissible forcing amplitude $f_{\text{SOS}}$ than that obtained by LMI $f_{\text{LMI}}$. This means SOS provides a less conservative estimation of input-output gain and permissible forcing amplitude, which is because SOS uses full nonlinearity while LMI describes nonlinearity using quadratic constraints as equations \eqref{eq:local_upper_bound_upsilon_m}-\eqref{eq:orthogonal_complement}. Similar observations that SOS provides less conservative estimation were also shown when computing the permissible perturbation amplitudes of shear flow models \citep{Liu2020PRE}.

\subsection{\label{sec:analysis_SSFG} Comparison of nonlinear SSFG $\mathcal{L}_p$ stability analysis with numerical simulations}

To support inequality \eqref{eq:gamma_upper_bound} predicted in Theorem \ref{thm:Liu2020PRE}, we conduct numerical simulations to demonstrate its validity for $p=\infty$ at selected Reynolds numbers and $\delta$. 
The numerical simulation employs a fifth-order Runge–Kutta–Fehlberg method (RKF45) \cite{fehlberg1969low} over the time interval $t\in [0,\;7\times 10^{4}]$ (i.e., $\tau=7\times 10^4$) with a fixed time step $0.07$, and this time interval is long enough to damp out initial transient and reach statistically steady states as shown later. Given each $\text{Re}$ and $\delta$ pair, we first solve LMI in \eqref{eq:sdp_P} to find a $\boldsymbol{P}$ matrix such that \eqref{eq:sdp_P} is feasible in Theorem \ref{thm:Liu2020PRE}. From Theorem \ref{thm:Liu2020PRE}, we can obtain the upper bound on the initial condition amplitude $\delta_f$ in \eqref{eq:delta_f_LMI}, the upper bound on forcing amplitude $f_{\text{LMI}}$ in \eqref{eq:forcing_upper_bound}, and the upper bound on input-output amplification $\gamma$ in \eqref{eq:gamma_value}. We then set up numerical simulations of the nonlinear input-output system in equation \eqref{eq:simp_sys} with a random initial disturbance with amplitude as $\|\boldsymbol{a}_0\|=\delta_f$ in \eqref{eq:delta_f_LMI} and random forcing with amplitude as $\|\boldsymbol{f}(t)\|=f_{\text{LMI}}$ in \eqref{eq:forcing_upper_bound}. Here, all random numbers are normally distributed and generated by the \texttt{randn} command in MATLAB with zero mean and unit variance, and then we normalize the vector norm of $\boldsymbol{a}_0$ and $\boldsymbol{f}(t)$ as $\delta_f$ and $f_{\text{LMI}}$, respectively. The random forcing $\boldsymbol{f}(t)$ is generated and normalized to $f_{\text{LMI}}$ amplitude at each time step. In this subsection, we only compare LMI prediction in Theorem \ref{thm:Liu2020PRE} with numerical simulations, because SOS provides similar results as shown in Figure~\ref{fig:gamma_forcing_upper_bound}.

With this setup in numerical simulations, we have $ \|\boldsymbol{f}_\tau(t)\|_{\mathcal{L}_\infty}=\underset{0 \leq t \leq \tau}{\sup}\|\boldsymbol{f}_\tau(t)\| =f_{\text{LMI}}$, and thus the right-hand side (RHS) of inequality \eqref{eq:gamma_upper_bound} in Theorem \ref{thm:Liu2020PRE} associated with $p=\infty$ is 
\begin{align}
    \xi:=\gamma f_{\text{LMI}}+\beta_\infty,
    \label{eq:thupper_output}
\end{align}
where 
\begin{align}
\beta_{\infty} = \|\boldsymbol{a}_0\| \sqrt{\frac{\lambda_{\max}(\boldsymbol{P})}{\lambda_{\min}(\boldsymbol{P})}}=\delta_f\sqrt{\frac{\lambda_{\max}(\boldsymbol{P})}{\lambda_{\min}(\boldsymbol{P})}}=\delta,
\label{eq:beta_infty}
\end{align}
according to \eqref{eq:beta_value}, \eqref{eq:delta_f_LMI}, and our setup of $\|\boldsymbol{a}_0\|$. Based on inequality \eqref{eq:gamma_upper_bound}, we have
\begin{align}
\|\boldsymbol{y}_\tau(t)\|_{\mathcal{L}_\infty}\leq \xi.
\end{align}
Using the definition of $\mathcal{L}_\infty$ norm in \eqref{eq:L_infty}, we have 
\begin{align}
    \|\boldsymbol{y}_\tau(t)\|\leq \xi, \;\;\forall t\in [0,\tau]. 
    \label{eq:y_output_norm_xi}
\end{align}

\begin{figure}[!htbp]
\centering
\begin{subfigure}[b]{0.496\textwidth}
  \includegraphics[width=\linewidth]{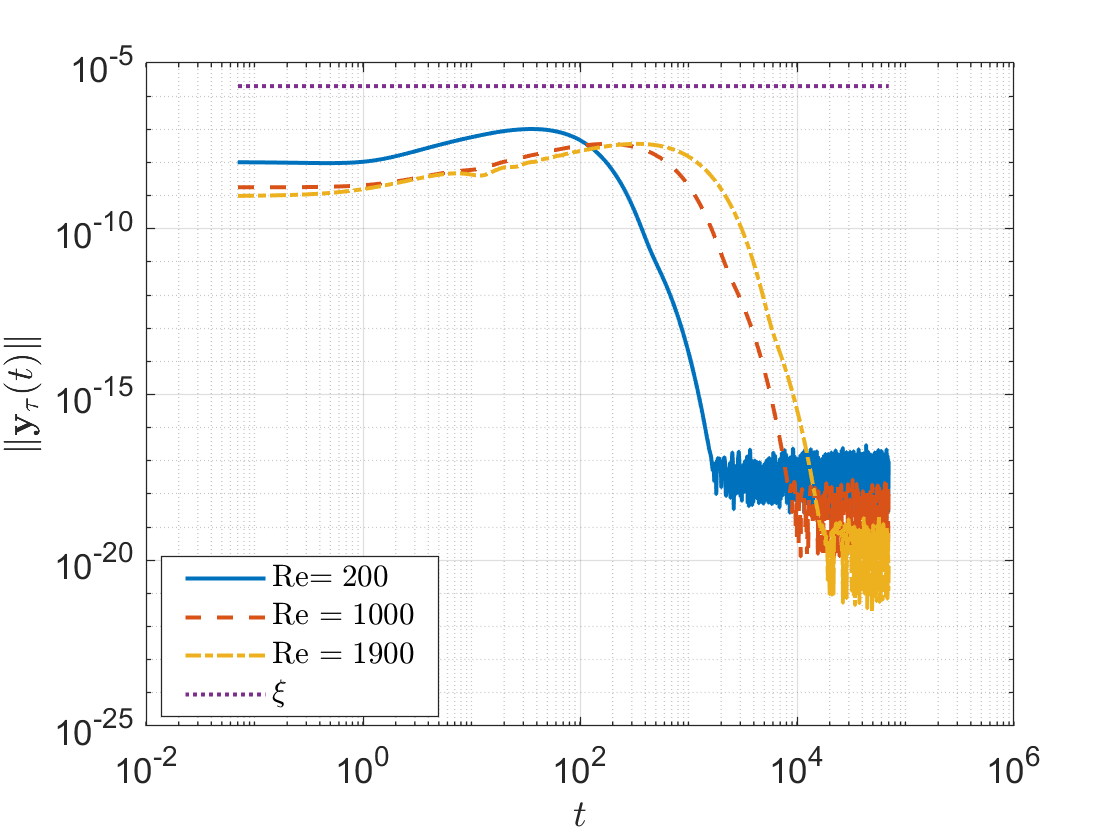}
  \caption{}
  \label{fig:upper_bound} 
\end{subfigure}
\hfill
\begin{subfigure}[b]{0.496\textwidth}
  \includegraphics[width=\linewidth]{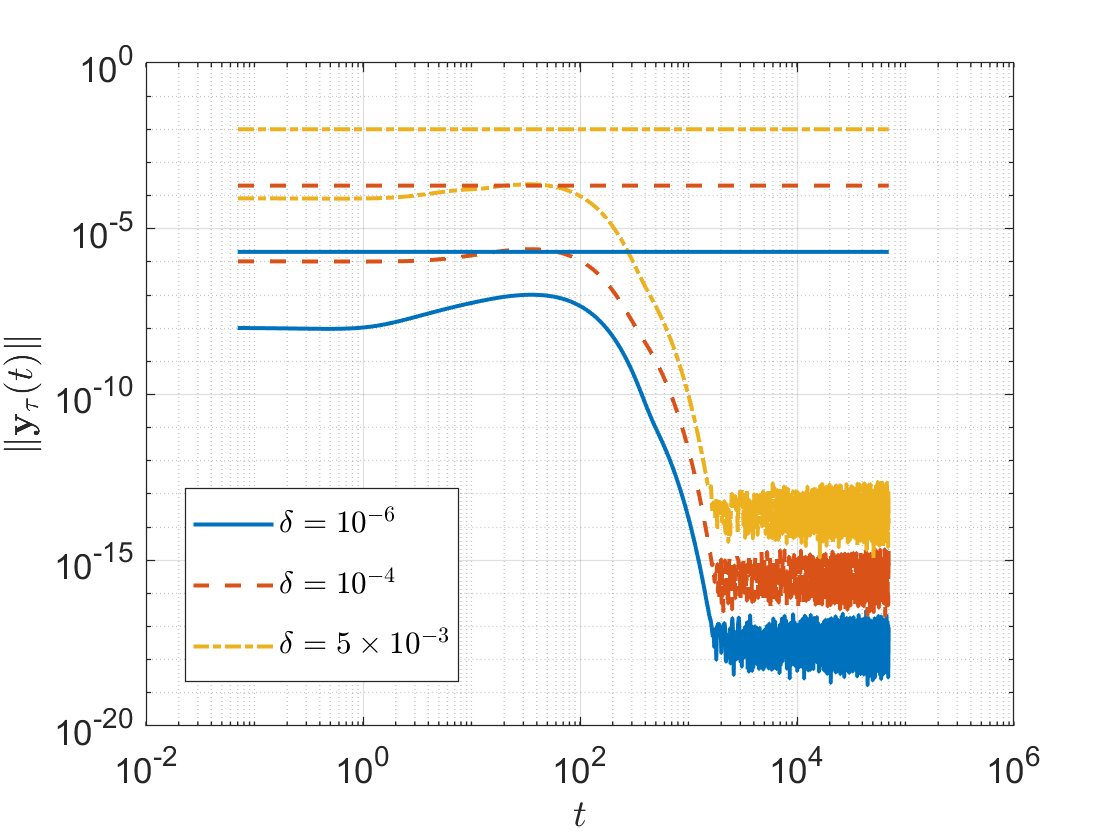}
  \caption{}
  \label{fig:upper_bound_delta} 
\end{subfigure}
\caption{\justifying (a) The norm of output $\|\boldsymbol{y}_\tau(t)\|$ from simulations at Re = 200, 1000, 1900, and $\delta = 10^{-6}$. (b) $\|\boldsymbol{y}_\tau(t)\|$ at Re = 200, and $\delta = 10^{-6}, 10^{-4},$ and $ 5\times10^{-3}$. All simulations are driven by initial disturbance amplitude as $\|\boldsymbol{a}_0\|=\delta_f$ based on \eqref{eq:delta_f_LMI} and random forcing amplitude as $\|\boldsymbol{f}(t)\|=f_{\text{LMI}}$ based on \eqref{eq:forcing_upper_bound}. In both panels, horizontal lines are the theoretical upper bound $\xi$ of the $\mathcal{L}_\infty$ norm of the output as defined in equation \eqref{eq:thupper_output}.}
\label{fig:combined_upper_bounds} 
\end{figure}

Figure \ref{fig:upper_bound} demonstrates that the simulated output norm $\|\boldsymbol{y}_\tau(t)\|$ at $\text{Re}=200,\;1000,\;1900$ and $\delta=10^{-6}$ remains bounded by the theoretical $\mathcal{L}_{\infty}$ upper bound $\xi$ in inequality~\eqref{eq:y_output_norm_xi}, demonstrating that the SSFG $\mathcal{L}_p$ stability predicted by Theorem \ref{thm:Liu2020PRE} is consistent with numerical simulations. The upper bound of output $\xi$ (horizontal line in Figure \ref{fig:upper_bound}) is the same for different \rev{Reynolds} numbers. This independence over Reynolds number can be shown by 
\begin{subequations}
\begin{align}
\gamma f_{\text{LMI}}=&\frac{\lambda_{\text{max}}(\boldsymbol{P}) \|2\boldsymbol{P}\|}{\lambda_{\text{min}}(\boldsymbol{P}) \varepsilon}  \frac{\lambda_{\text{min}}(\boldsymbol{P}) \varepsilon \delta}{\lambda_{\text{max}}(\boldsymbol{P}) \|2\boldsymbol{P}\| }=\delta,\label{eq:xi_simplify_a}\\
    \xi=&\gamma f_{\text{LMI}}+\beta_\infty=2\delta \label{eq:xi_simplify_b},
\end{align}
\end{subequations}
where we substitute the expressions of $f_{\text{LMI}}$, $\gamma$, and $\beta_\infty$ in equations \eqref{eq:forcing_upper_bound}, \eqref{eq:gamma_value}, \eqref{eq:beta_infty}, respectively. Figure \ref{fig:upper_bound} also shows that, as the Reynolds number increases, the system requires more time to dampen out the transient response due to initial conditions and stabilize into the final statistically steady state driven by the random forcing. This behavior aligns with the physical intuition that higher Reynolds numbers introduce weaker viscous damping effects and thus a longer time to dampen out the transient response.

Figure~\ref{fig:upper_bound_delta} illustrates the norm of output $\|\boldsymbol{y}_\tau(t)\|$ for Reynolds number $\text{Re} = 200$ and $\delta = 10^{-6}, 10^{-4}, 5\times10^{-3}$ obtained in the same manner as Figure \ref{fig:upper_bound}. The results indicate that the overall trend of the output norm remains lower than the theoretical upper bound in Theorem \ref{thm:Liu2020PRE} across different $\delta$, which supports that the SSFG $\mathcal{L}_p$ stability theorem can provide a valid upper bound of the output norm. Lower values of $\delta$ result in proportionally lower theoretical upper bound $\xi$ consistent with equation \eqref{eq:xi_simplify_b}. Figure \ref{fig:upper_bound_delta} also suggests that reducing the value of $\delta$ does not alter the qualitative behavior of the output response, but only scales down the output magnitude. 

\begin{figure}[!htbp]
\centering
\begin{subfigure}[b]{0.496\textwidth}
  \includegraphics[width=\linewidth]{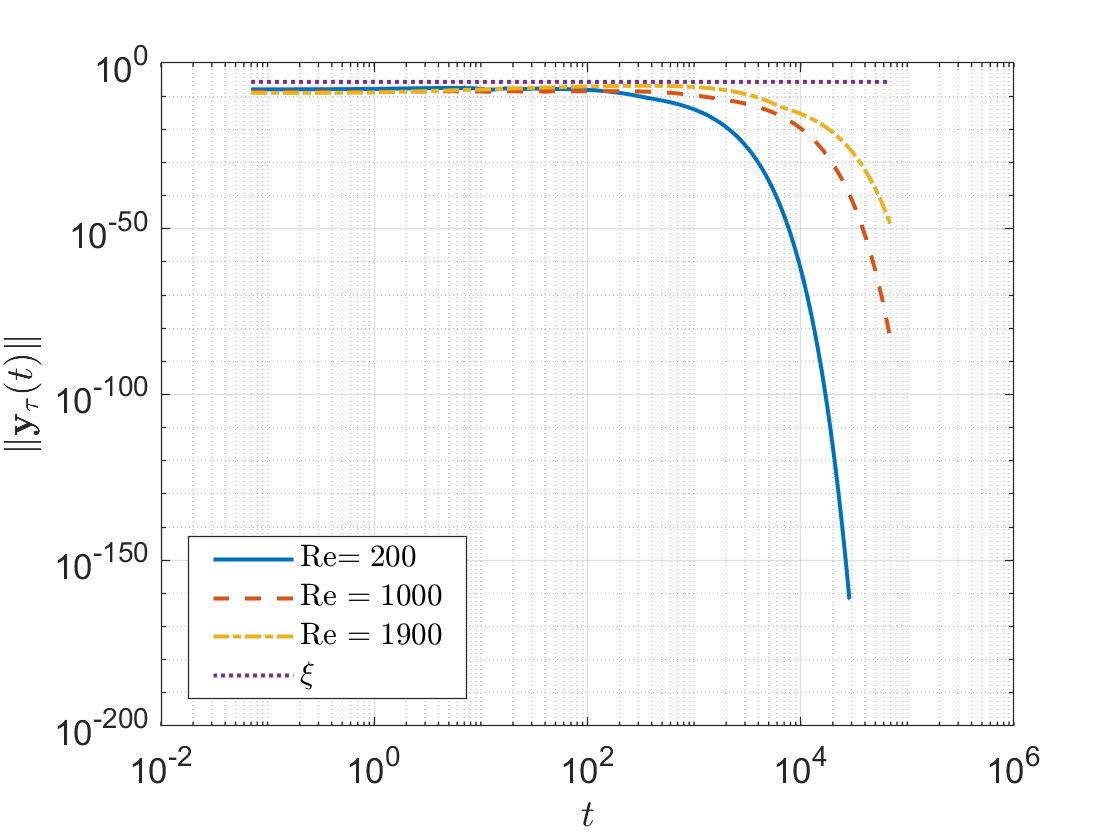}
  \caption{}
  \label{fig:no_forcing_norm_analysis} 
\end{subfigure}
\hfill
\begin{subfigure}[b]{0.496\textwidth}
  \includegraphics[width=\linewidth]{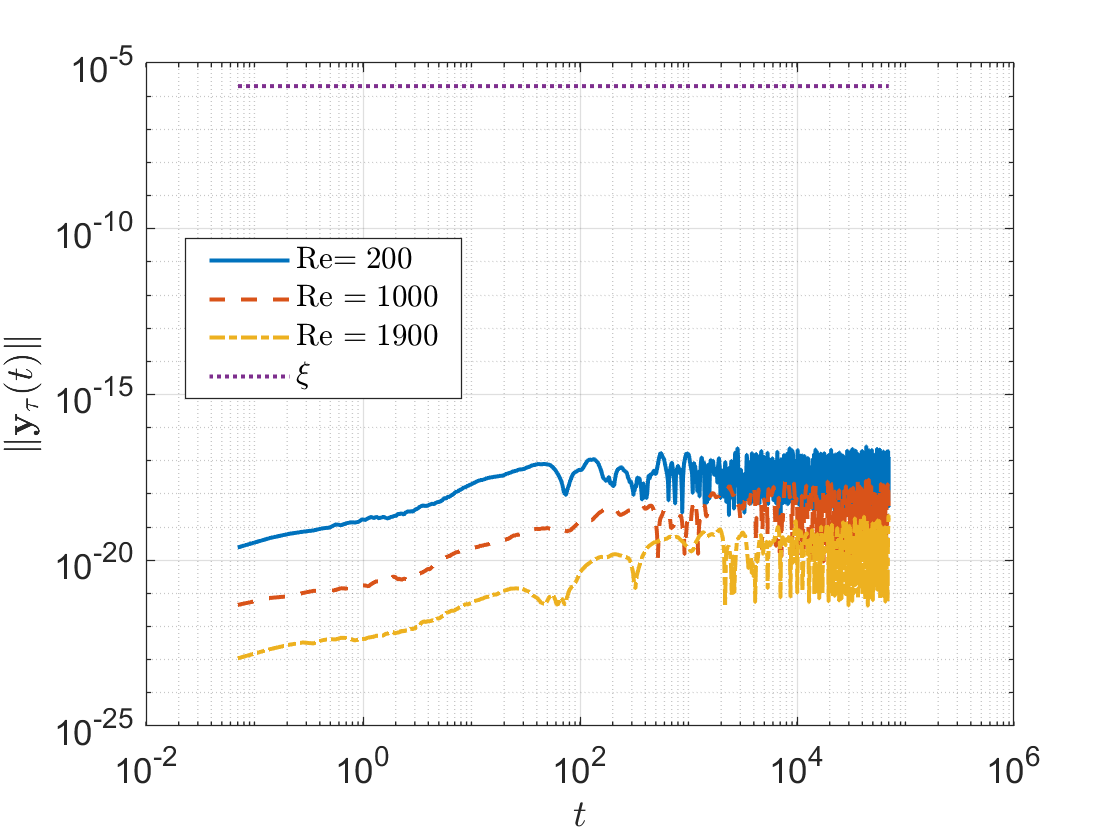}
  \caption{}
  \label{fig:no_initial_norm_anysis}
\end{subfigure}
\caption{\justifying (a) The norm of output $\|\boldsymbol{y}_\tau(t)\|$ from simulations at Re = 200, 1000, 1900, $\delta = 10^{-6}$ driven by initial condition ($\boldsymbol{a}_0\neq\boldsymbol{0}$ with $\|\boldsymbol{a}_0\|=\delta_f$ based on \eqref{eq:delta_f_LMI}) without input forcing ($\boldsymbol{f}=\boldsymbol{0}$). (b) The norm of output $\|\boldsymbol{y}_\tau(t)\|$ from simulations at Re = 200, 1000, 1900, $\delta = 10^{-6}$ driven by input forcing ($\boldsymbol{f}\neq \boldsymbol{0}$ and $\|\boldsymbol{f}(t)\|=f_{\text{LMI}}$ based on \eqref{eq:forcing_upper_bound}) without initial condition ($\boldsymbol{a}_0=0$). }
\label{fig:norm_analysis_noinitial_noforcing} 
\end{figure}

To determine whether the final state in Figures \ref{fig:upper_bound} and \ref{fig:upper_bound_delta} are driven by random forcing, we also conduct numerical simulations driven by initial condition ($\boldsymbol{a}_0\neq \boldsymbol{0}$) without forcing ($\boldsymbol{f}=\boldsymbol{0}$, Fig.~\ref {fig:no_forcing_norm_analysis}) and simulations driven by forcing ($\boldsymbol{f}\neq \boldsymbol{0}$) without initial condition ($\boldsymbol{a}_0=\boldsymbol{0}$, Fig.~\ref{fig:no_initial_norm_anysis}). The non-zero initial condition $\boldsymbol{a}_0$ or non-zero forcing $\boldsymbol{f}$ is based on the same setup as Fig.~\ref{fig:upper_bound}. Figure~\ref{fig:no_forcing_norm_analysis} shows that when we have zero forcing term, the response will decay to the laminar base state ($\boldsymbol{y}=\boldsymbol{a}=\boldsymbol{0}$), and the value of $\|\boldsymbol{y}_\tau(t)\|$ can decay to $\sim O(10^{-150})$. This value, although much smaller than machine precision, is still larger than the smallest normalized floating-point number of double precision in MATLAB $\sim O(10^{-308})$ that can be checked by \texttt{realmin} command. Figure~\ref{fig:no_initial_norm_anysis} shows that when we have zero initial condition, the output norm $\|\boldsymbol{y}_\tau(t)\|$ will build up initially and then saturate into a statistically steady state in a magnitude similar to Figure~\ref{fig:upper_bound}. As a result, we can conclude that the final state in Figure~ \ref{fig:upper_bound} is driven by the random input forcing. 

\begin{figure}[!htbp]
\centering
\includegraphics[width=0.8\linewidth]{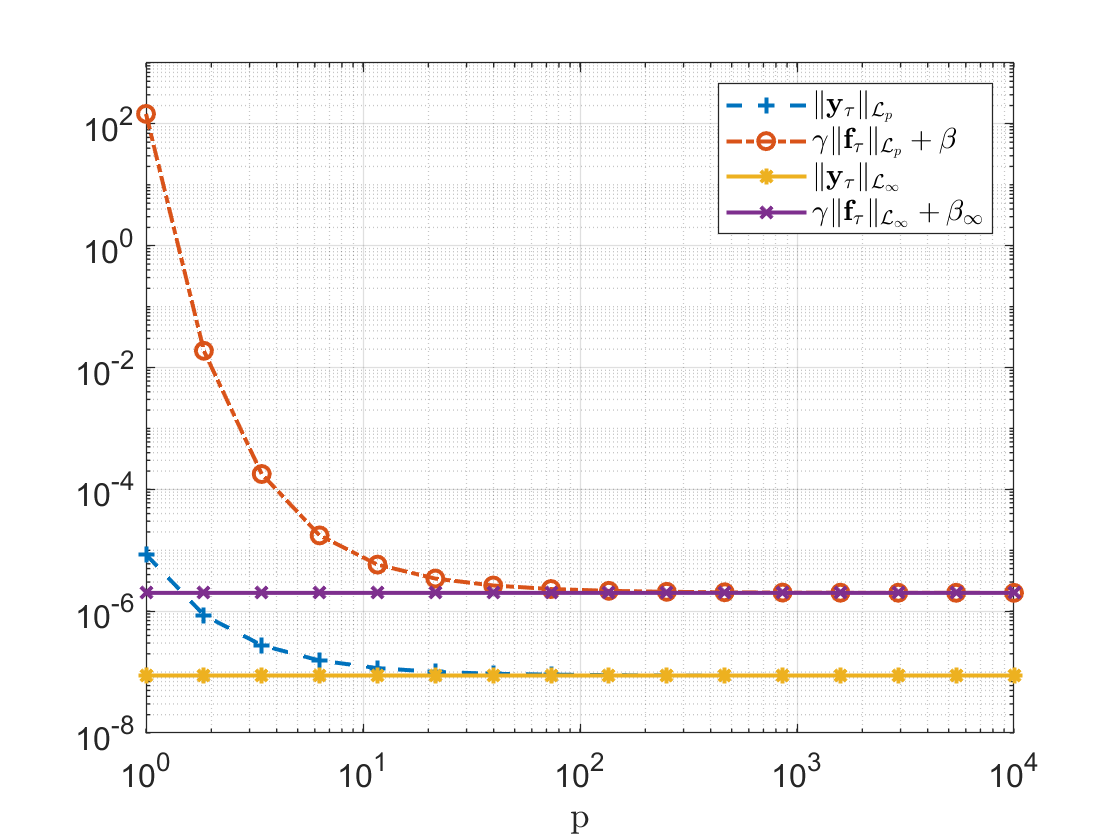} 
    \caption{\justifying The left-hand side (LHS) of inequality \eqref{eq:gamma_upper_bound} ($ \left\|\boldsymbol{y}_\tau\right\|_{\mathcal{L}_p} $) and the right-hand side (RHS) of inequality \eqref{eq:gamma_upper_bound} ($ \gamma\left\|\boldsymbol{f}_\tau\right\|_{\mathcal{L}_p}+\beta $) versus $p$ at Re = 200 and $\delta$ = $10^{-6}$. The gap between the LHS and RHS converges to a constant value as $p$ increases. The horizontal lines correspond to the LHS ($\left\|\boldsymbol{y}_{\tau}\right\|_{\mathcal{L}_\infty}$) and RHS ($\xi=\gamma\left\|\boldsymbol{f}_{\tau}\right\|_{\mathcal{L}_\infty} + \beta_\infty$ in \eqref{eq:thupper_output}) of inequality \eqref{eq:gamma_upper_bound} at $ p $ = $\infty$. }
\label{fig:inequality}
\end{figure}

We then investigate the validity of inequality \eqref{eq:gamma_upper_bound} in Theorem \ref{thm:Liu2020PRE} over different $p$ values. Here, we conduct the numerical simulations in the same manner as Figures \ref{fig:upper_bound} and \ref{fig:upper_bound_delta} with initial disturbance amplitude as $\delta_f$ in equation \eqref{eq:delta_f_LMI} and random forcing $\boldsymbol{f}$ amplitude as $f_{\text{LMI}}$ in equation \eqref{eq:forcing_upper_bound}. Then, we obtain the output $\boldsymbol{y}$ from numerical simulation to compute the left-hand side (LHS) of inequality \eqref{eq:gamma_upper_bound} ($ \left\|\boldsymbol{y}_\tau\right\|_{\mathcal{L}_p} $). The right-hand side (RHS) of the inequality \eqref{eq:gamma_upper_bound} ($ \gamma\left\|\boldsymbol{f}_\tau\right\|_{\mathcal{L}_p}+\beta $) is computed based on the forcing $\boldsymbol{f}$ with $\gamma$ and $\beta$ obtained from equation \eqref{eq:gamma_beta} in Theorem \ref{thm:Liu2020PRE}. The input-output amplification $\gamma$ in equation \eqref{eq:gamma_value} is independent of $p$, while the bias $\beta$ depends on $p$ as shown in \eqref{eq:beta_value}. The computation of $\|\cdot\|_{\mathcal{L}_p}$ norm as \eqref{eq:Lpnorm} also introduces the dependence on $p$ in both LHS and RHS of inequality \eqref{eq:gamma_upper_bound}. This comparison between LHS and RHS of the inequality \eqref{eq:gamma_upper_bound} is comparing the difference between the theoretical prediction from Theorem \ref{thm:Liu2020PRE} (RHS of \eqref{eq:gamma_upper_bound}) and numerical simulations (LHS of \eqref{eq:gamma_upper_bound}). Figure~\ref{fig:inequality} compares the LHS and RHS of inequality~\eqref{eq:gamma_upper_bound} across varying values of $p$. Throughout all values of $p$, the LHS of ~\eqref{eq:gamma_upper_bound} remains below the lines representing the RHS of ~\eqref{eq:gamma_upper_bound}, which demonstrates that the inequality~\eqref{eq:gamma_upper_bound} is satisfied for all values of $p$. As $p$ increases, the gap between the LHS and RHS decreases, indicating that the two sides of inequality~\eqref{eq:gamma_upper_bound} become closer. Eventually, the gap converges to an asymptotic value associated with $ p=\infty$.  This further supports that the prediction of the SSFG $\mathcal{L}_p$ stability theorem is consistent with our observations from numerical simulations across all $p\in [1,\infty]$ as stated in Theorem \ref{thm:Liu2020PRE}.

\begin{figure}[!htbp]
    \centering
    \begin{subfigure}{0.495\textwidth}
        \centering
        \includegraphics[width=\textwidth]{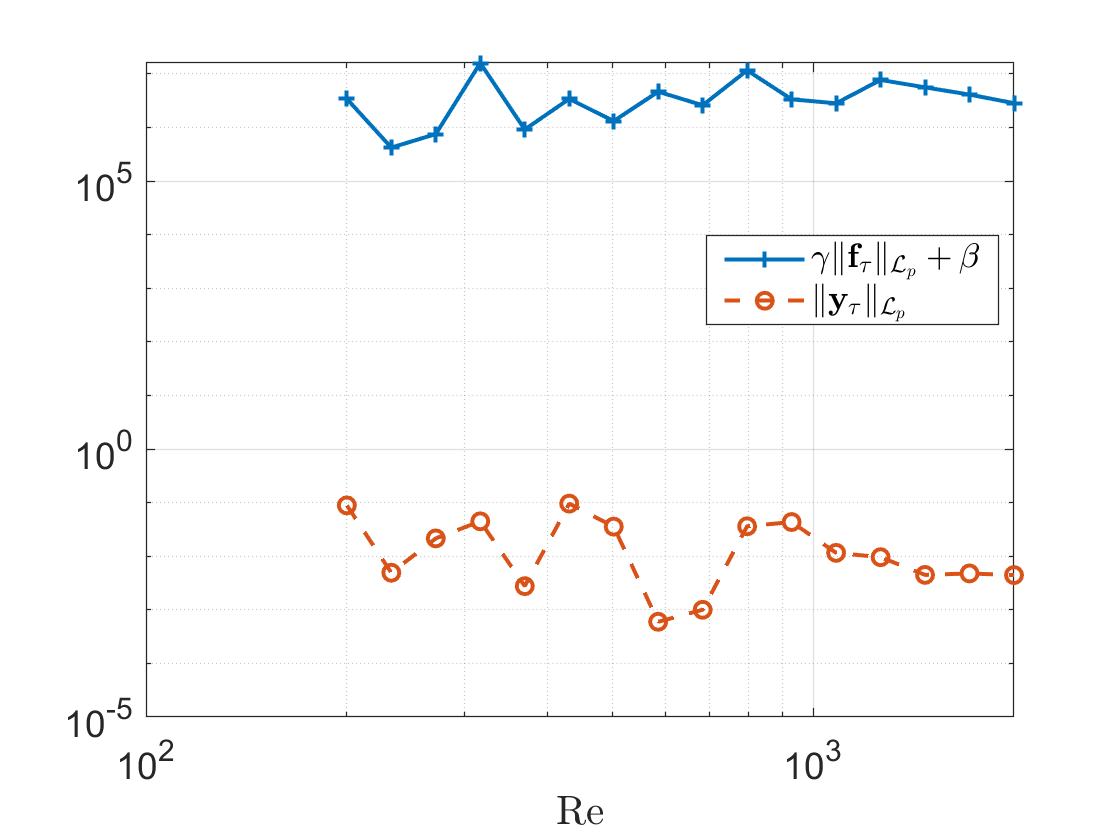}
        \caption{when p = 1}
        \label{fig:1}
    \end{subfigure}
    \begin{subfigure}{0.495\textwidth}
        \centering
        \includegraphics[width=\textwidth]{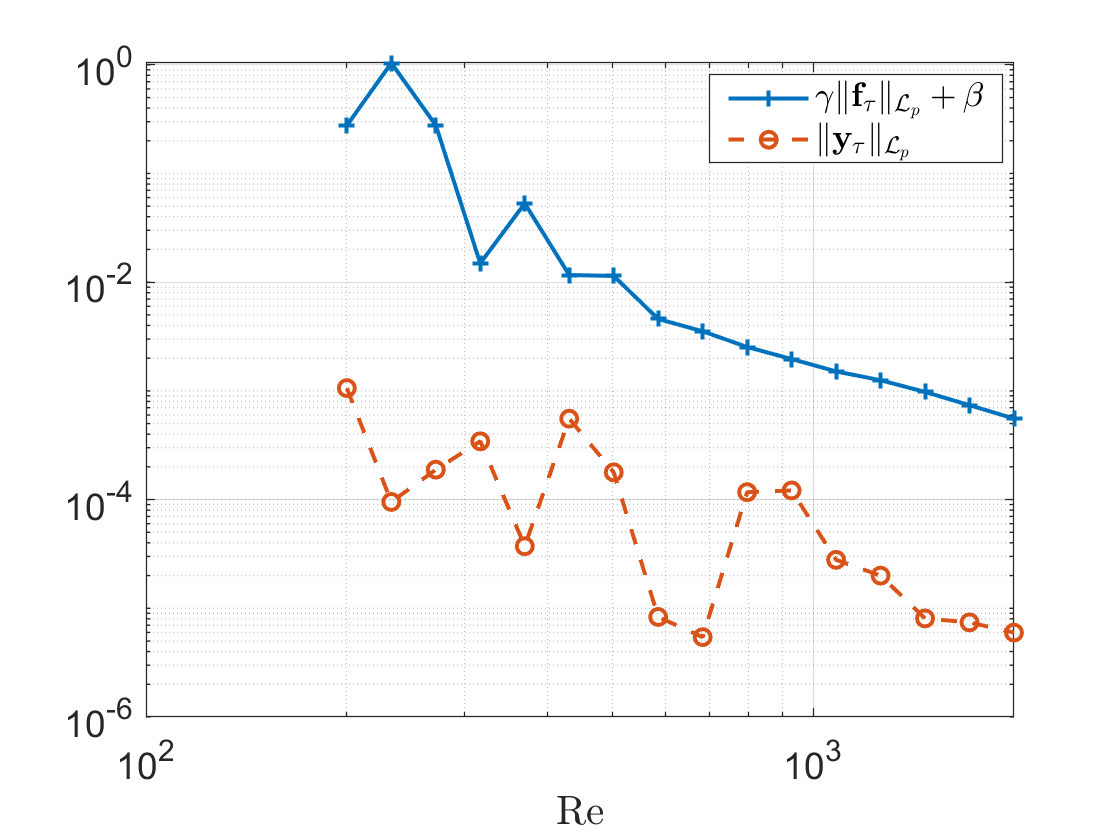}
        \caption{when p = 6.3096}
        \label{fig:6.3096}
    \end{subfigure}
    \begin{subfigure}{0.5\textwidth}
        \centering
        \includegraphics[width=\textwidth]{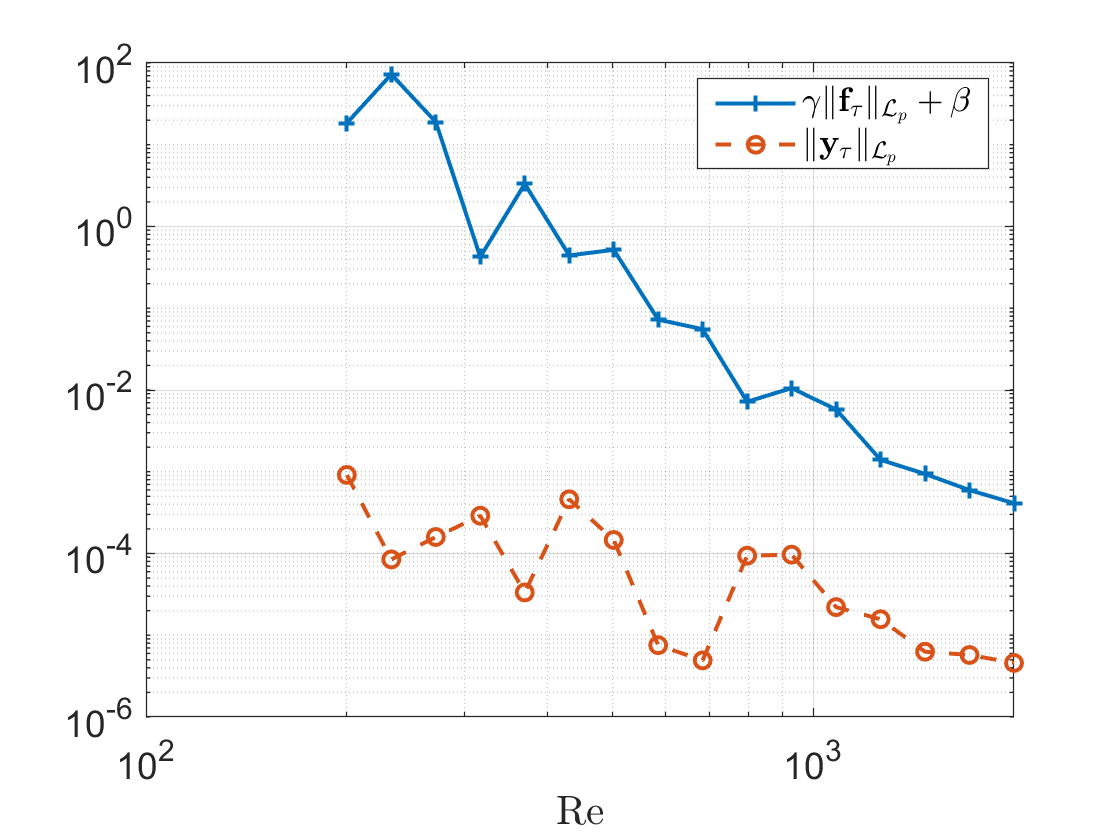}
        \caption{when p = $10^4$}
        \label{fig:10000}
    \end{subfigure}
    \begin{subfigure}{0.49\textwidth}
    \centering
    \includegraphics[width=\linewidth]{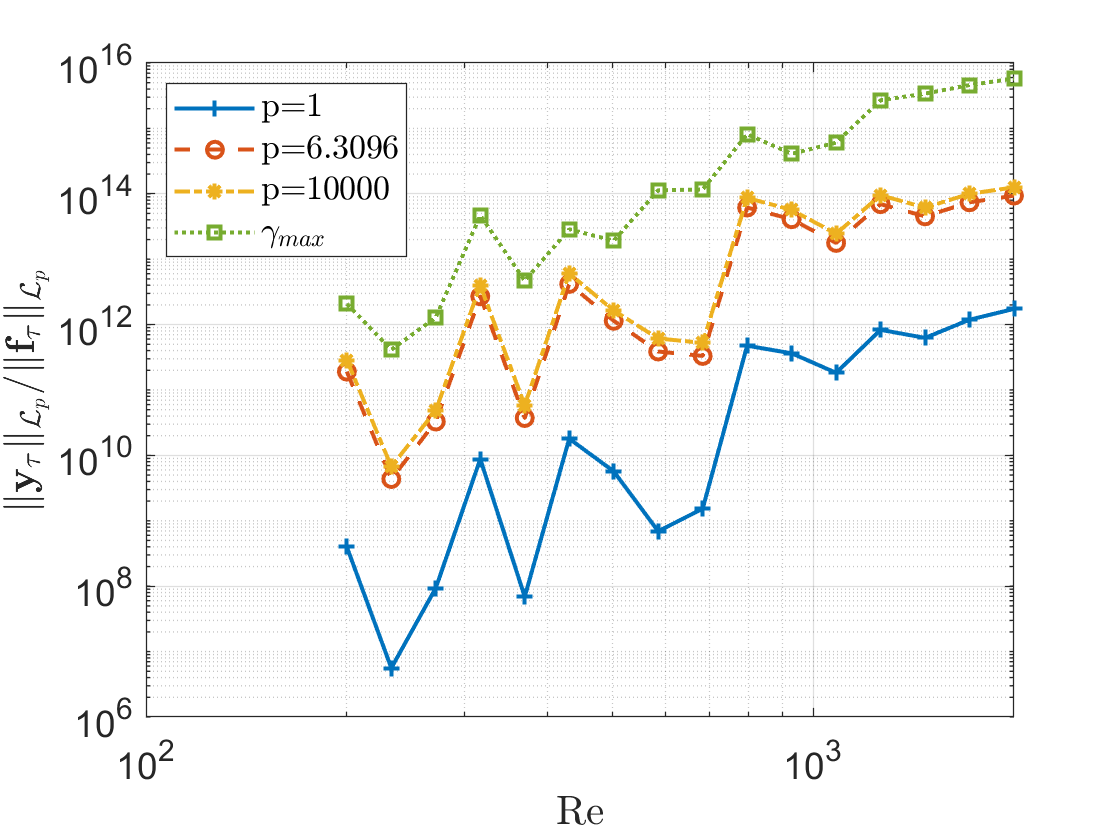} 
    \caption{}
    \label{fig:prop}
    \end{subfigure}
    \caption{\justifying  Variation of \( \gamma \|\boldsymbol{f}_\tau\|_{\mathcal{L}_p} + \beta  \) and \(\|\boldsymbol{y}_\tau\|_{\mathcal{L}_p} \) in inequality \eqref{eq:gamma_upper_bound} over the Reynolds number (\(\text{Re}\)) for (a) $p=1$, (b) $p=6.3096$, and (c) $p=10^4$.  Panel (d) shows the input-output amplification $\|\boldsymbol{y}_\tau\|_{\mathcal{L}_p}/\|\boldsymbol{f}_\tau\|_{\mathcal{L}_p}$ over Re for $p = 1$, $p = 6.3096$, $p = 10^4$, and the input-output gain $\gamma_\text{max}$ as defined in equation \eqref{eq:gamma_max}. }
    \label{fig:GapvarypoverRe}
\end{figure}

We further examine the gap between the LHS and RHS of inequality \eqref{eq:gamma_upper_bound} over different Reynolds numbers. After obtaining the $\gamma(\text{Re},\delta)$ from Theorem \ref{thm:Liu2020PRE}, we then select the largest $\delta$ such that Theorem \ref{thm:Liu2020PRE} is feasible at each Reynolds number to obtain the associated $\gamma$ and $\beta$. We then conduct numerical simulations at this $\text{Re}$ and $\delta$ pair to compute the $\mathcal{L}_p$ norm of input and output. Figures \ref{fig:1}-\ref{fig:10000} demonstrate how higher \( p \)-values reduce the gap between LHS and RHS of inequality \eqref{eq:gamma_upper_bound} in Theorem~\ref{thm:Liu2020PRE}, which is similar to our observation in Figure \ref{fig:inequality}. For $p = 1$ (Figure~\ref{fig:1}), the gap remains large and nearly constant over $\text{Re}$. For $p=10^4$ in Figure~\ref{fig:10000}, the gap between LHS and RHS of inequality \eqref{eq:gamma_upper_bound} also decreases as $\text{Re}$ increases.

To analyze how input-output amplification $\gamma$ predicted by Theorem \ref{thm:Liu2020PRE} compared with numerical results, Figure~\ref{fig:prop} illustrates the input-output amplification from numerical simulations $\|\boldsymbol{y}_{\tau}\|_{\mathcal{L}_p} / \|\boldsymbol{f}_{\tau}\|_{\mathcal{L}_p}$ (corresponding to $\gamma$ in Theorem \ref{thm:Liu2020PRE}) versus the $\mathrm{Re}$ with different $p$. For higher $\mathrm{Re}$, lower viscous damping keeps the input-output amplification comparatively high, and thus, the ratio of output to input increases with $\mathrm{Re}$. The input-output amplification from simulations are then compared with $\gamma_{\max}$, which is maximized over different $\delta$ after solving Theorem \ref{thm:Liu2020PRE} over a range of $(\text{Re}, \delta)$ as defined below:
\begin{align}
\gamma_{\text{max}} = \max_{\delta} {\gamma}(\text{Re},\delta).  \label{eq:gamma_max}
\end{align}
Figure~\ref{fig:prop} shows that $\gamma_{\text{max}}$ provides an upper bound across all $p$ values, illustrating that Theorem \ref{thm:Liu2020PRE} can predict a valid upper bound of input-output amplification consistent with numerical simulations across different $\mathrm{Re}$ and $p$. Moreover, increasing rate of $\gamma_{\text{max}}$ over $\text{Re}$ also closely agree with input-output amplification from simulations $\|\boldsymbol{y}_{\tau}\|_{\mathcal{L}_p} / \|\boldsymbol{f}_{\tau}\|_{\mathcal{L}_p}$; i.e., these lines are nearly parallel in log-log plot as shown in Figure~\ref{fig:prop}. The input-output amplification $\gamma$ predicted from Theorem \ref{thm:Liu2020PRE} is valid for any $p\in [1,\infty]$, and Figure~\ref{fig:prop} shows that higher $p$ values lead to simulation results closer to $\gamma_{\text{max}}$, which is consistent with the trend in Figures~\ref{fig:1}-\ref{fig:10000} that higher $p$ leads to smaller gap in LHS and RHS of inequality~\eqref{eq:gamma_upper_bound}. The Re-dependence further highlights how nonlinear systems can significantly amplify the input forcing, and this input-output amplification increases significantly as we increase $\text{Re}$.

\subsection{\label{sec:comp linear and nonlinear}Comparison with linear finite-gain $\mathcal{L}_p$ stability and linear $\mathcal{L}_2$ stability}

In this subsection, we will then compare the nonlinear SSFG $\mathcal{L}_p$ stability against linear finite-gain $\mathcal{L}_p$ stability and linear $\mathcal{L}_2$ stability. We define the corresponding linear time-invariant system as shown below by dropping the nonlinear term $\boldsymbol{\Upsilon}$ in \eqref{eq:simp_eq}: 
\begin{subequations}\label{eq:LTI}
\begin{align} 
\dot{\boldsymbol{a}}=& \boldsymbol{L} \boldsymbol{a} + \boldsymbol{B}\boldsymbol{f},
\label{eq:LTI_1}
\\
\boldsymbol{y} =& \boldsymbol{C} \boldsymbol{a} + \boldsymbol{D} \boldsymbol{f},
\label{eq:LTI_2}
\end{align}
\end{subequations}
where $\boldsymbol{B} \in \mathbb{R}^{n\times n}$  and $\boldsymbol{C} \in \mathbb{R}^{n\times n}$ are identity matrices, and $\boldsymbol{D} \in \mathbb{R}^{n\times n}$ is the zero matrix. \rev{These $\boldsymbol{B}$, $\boldsymbol{C}$, and $\boldsymbol{D}$ matrices are chosen to be consistent with the form of the nonlinear input-output system defined in \eqref{eq:simp_eq}, and the following results hold for general $\boldsymbol{B}$, $\boldsymbol{C}$, and $\boldsymbol{D}$ matrices.} Then, we present the finite-gain $\mathcal{L}_p$ stability for a linear time-invariant system, which is a direct application of Theorem \ref{thm:small_signal_stability} to linear time-invariant system with a Lyapunov function $V=\boldsymbol{a}^\text{T}\boldsymbol{P}_{cor}\boldsymbol{a}$. 

\begin{corollary} \cite[Corollary 5.2]{Khalil2002}
\label{cor:linear}
The linear time-invariant system \eqref{eq:LTI_1}-\eqref{eq:LTI_2} is finite-gain $\mathcal{L}_p$ stable for each $p \in[1, \infty]$ if $\boldsymbol{L}$ is Hurwitz. Moreover, we have 
\begin{equation}
\left\|\boldsymbol{y}_\tau\right\|_{\mathcal{L}_p} \leq \gamma_{cor}\left\|\boldsymbol{f}_\tau\right\|_{\mathcal{L}_p}+\beta_{cor}
\label{eq:ineq_output_corollary}
\end{equation}
is satisfied with
\begin{subequations}
\begin{align}
\gamma_{cor}&=\|\boldsymbol{D}\|_2+\frac{2 \lambda_{\max }^2(\boldsymbol{P}_{cor})\|\boldsymbol{B}\|_2\|\boldsymbol{C}\|_2}{\lambda_{\min }(\boldsymbol{P}_{cor})}, \\ \quad \beta_{cor}&=\rho\|\boldsymbol{C}\|_2\left\|\boldsymbol{a}_0\right\| \sqrt{\frac{\lambda_{\max }(\boldsymbol{P}_{cor})}{\lambda_{\min }(\boldsymbol{P}_{cor})}},\;\;\;
\rho= \begin{cases}1, & \text { if } p=\infty \\ \left(\frac{2 \lambda_{\max }(\boldsymbol{P}_{cor})}{p}\right)^{1 / p}, & \text { if } p \in[1, \infty),\end{cases}
\end{align}
\end{subequations}
and $\boldsymbol{P}_{cor}$ is the solution of the Lyapunov equation $\boldsymbol{P}_{cor} \boldsymbol{L} + \boldsymbol{L}^\text{T} \boldsymbol{P}_{cor} = -\boldsymbol{I}$.
\end{corollary}

We then present the finite $\mathcal{L}_2$ gain of a linear time-invariant system. 
\begin{theorem} \cite[Theorem 5.4]{Khalil2002}\label{thm:linear}
    Consider the linear time-invariant system \eqref{eq:LTI_1}-\eqref{eq:LTI_2}
where $\boldsymbol{L}$ is Hurwitz. Let $\mathcal{\boldsymbol{H}}(s):=\boldsymbol{C}(s\boldsymbol{I}-\boldsymbol{L})^{-1} \boldsymbol{B}+\boldsymbol{D}$. Then, the $\mathcal{L}_2$ gain of the system is $\gamma_{\mathcal{L}_2}:=\underset{\omega \in R}{\sup}\;\|\mathcal{\boldsymbol{H}}(j \omega)\|_2=\underset{\omega \in R}{\sup}\;\bar{\sigma}[\mathcal{H}(j \omega)]$, i.e., $\|\boldsymbol{y}\|_{\mathcal{L}_2}\leq \gamma_{\mathcal{L}_2}\|\boldsymbol{f}\|_{\mathcal{L}_2}$.
\end{theorem}
In Theorem \ref{thm:linear}, $\text{j}:=\sqrt{-1}$ is imaginary unit, $\omega\in \mathbb{R}$ is temporal frequency, $s\in \mathbb{C}$ is a complex frequency-domain parameter, and $\bar{\sigma}[\cdot]$ is the largest singular value. The $\mathcal{L}_2$ gain $\gamma_{\mathcal{L}_2}$ is also known as $\mathcal{H}_\infty$ norm of the transfer matrix $\mathcal{H}(s)$ \citep{Boyd1994} \rev{and is} widely used to analyze transition to turbulence \citep{jovanovic2004modeling,schmid2007nonmodal,Hwang2010b,illingworth2020streamwise}. Here, we compute $\mathcal{L}_2$ gain $\gamma_{\mathcal{L}_2}$ by \texttt{hinfnorm} command in MATLAB. The singular value-based analysis is also closely related to resolvent analysis widely used in wall-bounded shear flows; e.g., \citep{mckeon2010critical,mckeon2017engine,illingworth2018estimating,symon2021energy}. Due to the linear superposition principle, the input-output gain of linear systems is independent of the forcing amplitude, and therefore, Corollary \ref{cor:linear} and Theorem \ref{thm:linear} do not require \emph{small-signal} as Theorems \ref{thm:small_signal_stability}-\ref{thm:SOS}; i.e., they do not require the forcing amplitude to be smaller than a threshold value (e.g., \eqref{eq:forcing_upper_bound} and \eqref{eq:forcing_upper_bound_SOS}) to sustain a finite input-output gain.

\begin{figure}[!htbp]
        \centering
        \includegraphics[width=0.8\textwidth]{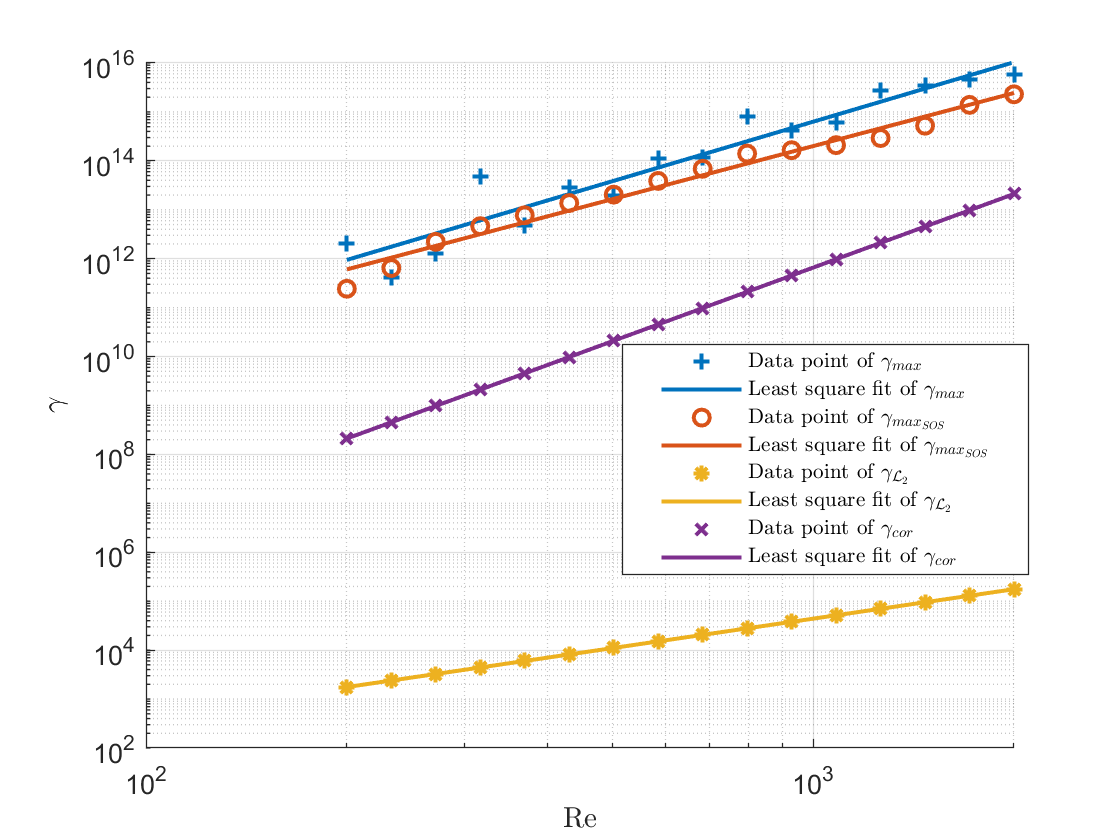}
        \caption{\justifying  A comparison between nonlinear $\mathcal{L}_p$ gain including $\gamma_{max}$ using LMI in Theorem \ref{thm:Liu2020PRE} and nonlinear $\mathcal{L}_p$ gain $\gamma_{max_{\text{SOS}}}$ using SOS in Theorem \ref{thm:SOS}, linear $\mathcal{L}_p$ gain $\gamma_{cor}$ in Corollary \ref{cor:linear}, and linear $\mathcal{L}_2$ gain $\gamma_{\mathcal{L}_2}$ in Theorem \ref{thm:linear}. Markers are input-output gain from the computation of these Theorems and Corollary, and lines are the least-squares fitting in the form of $\gamma=10^A Re^{\zeta}$ with values of $A$ and $\zeta$ in Table \ref{tab:scaling_exponents}.}
        \label{fig:leastsqft}
    \end{figure}

Figure \ref{fig:leastsqft} compares the input-output amplification $\gamma$ as a function of Re using four methodologies: (i) nonlinear $\mathcal{L}_p$ gain $\gamma_{max}$ in \eqref{eq:gamma_max} obtained from the nonlinear SSFG $\mathcal{L}_p$ stability in Theorem \ref{thm:Liu2020PRE} using LMI, (ii) nonlinear $\mathcal{L}_p$ gain $\gamma_{max_{\text{SOS}}}:= \underset{\delta}{\max}\;{\gamma_{\text{SOS}}}(\text{Re},\delta)$ where $\gamma_{\text{SOS}}$ is computed based on Theorem \ref{thm:SOS} using SOS, (iii) linear $\mathcal{L}_p$ gain $\gamma_{cor}$ from Corollary \ref{cor:linear}, and (iv) linear $\mathcal{L}_2$ gain $\gamma_{\mathcal{L}_2}$ from Theorem \ref{thm:linear}. At all Reynolds numbers, we found that nonlinear $\mathcal{L}_p$ gain ($\gamma_{max}$ using LMI in Theorem \ref{thm:Liu2020PRE} and $\gamma_{max_{\text{SOS}}}$ using SOS in Theorem \ref{thm:SOS}) are in the same order of magnitude, and $\gamma_{max}$ is slightly larger than $\gamma_{max_{\text{SOS}}}$. This indicates that SOS provides a less conservative estimation of $\mathcal{L}_p$ input-output gain, which is similar to the region of attraction analysis where SOS can systematically improve the prediction compared with LMI \citep{Liu2020PRE}. The nonlinear $\mathcal{L}_p$ gain computed by either LMI or SOS is several orders of magnitude higher than the linear $\mathcal{L}_p$ gain ($\gamma_{cor}$ from Corollary \ref{cor:linear}), which indicates that nonlinear effects can significantly increase the input-output amplification. Moreover, both nonlinear and linear $\mathcal{L}_p$ gains are also much higher than linear $\mathcal{L}_2$ gain ($\mathcal{H}_\infty$ norm) from Theorem \ref{thm:linear}, which is likely because our $\mathcal{L}_p$ gain (either linear or nonlinear) is valid for each $p\in [1,\infty]$, while $\mathcal{L}_2$ gain only corresponds to a special value of $p=2$.

We also fit all these four different input-output gain as a function \(\gamma = 10^{A} \text{Re}^{\zeta}\) via the least-squares fitting with coefficients \(A\) and \(\zeta\) reported in Table~\ref{tab:scaling_exponents}, and fitted lines are also shown in Figure \ref{fig:leastsqft}. Here, we found that the nonlinear $\mathcal{L}_{p}$ gain from LMI scales as $\sim \text{Re}^{4.04}$, which has a scaling exponent close to but lower than the linear $\mathcal{L}_p$ gain with a scaling of $\sim \text{Re}^{5.00}$. Moreover, the nonlinear $\mathcal{L}_p$ gain from SOS scales as $\sim \text{Re}^{3.61}$ with a scaling exponent lower than that obtained by LMI. The linear $\mathcal{L}_2$ gain scales as $\sim \text{Re}^{2.00}$, which is the scaling law of $\mathcal{H}_\infty$ norm for transitional wall-bounded shear flows \citep{trefethen1993hydrodynamic,jovanovic2004modeling,liu2021structured}. Such a scaling exponent is much lower than either linear $\mathcal{L}_p$ gain or nonlinear $\mathcal{L}_p$ gain. This comparison shows that when measuring the input-output gain based on general $\mathcal{L}_p$ ($p\in [1,\infty]$) space, the input-output gain will increase with the Reynolds number much faster than input-output amplification measured by $\mathcal{L}_2$ gain ($\mathcal{H}_\infty$ norm).

\begin{table}[!htbp]
    \centering
    \caption{ A and $\zeta$ obtained from a least-squares fitting to $\gamma$ = $10^A$$\text{Re}^\zeta$ with four different type of $\gamma$ in Figure \ref{fig:leastsqft}.}
    \label{tab:scaling_exponents}
    \begin{tabular}{@{}lcc@{}}
    \toprule
    \text{Input-output gain $\gamma$} & \(\zeta\) & \(A\) \\
    \midrule
    $\gamma_{max}$ & 4.0410 & 2.6771 \\
    $\gamma_{max_{\text{SOS}}}$ & 3.6058 & 3.4844 \\
    $\gamma_{cor}$ & 4.9991 & -3.1790 \\
    $\gamma_{\mathcal{L}_2}$ & 1.9999 & -1.3562 \\
    \bottomrule
    \end{tabular}
\end{table}

\subsection{\label{sec:comp SOS and Bi} Permissible forcing amplitude compared with bisection}

In this subsection, we will further compare the permissible forcing amplitude $f_{\text{LMI}}$ and $f_{\text{SOS}}$ against the permissible forcing amplitude we determine from numerical simulations. We modify the forcing amplitude in numerical simulations and use bisection method to determine the permissible forcing amplitude $f_{\Pi}(\text{Re},\delta)$, below which $\|\boldsymbol{y}_\tau(t)\|_{\mathcal{L}_\infty}\leq \delta$ and above which certain random forcing will lead to $\|\boldsymbol{y}_\tau(t)\|_{\mathcal{L}_\infty}>\delta$. As we choose output the same as state variable (i.e., $\boldsymbol{y}=\boldsymbol{a}$ in \eqref{eq:simp_sys_2}), this bisection criteria directly corresponds to the state variable condition of equation \eqref{eq:upper_bound_state_variable_LMI} in Theorem \ref{thm:Liu2020PRE} and equation \eqref{eq:upper_bound_state_variable_SOS} in Theorem \ref{thm:SOS}. We start with \(f_{\text{min}} = 0\) and \(f_{\text{max}} = 10^{-2}\) as the initial bounds of the forcing amplitude in the bisection search. We can then compute the midpoint forcing amplitude at iteration \(k\) defined as: 
$f_k = \frac{f_{\text{min}} + f_{\text{max}}}{2}$. For each \(f_k\), the random forcing $\boldsymbol{f}(t)$ is normalized to have amplitude $\|\boldsymbol{f}(t)\|=f_k$ in each time step, and \(M \) simulations were performed with random initial conditions \(\boldsymbol{a}_0\) with amplitude $\|\boldsymbol{a}_0\|=\delta_f$ according to equation \eqref{eq:delta_f_LMI} in Theorem \ref{thm:Liu2020PRE}. Here, we set up the initial condition as $\delta_f$ to be consistent with Theorem \ref{thm:Liu2020PRE}, which also introduces the $\delta$-dependence in the permissible forcing amplitude $f_\Pi(\text{Re},\delta)$. We compute the $\mathcal{L}_\infty$ norm \( \|\boldsymbol{y}_\tau(t)\|_{\mathcal{L}_\infty} \) over \(t \in[0, \tau]\) as the criteria in bisection:
\begin{equation}
    \begin{cases}
        f_{\text{min}} \leftarrow f_k, & \text{If } \forall\; M\; \text{random $\boldsymbol{a}_0$ with}\; \|\boldsymbol{a}_0\|=\delta_f\;\text{leading to } \|\boldsymbol{y}_\tau(t)\|_{\mathcal{L}_\infty} \leq \delta,\\
        f_{\text{max}} \leftarrow f_k, & \text{If }\exists\; \boldsymbol{a}_0 \text{ with } \|\boldsymbol{a}_0\|=\delta_f \text{ within $M$ random $\boldsymbol{a}_0$ leading to } \|\boldsymbol{y}_\tau(t)\|_{\mathcal{L}_\infty} > \delta. 
    \end{cases}
    \label{eq:update_rule}
\end{equation}
Here, $\tau=7\times 10^{4}$ is the final time of our simulations. We will repeat the above procedure in \eqref{eq:update_rule} to update $f_{\text{min}}$ and $f_{\text{max}}$ until the convergence of the bisection search is achieved when the interval width satisfies:
\begin{equation}
    |f_{\text{max}} - f_{\text{min}}| < 10^{-12}. 
    \label{eq:convergence}
\end{equation}

\begin{figure}[!htbp]
\begin{subfigure}{0.49\linewidth}
        \centering
        \includegraphics[width=\linewidth]{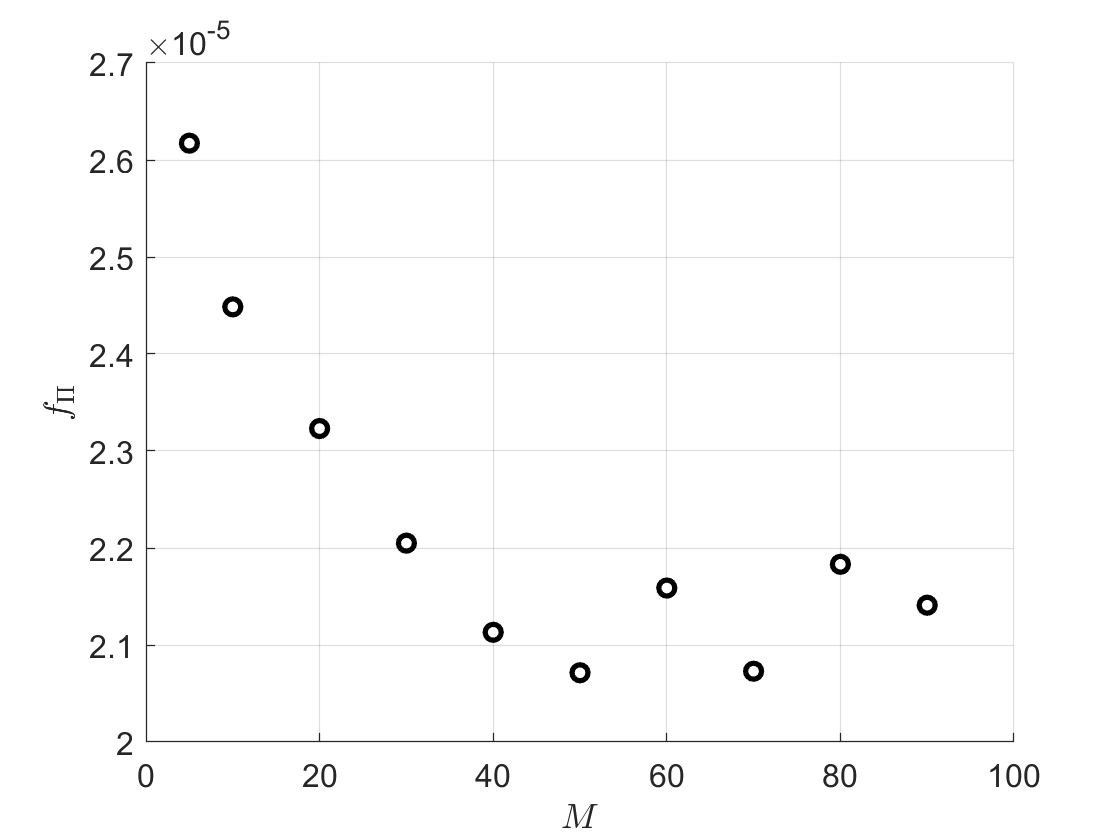} 
        \caption{\label{fig:convergence} Convergence analysis for a fixed $(\text{Re},\delta)$.}
\end{subfigure}
\begin{subfigure}{0.496\linewidth}
        \centering
        \includegraphics[width=\linewidth]{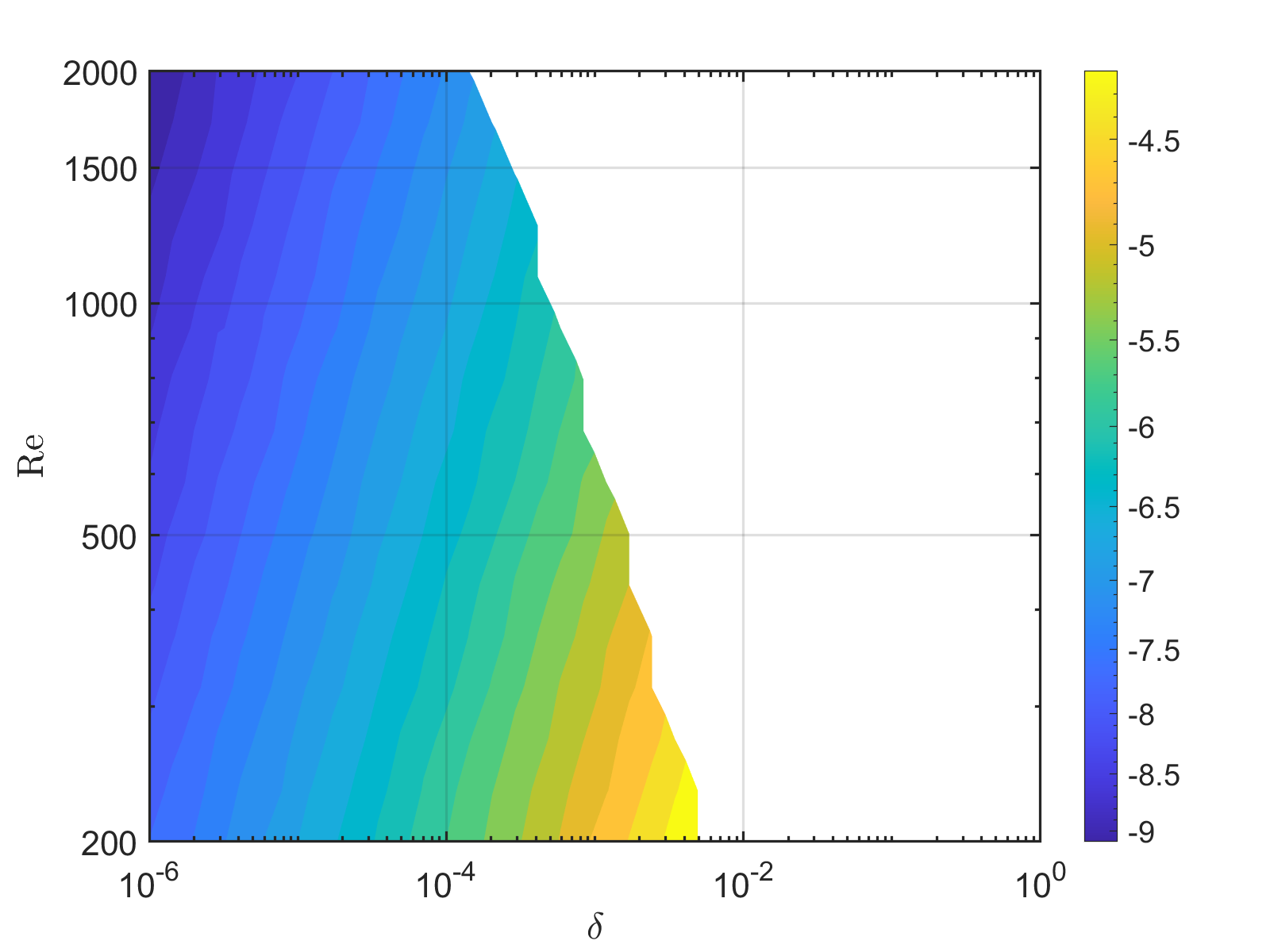} 
        \caption{Dependence of $\log_{10}[f_\Pi(\text{Re},\delta)]$ on $\delta$  and $\text{Re}$.}
        \label{fig:forcing_bi_M_40_delta_40}
\end{subfigure}
\caption{\justifying \rev{(a) $f_\Pi$ determined from bisection search for $\text{Re}=200$ and $\delta=10^{-3}$, where $M$ simulations with random initial conditions are conducted for each forcing amplitude during bisection search. (b) Permissible forcing amplitude $\log_{10}[f_{\Pi}(\text{Re},\delta)]$ obtained by numerical simulations and bisection search in equation \eqref{eq:update_rule}, where we sample 40 logarithmically spaced $\delta$ within $ \delta\in [10^{-6}, 1]$ and choose $M = 40$.}}
\label{fig:Comparison}
\end{figure}

\rev{To examine the impact of increasing $M$, Fig.~\ref{fig:convergence} demonstrates that for the case of $\text{Re}=200$ and $\delta=10^{-3}$, the permissible forcing amplitude $f_\Pi$ converges at approximately $M = 40$ random initial conditions.} Figure~\ref{fig:forcing_bi_M_40_delta_40} then shows the permissible forcing amplitude $f_{\Pi}(\text{Re},\delta)$ obtained by numerical simulations and bisection search \rev{with $M=40$ and 40 logarithmically spaced $\delta$ points. Here, we only consider 40 instead of 400 logarithmically spaced $\delta$ in bisection search to save computational resources, and the number of sample points over $\delta$ does not influence the results at each $\delta$.} Figure~\ref{fig:forcing_bi_M_40_delta_40} shows that $f_\Pi(\text{Re},\delta)$ will decrease as $\text{Re}$ increases because of a higher input-output amplification in higher Re regimes (Figs. \ref{fig:prop} and \ref{fig:leastsqft}) that constrains a smaller permissible forcing amplitude to maintain $\|\boldsymbol{y}_\tau(t)\|_{\mathcal{L}_\infty}\leq \delta$. Moreover, $f_{\Pi}$ will increase at larger $\delta$, which is because a larger $\delta$ allows a higher output norm (i.e., $\|\boldsymbol{y}_\tau(t)\|_{\mathcal{L}_\infty}\leq \delta$), and thus a larger forcing amplitude will be allowed. Such a trend over Re and $\delta$ of permissible forcing amplitude $f_\Pi$ obtained from bisection search is similar to what we found from nonlinear SSFG $\mathcal{L}_p$ stability using either LMI ($f_{\text{LMI}}$ in Figure \ref{fig:contour_lmi}) or SOS ($f_{\text{SOS}}$ in Figure \ref{fig:contour_SOS}), although $f_\Pi$ is much larger than either $f_{\text{LMI}}$ or $f_{\text{SOS}}$. This shows that SSFG $\mathcal{L}_p$ stability provides a conservative estimation of the permissible forcing amplitude compared with simulations. The nonlinear SSFG $\mathcal{L}_p$ stability theorem instead can guarantee a finite input-output gain as long as the initial disturbance and forcing satisfy given requirements in Theorem \ref{thm:Liu2020PRE} or Theorem \ref{thm:SOS}. 

\begin{figure}[!htbp]
\begin{subfigure}{0.495\linewidth}
\centering
\includegraphics[width=\linewidth]{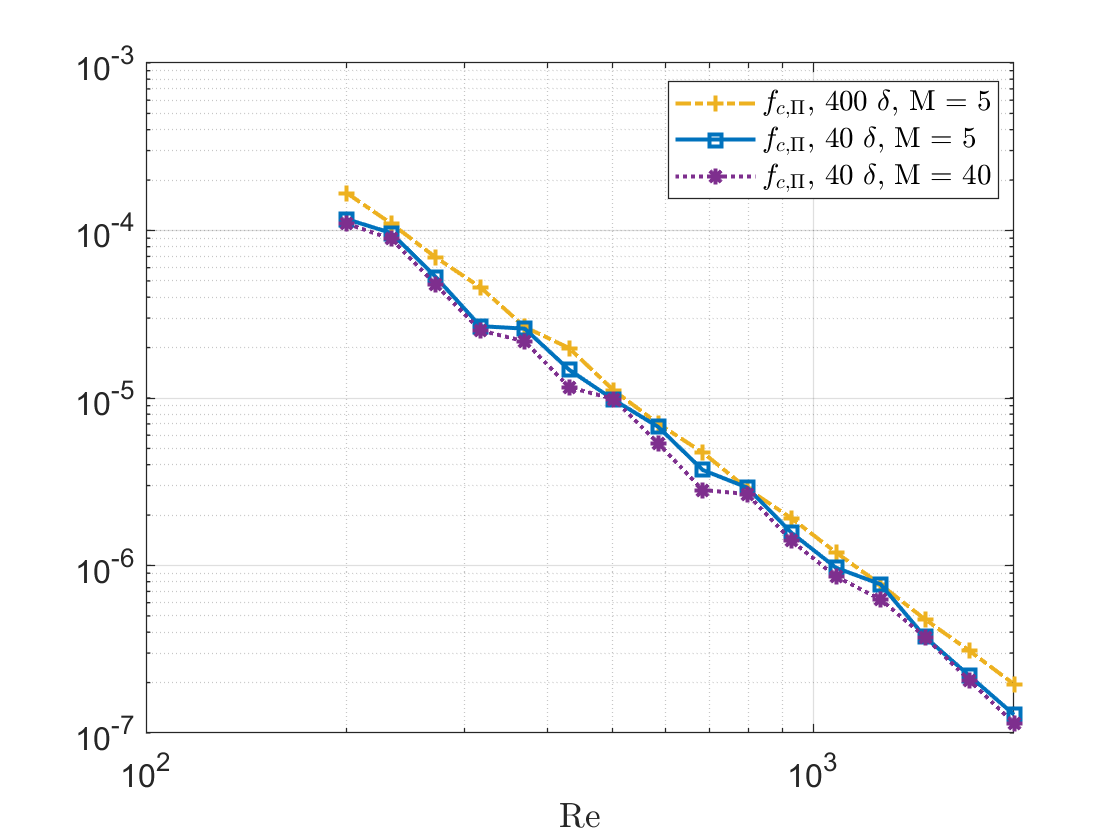}
\caption{
}
\label{fig:comprasion_M_delta}
\end{subfigure}
\begin{subfigure}{0.495\linewidth}
\centering
\includegraphics[width=\linewidth]{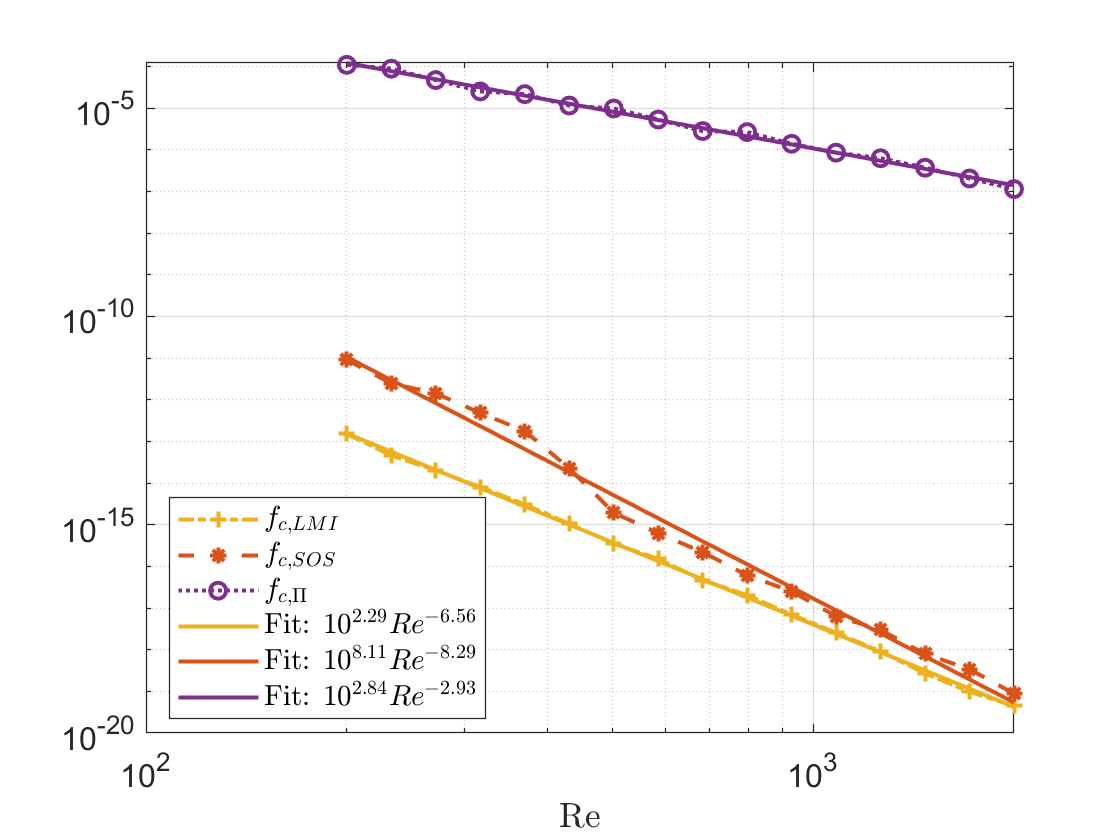} 
    \caption{
    } 
\label{fig:forcing_e_8910}
\end{subfigure}
\caption{\justifying \rev{(a) The comparison of permissible forcing amplitude $f_{c,\Pi}$ under different number of logarithmically sampled $\delta\in [10^{-6},1]$ and $M$.} (b) The comparison of permissible forcing amplitude below which the output norm is bounded as $\|\boldsymbol{y}_\tau(t)\|_{\mathcal{L}_p}\leq \delta$, including \(f_{ c,\Pi}\) in \eqref{eq:f_c_Pi} obtained by bisection search \rev{with $M=40$ and 40 logarithmically spaced $\delta$ points}, \(f_{c,\text{LMI}}\) in \eqref{eq:f_c_LMI} obtained by LMI in Theorem \ref{thm:Liu2020PRE}, and \(f_{c,\text{SOS}}\) in \eqref{eq:f_c_SOS} obtained by SOS in Theorem \ref{thm:SOS}.}
\end{figure}

We then compare the Reynolds number dependence of these permissible forcing amplitudes. We compute the maximal value of $f_\Pi$ over $\delta$ such that Theorem \ref{thm:Liu2020PRE} is feasible:
\begin{align}
\label{eq:f_c_Pi}
    f_{c,\Pi}(\text{Re}) = \max_{\delta} \ f_{\Pi}(\text{Re},\delta).  
\end{align}
Similarly, we also define 
 \begin{subequations}
\begin{align}
f_{c,\text{LMI}}(\text{Re}) =& \max_{\delta} \ f_{\text{LMI}}(\text{Re},\delta),\label{eq:f_c_LMI} \\
f_{c,\text{SOS}}(\text{Re}) = &\max_{\delta} \ f_{\text{SOS}}(\text{Re},\delta),\label{eq:f_c_SOS}
\end{align}
\end{subequations}
which maximize the permissible forcing amplitude determined from Theorems \ref{thm:Liu2020PRE} and \ref{thm:SOS} over all feasible $\delta$. \rev{Figure~\ref{fig:comprasion_M_delta} shows that $f_{c,\Pi}$ with $M = 40$ is slightly lower than that using $M=5$ random simulations, consistent with the trend in Fig.~\ref{fig:convergence}. This is because more simulations with random initial conditions will be more likely to reach the worst-case initial condition, leading to a lower permissible forcing amplitude $f_{c,\Pi}$. Using 400 logarithmically spaced points of $\delta$ will lead to a slightly larger $f_{c,\Pi}$ than that with 40 sample points of $\delta$ (Fig.~\ref{fig:comprasion_M_delta}), but they are still within the same order of magnitude.} Figure~\ref{fig:forcing_e_8910} compares $f_{c,\text{SOS}}$, $f_{c,\text{LMI}}$, and $f_{c,\Pi}$ \rev{(with 40 logarithmically sampled $\delta$ and $M=40$)} over Re for estimating the permissible forcing amplitude below which the output norm is bounded as $\|\boldsymbol{y}_\tau(t)\|_{\mathcal{L}_\infty}\leq \delta$. The permissible forcing amplitude determined from three methods, including $f_{c,\text{SOS}}$ (SOS-based analysis), $f_{c,\text{LMI}}$ (LMI-based analysis), and $f_{c,\Pi}$ (simulations with bisection search), all decreases as Re increases, reflecting stricter constraints on forcing amplitude at higher $\mathrm{Re}$. This trend aligns with the physical expectation that increased Re requires a smaller forcing amplitude to maintain the laminar state. For example, the onset of transition to turbulence in pipe flow experiments can be shifted from $\text{Re}=2000$ to $\text{Re}=13,000$ by reducing the inlet disturbance \citep{reynolds1883xxix}, and the critical Reynolds number at which transition to turbulence occurs is increasing as turbulence intensity decreases \citep{brandt2004transition,fransson2005transition,he2015transition,mathur2018temporal}. Moreover, Figure~\ref{fig:forcing_e_8910} shows that $f_{c,\text{LMI}}$ and $f_{c,\text{SOS}}$ are both lower than the permissible forcing amplitude $f_{c,\Pi}$ identified from bisection search for $\text{Re}\in[200,2000]$, which shows that our SSFG $\mathcal{L}_p$ stability theorem identifies a conservative permissible forcing amplitude yet consistent with numerical simulations.

\begin{table}[!htbp]
    \centering
    \caption{ A and $\zeta$ obtained from a least-squares fitting to $f_c$ = $10^A$$\text{Re}^\zeta$ with three different type of $f_c$ in Figure \ref{fig:forcing_e_8910}.}
    \label{tab:scaling_exponents_forcing}
    \begin{tabular}{@{}lcc@{}}
    \toprule
    \text{Permissible forcing amplitude $f_c$} & \(\zeta\) & \(A\) \\
    \midrule
    $f_{c,\text{LMI}}$ & -6.5635 & 2.2939 \\
    $f_{c,{\text{SOS}}}$ & -8.2921 & 8.1055 \\
    $f_{c,\Pi}$ & \rev{-2.9341} & \rev{2.8368} \\
    \bottomrule
    \end{tabular}
\end{table}

We also fit all these three different permissible forcing amplitudes as functions \(f_c = 10^{A} \text{Re}^{\zeta}\) via a least-squares fitting with coefficients \(A\) and \(\zeta\) reported in Table~\ref{tab:scaling_exponents_forcing}, and fitted lines are also shown in Figure \ref{fig:forcing_e_8910}. Here, we found the permissible forcing amplitude from LMI $f_{c,\text{LMI}}$ scales as $\sim \text{Re}^{-6.56}$, while the permissible forcing amplitude from SOS $f_{c,\text{SOS}}$ scales as $\sim \text{Re}^{-8.29}$. The permissible forcing amplitude from bisection search $f_{c,\Pi}$ instead scales as $\sim \text{Re}^{-2.93}$. This comparison shows that the permissible forcing amplitude computed based on LMI or SOS will decrease with the Reynolds number much faster than the permissible forcing amplitude determined by the bisection search.

\begin{figure}[!htbp]
    \centering
    \begin{subfigure}{0.496\linewidth}
        \centering
        \includegraphics[width=\linewidth]{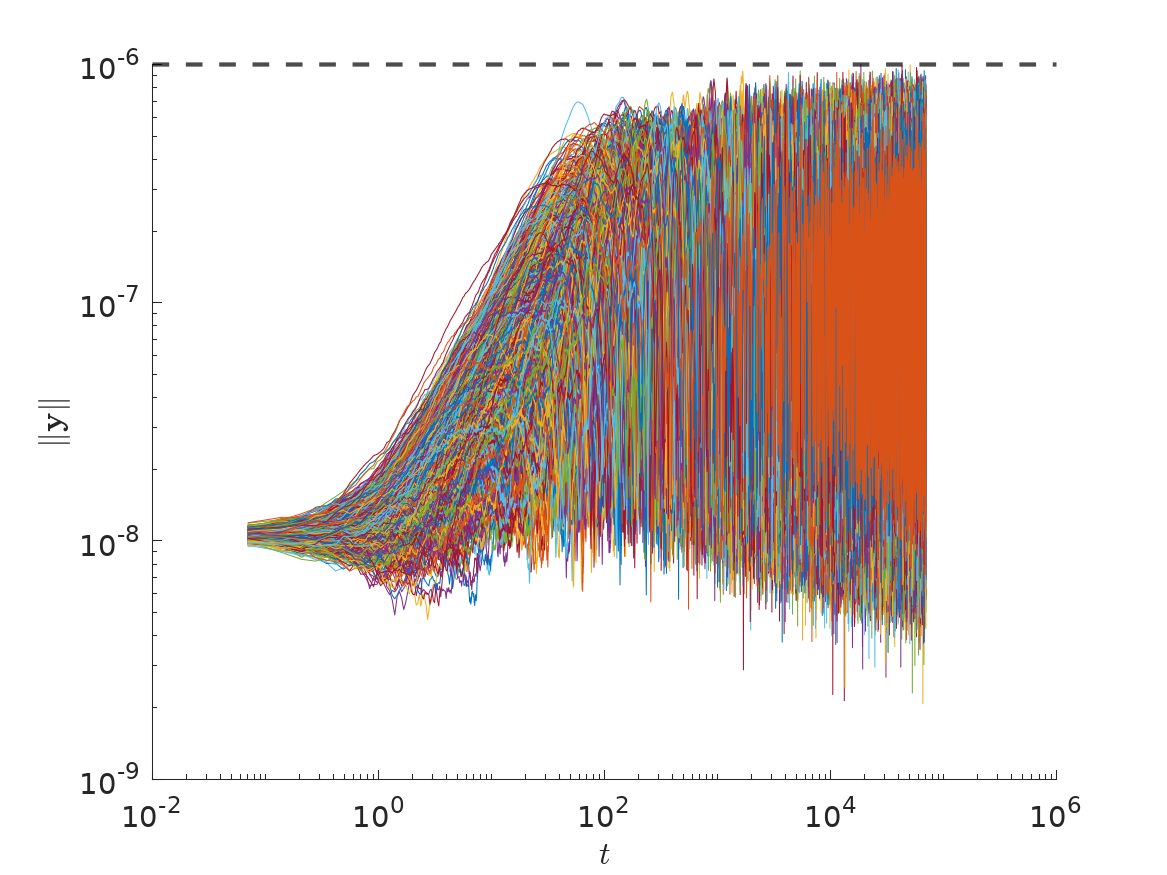} 
        \caption{Output norm of simulations $\boldsymbol{y}(t) \in Z_\Pi$ with forcing amplitude \(||\boldsymbol{f}(t)|| \leq f_{\Pi}\).}
        \label{fig:laminar_other_bi}
    \end{subfigure}
    \begin{subfigure}{0.496\linewidth}
        \centering
        \includegraphics[width=\linewidth]{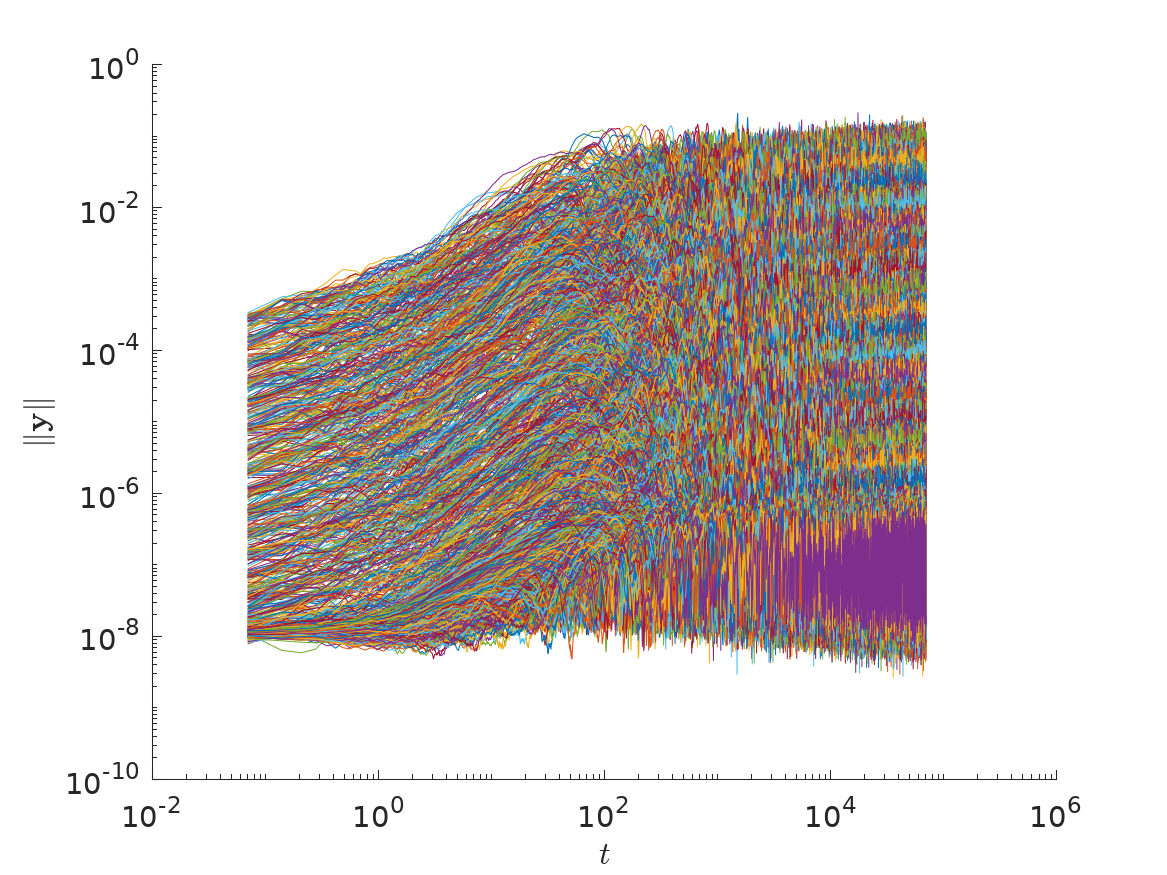} 
        \caption{Output norm of simulations \(\boldsymbol{y}(t)\) $\in Q_\Pi$ with forcing amplitude \(||\boldsymbol{f}(t)||> f_{\Pi}\).}
        \label{fig:turbulent_other_bi}
    \end{subfigure}
    \caption{\justifying Output norm $\|\boldsymbol{y}_\tau(t)\|$ of simulations from bisection search (a) $\boldsymbol{y}(t)\in Z_\Pi$ and (b) $\boldsymbol{y}(t)\in Q_\Pi$ at Re = 200, $\delta = 10^{-6}$ leading to initial disturbance amplitude $\delta_f = 1.0609\times10^{-8}$ from LMI in Theorem \ref{thm:Liu2020PRE}. The permissible forcing amplitude identified from bisection is \rev{$f_{\Pi}= 2.1307 \times 10^{-8}$}. The precise value of $f_\Pi$ may fluctuate in each run, and this figure only considers \rev{$M=40$} random initial conditions. 
    }
\label{fig:laminar and turbulent_another bisection}
\end{figure}

To demonstrate the role of permissible forcing amplitude $f_{\Pi}$ identified from bisection search, we then classify simulations conducted during bisection search into two sets (a) $\|\boldsymbol{f}(t)\|\leq f_\Pi$ and (b) $\|\boldsymbol{f}(t)\|> f_\Pi$. Below this permissible forcing amplitude ($||\boldsymbol{f}(t)|| \leq  f_{\Pi}$), the initial disturbances with amplitude $\delta_f$ lead to transient growth that eventually decays over time, leading to the maximal output norm close to but lower than the preset threshold value $\delta$; i.e., $\|\boldsymbol{y}_\tau(t)\|_{\mathcal{L}_\infty}\leq\delta$ as defined by a set of simulations $Z_\Pi$ in \eqref{eq:Z_Pi}. In contrast, when the forcing amplitude exceeds $f_{\Pi}$ ($||\boldsymbol{f}(t)|| > f_{\Pi}$), there exists certain initial disturbances $\boldsymbol{a}_0$ with amplitude $\|\boldsymbol{a}_0\|=\delta_f$ that can lead to $\|\boldsymbol{y}_\tau(t)\|_{\mathcal{L}_\infty}> \delta$ as defined by a set of simulations $Q_\Pi$ in \eqref{eq:Q_Pi}:
\begin{subequations}
\begin{align}
    Z_\Pi :=&\; \{ \boldsymbol{y}(t) \,|\, \forall\; M\; \text{random $\boldsymbol{a}_0$ with}\; \|\boldsymbol{a}_0\|=\delta_f\;\text{leading to }  \|\boldsymbol{y}_\tau(t)\|_{\mathcal{L}_\infty} \leq \delta \},\label{eq:Z_Pi} \\
    Q_\Pi :=&\; \{ \boldsymbol{y}(t) \,|\, \exists\; \boldsymbol{a}_0 \text{ with } \|\boldsymbol{a}_0\|=\delta_f \text{ within $M$ random $\boldsymbol{a}_0$ leading to } \|\boldsymbol{y}_\tau(t)\|_{\mathcal{L}_\infty} > \delta. \label{eq:Q_Pi}
\end{align}
\end{subequations}
These two sets contain simulation results obtained in the procedure of bisection search for the permissible forcing amplitude $f_\Pi$. Figure \ref{fig:laminar and turbulent_another bisection} shows the output norm $\|\boldsymbol{y}(t)\|$ with (a) $\boldsymbol{y}(t)\in Z_\Pi$ and (b) $\boldsymbol{y}(t)\in Q_\Pi$ \rev{with $\text{Re}=200$, $\delta=10^{-6}$, and $M=40$.} As expected, the output norm $\|\boldsymbol{y}(t)\|$ of all simulations with forcing amplitude $\|\boldsymbol{f}(t)\|\leq f_\Pi$ in Figure \ref{fig:laminar and turbulent_another bisection}a start from a magnitude close to $\delta_f$ followed by transient growth bounded as $ \|\boldsymbol{y}_\tau(t)\|_{\mathcal{L}_\infty}\leq \delta$. When forcing is larger than the threshold value $\|\boldsymbol{f}(t)\|\geq f_\Pi$, we can see from Figure \ref{fig:laminar and turbulent_another bisection}b that some simulations have final output norm larger than $\delta$; i.e., $\|\boldsymbol{y}_\tau(t)\|_{\mathcal{L}_\infty}> \delta$. For both panels, they have a random initial disturbance amplitude $\|\boldsymbol{a}_0\|=\delta_f$. The large initial amplitude of $\|\boldsymbol{y}(t)\|$ in Figure \ref{fig:laminar and turbulent_another bisection}b is induced by a large forcing amplitude in bisection search that can reach $f_{\text{max}}=10^{-2}$. Certain simulations in Figure \ref{fig:laminar and turbulent_another bisection}b still shows $\|\boldsymbol{y}_\tau(t)\|_{\mathcal{L}_\infty}\leq \delta$, which is possible because in the definition of $Q_\Pi$ in \eqref{eq:Q_Pi}, it only requires to exist some $\boldsymbol{a}_0$ (not necessarily all $\boldsymbol{a}_0$) leading to $\|\boldsymbol{y}_\tau(t)\|_{\mathcal{L}_\infty}> \delta$. The comparison in Figure \ref{fig:laminar and turbulent_another bisection} thus highlights $f_{\Pi}$ as a role of permissible forcing amplitude governing the system's response, which represents the maximum input forcing amplitude to maintain the solution close to the laminar state as measured by $\|\boldsymbol{y}_\tau(t)\|_{\mathcal{L}_\infty}\leq\delta$ in \eqref{eq:Z_Pi} and \eqref{eq:update_rule}. This result also demonstrates the sensitivity of the system to input forcing amplitude, which can potentially induce a transition to chaotic states.

\begin{table}
    \centering
    \caption{\justifying \rev{Computational cost comparison for LMI, SOS, and bisection search when Re = 200, $\delta = 10^{-3}$. We set $\epsilon=10^{-4}$ for LMI and $\epsilon=10^{-2}$ for SOS, respectively. For bisection search, we set simulation interval $\tau = 7 \times10^{4}$ and conduct $M = 40$ simulations with random initial conditions for each forcing amplitude. }}   
\label{tab:comp_cost}
    \begin{tabular}{@{}lccc@{}}
    \toprule
     Method & LMI & SOS & Bisection \\
    \midrule
    Time (s) & 49 & 244 & 24313 \\
    Memory (GB) & 1.42  & 2.39 & 25.12 \\
    \bottomrule
    \end{tabular}
\end{table}

\rev{Table \ref{tab:comp_cost} quantitatively compares the computational cost of the LMI, SOS, and the bisection search in determining the permissible forcing amplitude. The results reported in Table \ref{tab:comp_cost} correspond to the case when $\text{Re} = 200$ and $\delta = 10^{-3}$, and we set $\varepsilon = 10^{-4}$ for the LMI and $\varepsilon = 10^{-2}$ for SOS, respectively. For the bisection search, we use a simulation interval $\tau = 7 \times 10^4$ and conduct $M = 40$ random initial conditions for each forcing amplitude during the bisection search. The comparison in Table \ref{tab:comp_cost} demonstrates a critical trade-off between computational cost and the conservatism of the resulting bound. The LMI provides a rigorous upper bound on permissible forcing amplitude with the lowest computational cost, although it yields the most conservative estimate. The SOS offers the advantage of a less conservative bound, but at approximately five times the computational cost compared with LMI. This is because LMI uses quadratic constraints (equations (17)-(19)) to characterize the nonlinearity, leading to a reduced number of constraints and a smaller size of the largest positive semi-definite cone \citep{Liu2020PRE}. Compared to the LMI and SOS methods, the bisection search entails the highest computational cost, as it requires numerous long-time numerical simulations. This comparison underscores that the LMI and SOS offer a computationally advantageous pathway for obtaining certifiable permissible forcing amplitude compared with bisection search.}

\section{\label{sec:conclusion}Conclusions and future work}

This work performs nonlinear input–output analysis for a nine-mode transitional shear flow model \cite{moehlis2004low} using the Small-Signal Finite-Gain (SSFG) $\mathcal{L}_p$ stability theorem (Theorem \ref{thm:small_signal_stability} \citep[Theorem 5.1]{Khalil2002}). We construct quadratic Lyapunov functions of unforced nonlinear systems by formulating and solving Linear Matrix Inequalities (LMI, Theorem \ref{thm:Liu2020PRE}) and Sum-of-Squares (SOS, Theorem \ref{thm:SOS}). This SSFG $\mathcal{L}_p$ stability theorem allows us to obtain upper bounds on the permissible forcing amplitude ($f_{\text{LMI}}$ or $f_{\text{SOS}}$) below which a finite nonlinear input–output gains ($\gamma$ or $\gamma_{\text{SOS}}$) can be maintained, rigorously analyzing nonlinear input-output systems by Lyapunov type framework \cite{Khalil2002}. These quantities are computed as a function of the Reynolds number Re and the preset upper bound of state variable amplitude $\delta$ (i.e., $\|\boldsymbol{a}(t)\| \leq \delta$ as in \eqref{eq:upper_bound_state_variable}). The nonlinear input-output $\mathcal{L}_p$ gains $\gamma$ and $\gamma_{\text{SOS}}$ increase with the Reynolds number $\mathrm{Re}$ and remain nearly independent of $\delta$ at low $\delta$ until the $\delta$ value leading to the infeasibility of LMI or SOS. The permissible forcing amplitude, $f_{\text{LMI}}$ and $f_{\text{SOS}}$, decrease with $\mathrm{Re}$, reflecting the requirement for smaller forcing amplitudes to maintain a finite $\mathcal{L}_p$ gain at higher Re. Since larger $\delta$ permits larger state-variable amplitudes, the permissible forcing amplitude $f_{\text{LMI}}$ and $f_{\text{SOS}}$ increase with $\delta$.

To validate the prediction of the SSFG $\mathcal{L}_p$ theorem, we also conduct extensive numerical simulations with random initial disturbances and random forcing, where the final statistically steady state is driven by the random forcing after the decay of initial transient growth. We validate that the upper bound of the $\mathcal{L}_p$ norm ($p\in [1,\infty]$) of the output predicted by SSFG $\mathcal{L}_p$ theorem is consistent with the output from extensive numerical simulations over the Reynolds numbers $\text{Re}\in [200,2000]$. The gap between the theoretical prediction of SSFG $\mathcal{L}_p$ theorem and numerical simulations (i.e., the difference of the LHS and RHS of the inequality \eqref{eq:gamma_upper_bound}) decreases as we increase the $p$ value for computing $\mathcal{L}_p$ norm, and this gap converges to an asymptotic value associated with $\mathcal{L}_\infty$ norm. Higher \( p \)-values also reduce the gap between LHS and RHS of inequality \eqref{eq:gamma_upper_bound} at high Re. The input-output gain computed from simulations displays a trend over the Reynolds number similar to the theoretical prediction, and a higher $p$ value leads to a closer agreement between the theoretical prediction and numerical simulations. 

Moreover, we also compare the nonlinear SSFG $\mathcal{L}_p$ gains against the linear finite-gain $\mathcal{L}_p$ stability and linear $\mathcal{L}_2$ stability. The nonlinear input-output $\mathcal{L}_p$ gain from the SSFG $\mathcal{L}_p$ stability theorem (computed by either LMI or SOS) is higher than the linear input-output $\mathcal{L}_p$ gain by several orders of magnitude, underscoring that nonlinearity can amplify disturbances beyond linear prediction. The nonlinear input-output $\mathcal{L}_p$ gain exhibits a Reynolds-number scaling $\sim \text{Re}^{4.04}$ computed by the LMI and a scaling $\sim \text{Re}^{3.61}$ computed by SOS, while the linear $\mathcal{L}_p$ gain scales as $\sim \text{Re}^{5.00}$. Both of these $\mathcal{L}_p$ input-output gain predictions (either linear or nonlinear) have a Re scaling exponent much larger than linear $\mathcal{L}_2$ gain scaling as $\sim \text{Re}^{2.00}$, which is likely because our $\mathcal{L}_p$ input-output gain prediction is valid for general $\mathcal{L}_p$ norm with any $p\in [1,\infty]$. The $\mathcal{L}_2$ gain scaling $\sim \text{Re}^{2.00}$ for this nine-mode shear flow model is the same as the $\mathcal{L}_2$ gain scaling ($\mathcal{H}_\infty$ norm of transfer functions) of wall-bounded shear flows \cite{jovanovic2004modeling,liu2021structured,trefethen1993hydrodynamic}. Our comparison here highlights that when measuring input-output gain using the general $\mathcal{L}_p$ norm ($p\in [1,\infty]$), input-output amplification can be much higher than the widely used $\mathcal{L}_2$ gain that is limited to piecewise continuous, square-integrable input forcing.

Our nonlinear SSFG $\mathcal{L}_p$ stability theorem predicts a permissible forcing amplitude below which finite-gain input-output amplification can be sustained, which is an inherently nonlinear property. In contrast, linear systems are associated with an input-output gain independent of input forcing amplitude due to linear superposition, ignoring the amplitude-sensitive nature of the transition to turbulence phenomena. We then compare the permissible forcing amplitude predicted by the nonlinear SSFG $\mathcal{L}_p$ theorem with the permissible forcing amplitude $f_{\Pi}$ obtained by extensive numerical simulations and bisection search. Below this permissible forcing amplitude $(\|\boldsymbol{f}(t)\|\leq f_{\Pi})$, the output norm will be bounded by a preset value $\delta$ such that $\|\boldsymbol{y}_\tau(t)\|_{\mathcal{L}_\infty}\leq \delta$. We show that the permissible forcing amplitude determined by LMI or SOS is consistent with simulation results, but decreases faster over the Reynolds number than the permissible forcing amplitude obtained from extensive simulations and bisection search.

Therefore, these results demonstrate the applicability of the small-signal finite-gain $\mathcal{L}_p$ stability theorem for rigorous nonlinear input-output analysis of transitional shear flows that are sensitive to both initial disturbance and external disturbance (input forcing). This work also offers insights into determining a permissible forcing amplitude to constrain external disturbance for transition delay and suppression. \rev{One} future direction is to explore more general (non-quadratic) Lyapunov functions \rev{(see e.g., \citep{GOULART2012692,chernyshenko2014polynomial,huang2015sum,fuentes2022global})} through advanced SOS formulations or non-convex optimization methods, which could potentially reduce conservatism and provide tighter bounds on permissible forcing amplitude and nonlinear input-output gain. \rev{The conservatism of the LMI also arises from the fact that we use quadratic constraints (equations \eqref{eq:local_upper_bound_upsilon_m}-\eqref{eq:orthogonal_complement}) to characterize nonlinearity. This conservatism can be reduced by employing more general equalities or inequalities that the nonlinearity satisfies; see e.g., Ref.  \citep{kalur2022estimating} for region of attraction estimation.} 

\rev{Another future work direction is to} extend the SSFG $\mathcal{L}_p$ stability theorem to the Navier--Stokes equation or time-dependent (non-autonomous) system where Theorems \ref{thm:small_signal_stability}-\ref{thm:SOS} still apply. \rev{For example, we recently applied a closely related linear matrix inequalities formulation to analyze the upper bound of linear transient growth in accelerating and decelerating wall-driven channel flows \citep{wei2025upper}. This work \citep{wei2025upper} also demonstrates the feasibility of applying Lyapunov-type methods and LMI formulation directly to linearized Navier–Stokes operators with numerical discretization, and it is not necessary to construct a low-dimensional model based on Galerkin projection (e.g., the nine-mode model used here). We also show that a similar Lyapunov-type analysis can predict the growth rate the same as numerical simulations and Floquet theory for a linear time-periodic system \citep{kochnev2025stability}. We will extend the nonlinear SSFG $\mathcal{L}_p$ stability analysis introduced in this paper to time-varying wall-bounded shear flows in future work, thereby enabling nonlinear input–output analysis of a wider class of transitional shear flows.}

\begin{acknowledgments}
C.L. acknowledges the support from the University of Connecticut (UConn) Research Excellence Program (REP). Z.W. acknowledges the support from the UConn Summer Undergraduate Research Fund (SURF) Awards and the University Scholar Program. The computational work for this project was conducted using resources provided by the Storrs High-Performance Computing (HPC) cluster. We extend our gratitude to the UConn Storrs HPC and its team for their resources and support, which aided in achieving these results.
\end{acknowledgments}

\section*{Data Availability}
The data that support the findings of this article and the source code producing the data are openly available in Ref. \citep{wei_2025_15742363}. 

\appendix
\section{\label{appendix:a} The nine-mode shear flow model}

\begingroup
\allowdisplaybreaks
The dynamical equation of the total state variable, shown below, is obtained from the Galerkin projection of the Navier-Stokes equations; the derivation is detailed in \citep{moehlis2004low, Liu2020PRE}.
\begin{align}
     \frac{d \tilde{\boldsymbol{a}}}{d t} =& -\frac{\boldsymbol{\Xi}}{\operatorname{Re}} \tilde{\boldsymbol{a}}+\boldsymbol{J}(\tilde{\boldsymbol{a}}) \tilde{\boldsymbol{a}}+\boldsymbol{F},
\end{align}
where the matrix $\mathbf{\Xi}$ and the Jacobian terms $\boldsymbol{J}(\tilde{\boldsymbol{a}})\tilde{\boldsymbol{a}}$ are defined as Refs. \citep{moehlis2004low, Liu2020PRE}:
\begin{align}
\begin{split}    
\mathbf{\Xi} = \mathrm{diag}\left( \beta^2, \frac{4 \beta^2}{3} + \gamma^2, \kappa_{\beta\gamma}^2, \frac{3\alpha^2 + 4 \beta^2}{3}, 
\kappa_{\alpha\beta}^2, \frac{3 \alpha^2 + 4 \beta^2 + 3 \gamma^2}{3}, 
\kappa_{\alpha\beta\gamma}^2, \kappa_{\alpha\beta\gamma}^2, 9 \beta^2\right) ,
\end{split}
\end{align}

\begin{align}
[\boldsymbol{J}(\tilde{\boldsymbol{a}})\tilde{\boldsymbol{a}}]_1 &= \sqrt{\frac{3}{2}} \frac{\beta \gamma}{\kappa_{\beta \gamma}} \tilde{a}_2 \tilde{a}_3 - \sqrt{\frac{3}{2}} \frac{\beta \gamma}{\kappa_{\alpha \beta \gamma}} \tilde{a}_6 \tilde{a}_8, \\
[\boldsymbol{J}(\tilde{\boldsymbol{a}})\tilde{\boldsymbol{a}}]_2 &= \frac{10}{3\sqrt{6}}  \frac{\gamma^2}{\kappa_{\alpha \gamma}} \tilde{a}_4 \tilde{a}_6 - \frac{\gamma^2}{\sqrt{6}\kappa_{\alpha \gamma}} \tilde{a}_5 \tilde{a}_7 -\frac{\alpha \beta \gamma}{\sqrt{6} \kappa_{\alpha \gamma} \kappa_{\alpha \beta \gamma}} \tilde{a}_5 \tilde{a}_8 
 -\sqrt{\frac{3}{2}} \frac{\beta \gamma}{\kappa_{\beta \gamma}}\left(\tilde{a}_1 \tilde{a}_3+\tilde{a}_3 \tilde{a}_9\right), \\
[\boldsymbol{J}(\tilde{\boldsymbol{a}})\tilde{\boldsymbol{a}}]_3 &= \sqrt{\frac{2}{3}} \frac{\alpha \beta \gamma}{\kappa_{\alpha \gamma} \kappa_{\beta \gamma}} (\tilde{a}_5 \tilde{a}_6 + \tilde{a}_4 \tilde{a}_7) +\frac{\beta^2\left(3 \alpha^2+\gamma^2\right)-3 \gamma^2 \kappa_{\alpha \gamma}^2}{\sqrt{6} \kappa_{\alpha \gamma} \kappa_{\beta \gamma} \kappa_{\alpha \beta \gamma}} \tilde{a}_4 \tilde{a}_8, \\
[\boldsymbol{J}(\tilde{\boldsymbol{a}})\tilde{\boldsymbol{a}}]_4 &= -\frac{\alpha}{\sqrt{6}} (\tilde{a}_1 \tilde{a}_5 + \tilde{a}_5 \tilde{a}_9) - \frac{10}{3\sqrt{6}} \frac{\alpha^2}{\kappa_{\alpha \gamma}} \tilde{a}_2 \tilde{a}_6 -\sqrt{\frac{3}{2}} \frac{\alpha \beta \gamma}{\kappa_{\alpha \gamma} \kappa_{\beta \gamma}} \tilde{a}_3 \tilde{a}_7 -\sqrt{\frac{3}{2}} \frac{\alpha^2 \beta^2}{\kappa_{\alpha \gamma} \kappa_{\beta \gamma} \kappa_{\alpha \beta \gamma}} \tilde{a}_3 \tilde{a}_8, \\
[\boldsymbol{J}(\tilde{\boldsymbol{a}})\tilde{\boldsymbol{a}}]_5 &= \frac{\alpha}{\sqrt{6}} (\tilde{a}_1 \tilde{a}_4 + \tilde{a}_4 \tilde{a}_9) + \sqrt{\frac{2}{3}} \frac{\alpha \beta \gamma}{\kappa_{\alpha \gamma} \kappa_{\beta \gamma}} \tilde{a}_3 \tilde{a}_6 +\frac{\alpha^2}{\sqrt{6} \kappa_{\alpha \gamma}} \tilde{a}_2 \tilde{a}_7-\frac{\alpha \beta \gamma}{\sqrt{6} \kappa_{\alpha \gamma} \kappa_{\alpha \beta \gamma}} \tilde{a}_2 \tilde{a}_8, \\
[\boldsymbol{J}(\tilde{\boldsymbol{a}})\tilde{\boldsymbol{a}}]_6 &= \frac{10}{3\sqrt{6}} \frac{\alpha^2 - \gamma^2}{\kappa_{\alpha \gamma}} \tilde{a}_2 \tilde{a}_4 - \sqrt{\frac{2}{3}} \frac{2 \alpha \beta \gamma}{\kappa_{\alpha \gamma} \kappa_{\beta \gamma}} \tilde{a}_3 \tilde{a}_5 +\frac{\alpha}{\sqrt{6}}\left(\tilde{a}_1 \tilde{a}_7+\tilde{a}_7 \tilde{a}_9\right) +\sqrt{\frac{3}{2}} \frac{\beta \gamma}{\kappa_{\alpha \beta \gamma}}\left(\tilde{a}_1 \tilde{a}_8+\tilde{a}_8 \tilde{a}_9\right), \\
{[\boldsymbol{J}(\tilde{\boldsymbol{a}}) \tilde{\boldsymbol{a}}]_7} &=  \frac{\alpha \beta \gamma}{\sqrt{6} \kappa_{\alpha \gamma} \kappa_{\beta \gamma}} \tilde{a}_3 \tilde{a}_4+\frac{-\alpha^2+\gamma^2}{\sqrt{6} \kappa_{\alpha \gamma}} \tilde{a}_2 \tilde{a}_5 -\frac{\alpha}{\sqrt{6}}\left(\tilde{a}_1 \tilde{a}_6+\tilde{a}_6 \tilde{a}_9\right), \\
[\boldsymbol{J}(\tilde{\boldsymbol{a}})\tilde{\boldsymbol{a}}]_8 &= \frac{\gamma^2(3 \alpha^2 - \beta^2 + 3 \gamma^2)}{\sqrt{6}\kappa_{\alpha \gamma} \kappa_{\beta \gamma}\kappa_{\alpha \beta \gamma}} \tilde{a}_3 \tilde{a}_4 + \sqrt{\frac{2}{3}} \frac{\alpha \beta \gamma}{\kappa_{\alpha \gamma} \kappa_{\alpha \beta \gamma}} \tilde{a}_2 \tilde{a}_5, \\
[\boldsymbol{J}(\tilde{\boldsymbol{a}})\tilde{\boldsymbol{a}}]_9 &= \sqrt{\frac{3}{2}} \frac{\beta \gamma}{\kappa_{\beta \gamma}} \tilde{a}_2 \tilde{a}_3 - \sqrt{\frac{3}{2}} \frac{\beta \gamma}{\kappa_{\alpha \beta \gamma}} \tilde{a}_6 \tilde{a}_8,
\end{align}
where $[\boldsymbol{J}(\tilde{\boldsymbol{a}}) \tilde{\boldsymbol{a}}]_i:=\boldsymbol{e}_i^\text{T} \boldsymbol{J}(\tilde{\boldsymbol{a}}) \tilde{\boldsymbol{a}}, i=1,2,3, \ldots, 9$ is the $i^\text{th}$ component of $\boldsymbol{J}(\tilde{\boldsymbol{a}}) \tilde{\boldsymbol{a}}$, and 
\begin{align}
    \alpha := \frac{2\pi h}{L_{x_1}}, \quad \beta := \frac{\pi}{2}, \quad \gamma := \frac{2\pi h}{L_{x_3}},
\end{align}
\begin{align}
    \kappa_{\alpha\beta} := \sqrt{\alpha^2 + \beta^2}, \quad \kappa_{\beta\gamma} := \sqrt{\beta^2 + \gamma^2}, \quad 
    \kappa_{\alpha\beta\gamma} := \sqrt{\alpha^2 + \beta^2 + \gamma^2}.
\end{align}

 The laminar profile $U(y)$ in this model corresponds to a fixed point $\bar{\boldsymbol{a}} = [1 \ 0 \ 0 \ 0 \ 0 \ 0 \ 0 \ 0 \ 0]^\text{T}$, and it satisfies:
    \begin{equation}
    -\frac{\boldsymbol{\Xi}}{\mathrm{Re}} \bar{\boldsymbol{a}} + \boldsymbol{J}(\bar{\boldsymbol{a}}) \bar{\boldsymbol{a}} + \boldsymbol{F} = 0,
    \end{equation}
where $\boldsymbol{F}=\frac{\beta^2}{Re}\bar{\boldsymbol{a}}$ is the forcing term that drives the laminar base flow. We can perform a decomposition of the Galerkin coefficients similar to the Reynolds decomposition:
    \begin{equation}
    \tilde{\boldsymbol{a}} = \bar{\boldsymbol{a}} + \boldsymbol{a},
    \end{equation}
    so as to shift the laminar state to the origin of fluctuating coefficients $\boldsymbol{a}$. The resulting dynamical system for these fluctuating coefficients is:
    \begin{equation}
    \frac{d\boldsymbol{a}}{dt} = -\frac{\mathbf{\Xi}}{\mathrm{Re}} \boldsymbol{a} + \boldsymbol{J}(\bar{\boldsymbol{a}})\boldsymbol{a} + \boldsymbol{J}(\boldsymbol{a}) \bar{\boldsymbol{a}} + \boldsymbol{J}(\boldsymbol{a}) \boldsymbol{a}.
    \end{equation}
We then add a body force term $\boldsymbol{f}$ to obtain a forced dynamical system as shown below:
   \begin{equation}
    \frac{d\boldsymbol{a}}{dt} = -\frac{\mathbf{\Xi}}{\mathrm{Re}} \boldsymbol{a} + \boldsymbol{J}(\bar{\boldsymbol{a}})\boldsymbol{a} + \boldsymbol{J}(\boldsymbol{a}) \bar{\boldsymbol{a}} + \boldsymbol{J}(\boldsymbol{a}) \boldsymbol{a} + \boldsymbol{f},
    \end{equation}
    which is the nonlinear forced system in equation \eqref{eq:dynamical_eq_2}. The orthogonal complement satisfying $\boldsymbol{n}^\text{T}\boldsymbol{J}(\boldsymbol{a})\boldsymbol{a}=\boldsymbol{n}^{\text{T}}\boldsymbol{\Upsilon}(\boldsymbol{a})=0$ can be found as $\boldsymbol{n}^\text{T}=\begin{bmatrix}
        1 & 0 & 0 & 0 & 0 & 0 & 0 & 0 & -1
    \end{bmatrix}$.  

\section{\label{appendix:b} Componentwise upper bound of quadratic nonlinearity in \eqref{eq:local_upper_bound_upsilon_m}}

\begin{lemma}\citep[Lemma 1]{Liu2020PRE} Given a vector $\boldsymbol{\Upsilon} \in \mathbb{R}^n$ that can be decomposed into $\Upsilon_m:=\boldsymbol{e}_m^\text{T} \boldsymbol{\Upsilon}$ associated with a quadratic form $\Upsilon_m=$ $\boldsymbol{a}^\text{T} \boldsymbol{R}_m \boldsymbol{a}$ with a symmetric matrix $\boldsymbol{R}_m \in \mathbb{R}^{n \times n}$. In a local region $\|\boldsymbol{a}\|^2 \leqslant \delta^2$, each $\Upsilon_m^2$ is bounded as
\begin{align}
\Upsilon_m^2 \leqslant \delta^2 \boldsymbol{a}^\text{T} \boldsymbol{R}_m \boldsymbol{R}_m \boldsymbol{a}, \quad m=1,2, \ldots, n.
\end{align}
\end{lemma}

\begin{proof}
    
In a local region $\|\boldsymbol{a}\|^2 \leqslant \delta^2$, we have
\begin{align}
\Upsilon_m^2 & =\left(\boldsymbol{a}^\text{T} \boldsymbol{R}_m \boldsymbol{a}\right)\left(\boldsymbol{a}^\text{T} \boldsymbol{R}_m \boldsymbol{a}\right) \\
& =\|\boldsymbol{a}\|^2\left\|\boldsymbol{R}_m \boldsymbol{a}\right\|^2 \frac{\boldsymbol{a}^\text{T} \boldsymbol{R}_m \boldsymbol{a}}{\|\boldsymbol{a}\|\left\|\boldsymbol{R}_m \boldsymbol{a}\right\|} \frac{\boldsymbol{a}^\text{T} \boldsymbol{R}_m \boldsymbol{a}}{\|\boldsymbol{a}\|\left\|\boldsymbol{R}_m \boldsymbol{a}\right\|} \\
& =\|\boldsymbol{a}\|^2\left\|\boldsymbol{R}_m \boldsymbol{a}\right\|^2 \cos ^2 \theta_m \\
& \leqslant\|\boldsymbol{a}\|^2\left\|\boldsymbol{R}_m \boldsymbol{a}\right\|^2 \\
& \leqslant \delta^2 \boldsymbol{a}^\text{T} \boldsymbol{R}_m \boldsymbol{R}_m \boldsymbol{a}, \quad m=1,2, \ldots, n .
\end{align}
Here we use $\frac{\boldsymbol{a}^\text{T} \boldsymbol{R}_{\boldsymbol{m}} \boldsymbol{a}}{\|\boldsymbol{a}\|\left\|\boldsymbol{R}_m \boldsymbol{a}\right\|}=: \cos \theta_m$ and $\cos ^2 \theta_m \leqslant 1$ with $\theta_m$ representing the angle between vectors $\boldsymbol{a}$ and $\boldsymbol{R}_m \boldsymbol{a}$. The last step uses the bound on the local region $\|\boldsymbol{a}\|^2 \leqslant \delta^2$ to obtain the upper bound on $\Upsilon_m^2$ in equation \eqref{eq:local_upper_bound_upsilon_m}. 

\end{proof}

\section{\label{appendix:c} Influences of LMI result by the $\varepsilon$}
Here, we show the $\text{log}_{10}[\gamma(\text{Re},\delta)]$ and $\text{log}_{10}[f_{\text{LMI}}(\text{Re},\delta)]$ from LMI at different $\varepsilon$, which correspond to Figures \ref{fig:contour_gamma} and \ref{fig:contour_lmi}. Figure \ref{fig:epsilon_comparision_e_3} is associated with $\varepsilon=10^{-3}$, which shows more infeasible regions at high Reynolds numbers (white regions in Figure \ref{fig:epsilon_comparision_e_3}) compared with Figures \ref{fig:contour_gamma} and \ref{fig:contour_lmi} associated with $\varepsilon=10^{-4}$. Figure \ref{fig:epsilon_comparision_e_6} shows results associated with $\varepsilon=10^{-6}$, which is almost identical to Figures \ref{fig:contour_gamma} and \ref{fig:contour_lmi} associated with $\varepsilon=10^{-4}$. These results show that the value of $\varepsilon$ does not significantly influence results in this range. 

\begin{figure}[!htbp]
    \centering
    \begin{subfigure}{0.496\linewidth}
        \centering
        \includegraphics[width=\linewidth]{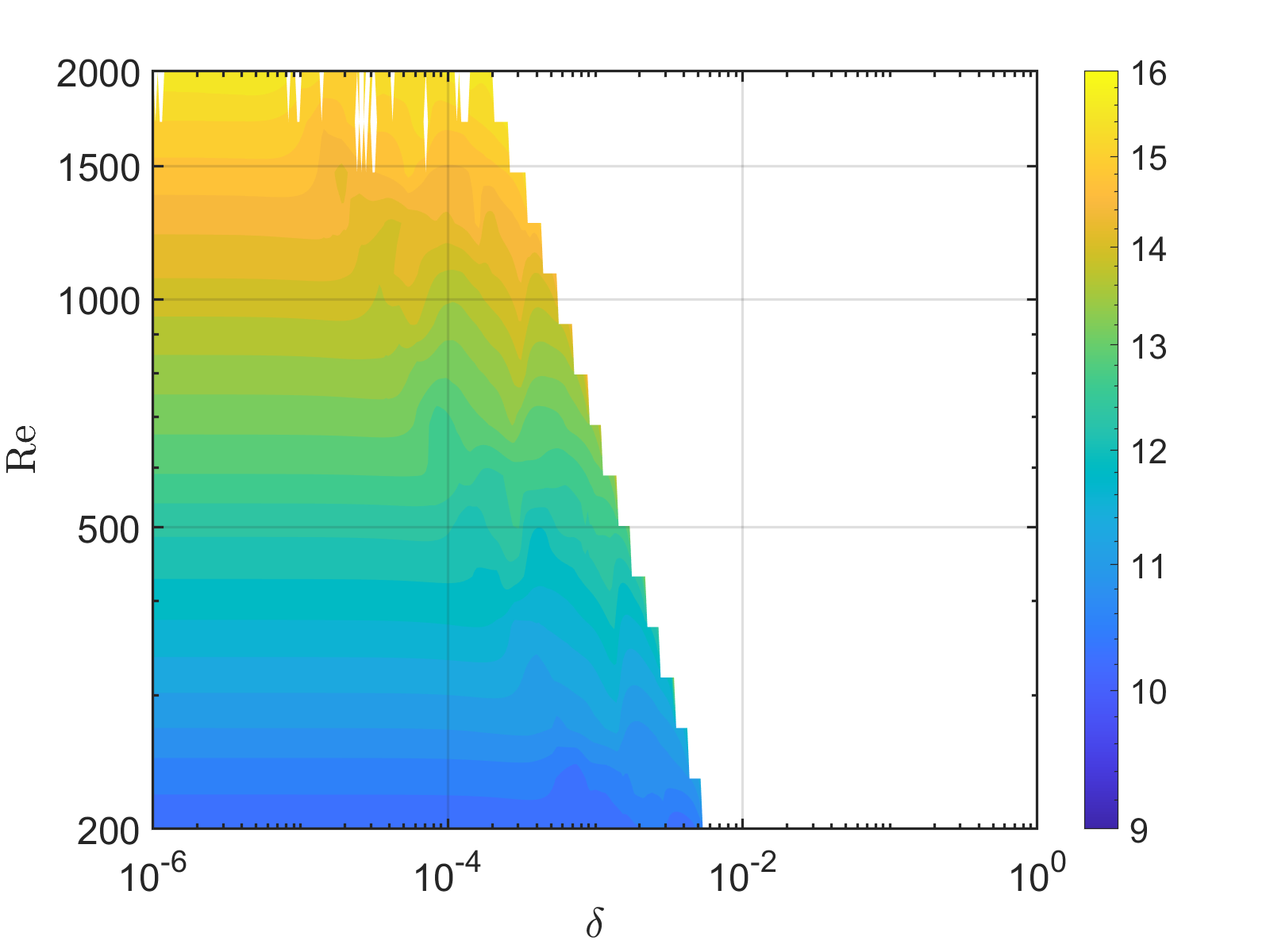} 
        \caption{$\text{log}_{10}[\gamma(\text{Re},\delta)]$ based on LMI method in Theorem \ref{thm:Liu2020PRE}.}
        \label{fig:turbulent_other_bi}
    \end{subfigure}
    \begin{subfigure}{0.496\linewidth}
        \centering
        \includegraphics[width=\linewidth]{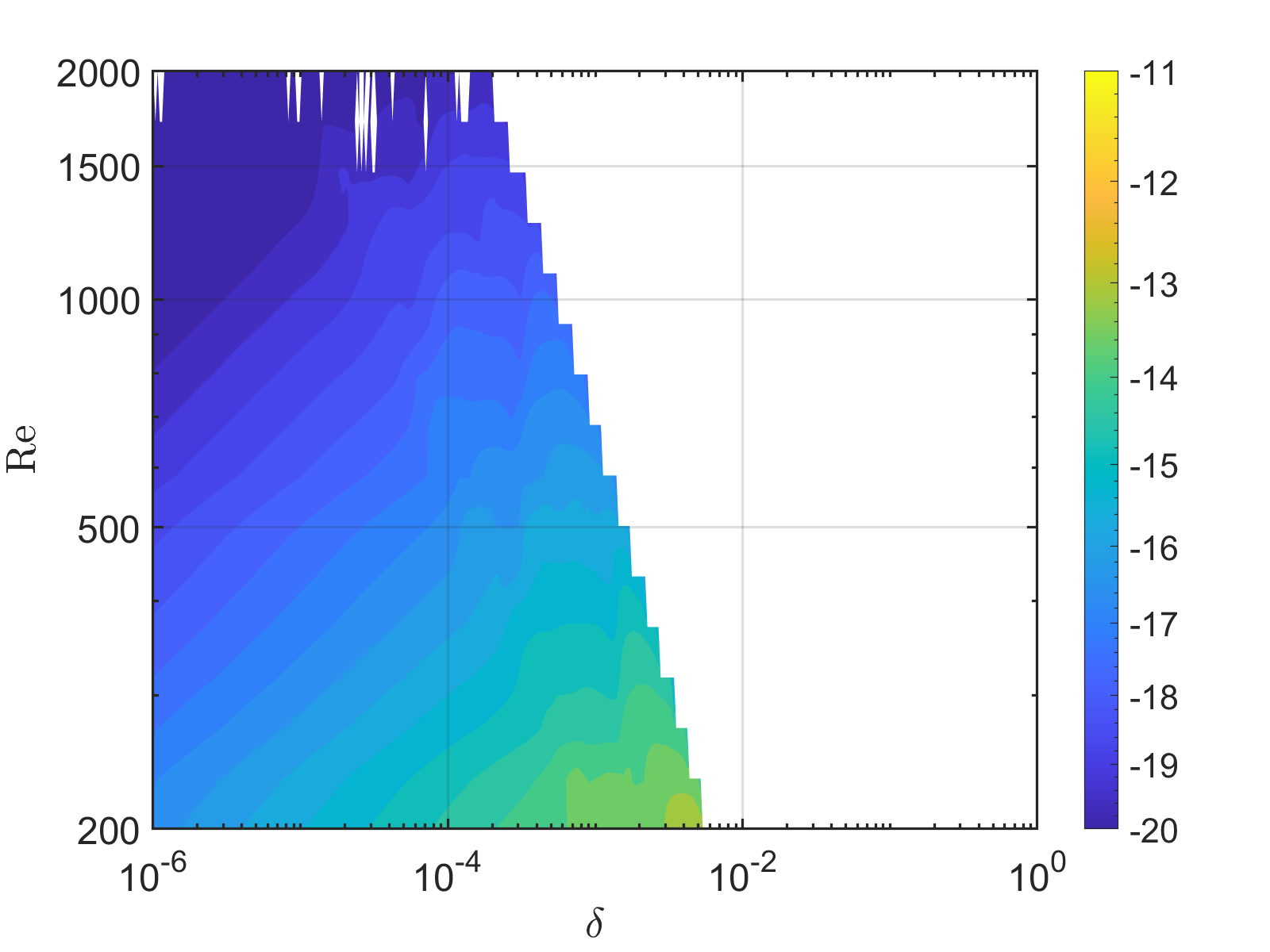} 
        \caption{$\text{log}_{10}[f_{\text{LMI}}(\text{Re},\delta)]$ based on LMI method in Theorem \ref{thm:Liu2020PRE}.}
        \label{fig:laminar_other_bi}
    \end{subfigure}
    \caption{\justifying (a) $\text{log}_{10}[\gamma(\text{Re},\delta)]$ and (b) $\text{log}_{10}[f_{\text{LMI}}(\text{Re},\delta)]$ when $\varepsilon = 10^{-3}$. The white region indicates the parameter regime where the LMI is infeasible. }
\label{fig:epsilon_comparision_e_3}
\end{figure}

\begin{figure}[!htbp]
    \centering
    \begin{subfigure}{0.496\linewidth}
        \centering
        \includegraphics[width=\linewidth]{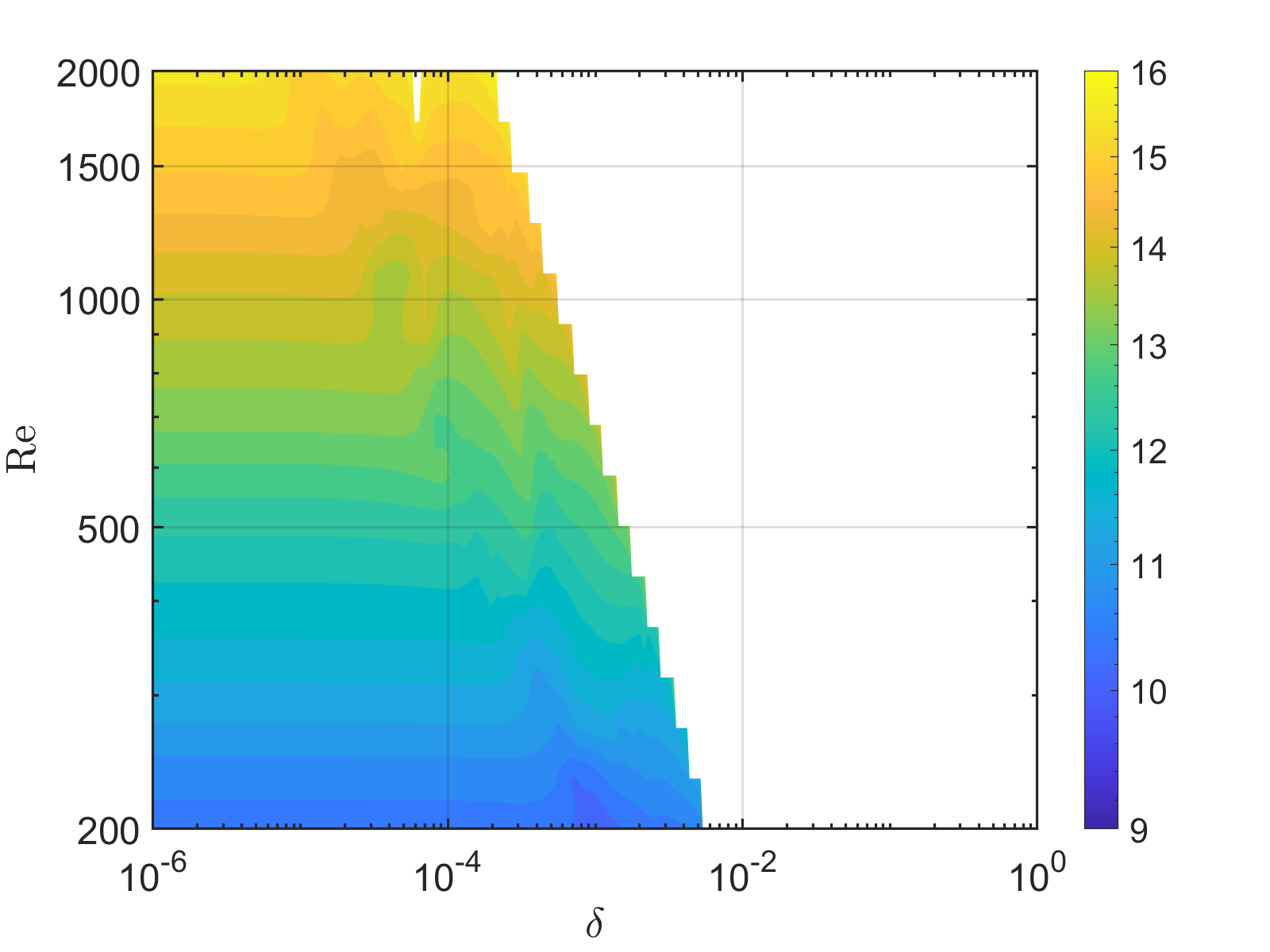} 
        \caption{$\text{log}_{10}[\gamma(\text{Re},\delta)]$ based on LMI method in Theorem \ref{thm:Liu2020PRE}.}
        \label{fig:turbulent_other_bi}
    \end{subfigure}
       \begin{subfigure}{0.496\linewidth}
        \centering
        \includegraphics[width=\linewidth]{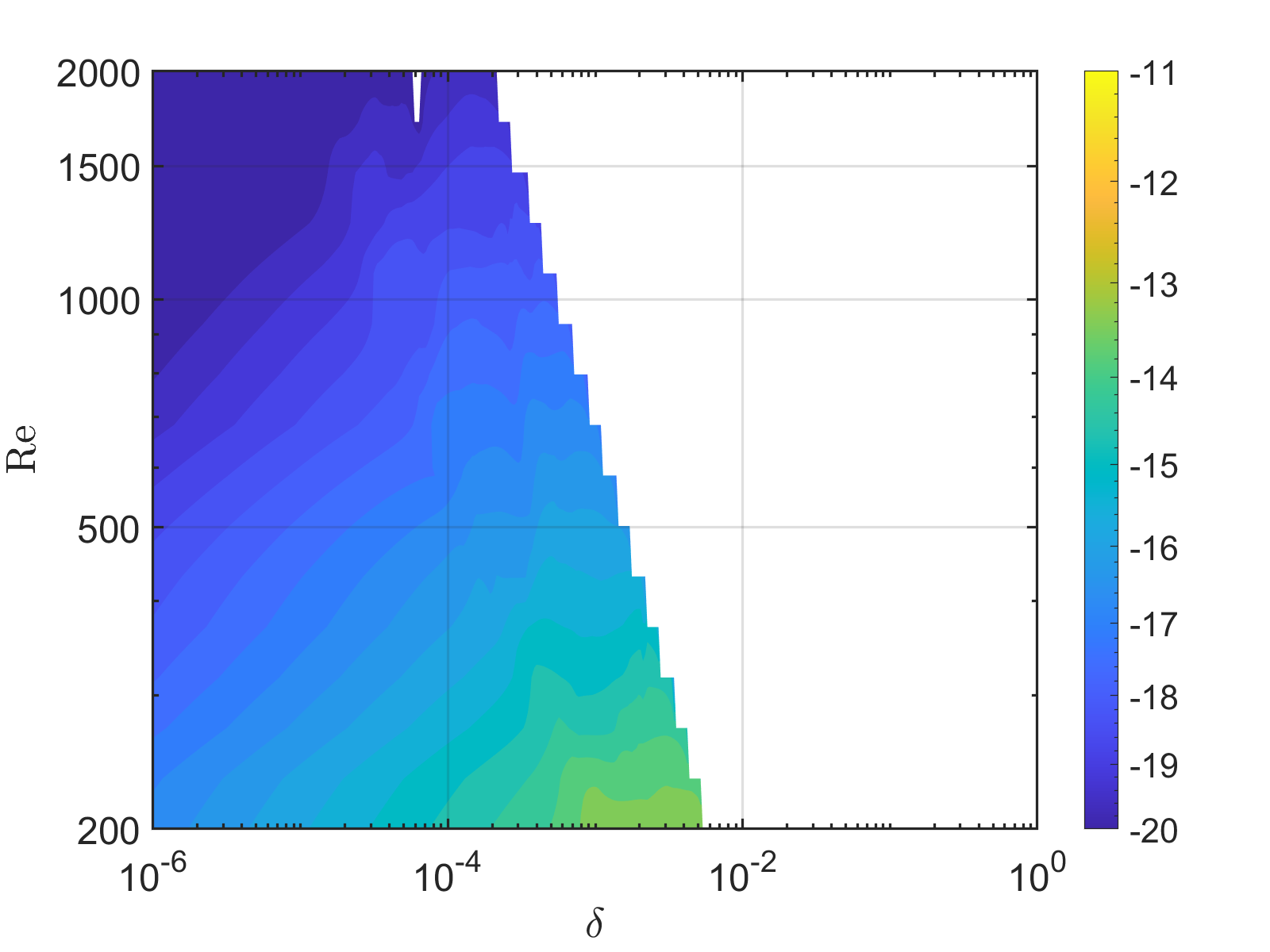} 
        \caption{$\text{log}_{10}[f_{\text{LMI}}(\text{Re},\delta)]$ based on LMI method in Theorem \ref{thm:Liu2020PRE}.}
        \label{fig:laminar_other_bi}
    \end{subfigure}
    \caption{\justifying (a) $\text{log}_{10}[\gamma(\text{Re},\delta)]$ and (b) $\text{log}_{10}[f_{\text{LMI}}(\text{Re},\delta)]$ when $\varepsilon = 10^{-6}$. The white region indicates the parameter regime where the LMI is infeasible. }
\label{fig:epsilon_comparision_e_6}
\end{figure}

\endgroup


\bibliography{apssamp}

\end{document}